\documentclass{lmcs}
\pdfoutput=1
\usepackage[utf8]{inputenc}

\usepackage{lastpage}
\lmcsdoi{21}{3}{20}
\lmcsheading{}{\pageref{LastPage}}{}{}%
{Jul.~09,~2024}{Sep.~03,~2025}{}

\title[The Size-Change Principle for Mixed Inductive and Coinductive types]%
      {The Size-Change Principle for Mixed\texorpdfstring{\\}{} Inductive and Coinductive types}

\author[P.~Hyvernat]{Pierre Hyvernat}
\address{
Universit\'e Savoie Mont Blanc, CNRS, LAMA, 73000
Chamb\'ery, France.
}
\email{pierre.hyvernat@univ-smb.fr}
\urladdr{\url{http://lama.univ-savoie.fr/~hyvernat/}}  % I added the url macro to get the ~ to render correctly

\thanks{This work was partially funded by ANR project RECIPROG, project
reference ANR-21-CE48-019-01}


\usepackage{alltt}

\usepackage{stmaryrd}
\usepackage{amssymb,amsmath}
\usepackage{cmll}
\usepackage{xspace}

\newif\ifUSEEXTERNALGRAPHICS
\USEEXTERNALGRAPHICStrue
\ifUSEEXTERNALGRAPHICS
  \usepackage{tikzexternal}
  \tikzsetexternalprefix{figures/}

\else
  \usepackage{tikz}
  \usetikzlibrary{matrix,arrows,calc,shapes.geometric,positioning,decorations.pathmorphing,decorations.pathreplacing}

  \usetikzlibrary{external}
  \tikzsetexternalprefix{figures/}
\fi

\let\nodeSize=\scriptsize
\let\transitionSize=\scriptsize
\newcommand\Set{\mathsf{Set}}

\newcommand\sleq{\sqsubseteq}
\newcommand\restrict{\upharpoonright}

\newcommand\N{\ensuremath{\mathbb{N}}\xspace}
\newcommand\Z{\ensuremath{\mathbb{Z}}\xspace}
\newcommand\Zi{\ensuremath{\mathbb{Z}_\infty}\xspace}

\newcommand\Coef{\ensuremath{\mathbb{W}}\xspace}
\renewcommand\P{\ensuremath{\mathbb{P}}\xspace}
\newcommand\V{\ensuremath{\mathcal{V}}\xspace}
\renewcommand\S{\ensuremath{\mathcal{S}}\xspace}
\newcommand\Op{\ensuremath{\mathcal{O}}\xspace}
\newcommand\A{\ensuremath{\mathcal{A}}\xspace}

\newcommand\StdSem{\Theta^{\text{\rm std}}}
\newcommand\NdtSem{\Theta^{\text{\rm ndt}}}

\newcommand\Sem[1]{\left\llbracket#1\right\rrbracket}
\newcommand\CSem[1]{\left\{\mskip-5mu\left|#1\right|\mskip-5mu\right\}}

\newcommand\cB[2]{{\left\lceil{#2}\right\rceil_{\scriptscriptstyle#1}}}
\newcommand\cD[2]{{{#2}_{\upharpoonright_{#1}}}} %amssymb
\newcommand\ccD[2]{{{#2}_{\downharpoonright_{#1}}}} %amssymb

\newcommand\comp{\circ}
\newcommand\ccomp{\diamond}

\newcommand\coef[1]{{\ensuremath{\langle#1\rangle}}}

\newcommand\Zero{{\ensuremath{\mathbf0}}\xspace}

\newcommand\Daimon{\ensuremath{\Omega}\xspace}

\newcommand\nf{\ensuremath{\mathsf{nf}}}

\newcommand\fix{\mathsf{fix}}

\newcommand\occ{\mathsf{occ}}

\newcommand\eqdef{\stackrel{\smash{\text{\sf def}}}{=}}
\newcommand\BLANK{{\texttt{\char"5F}}}

\newcommand\sqcoh{%
 \mathrel{\hbox{%
  \vrule width8pt height0pt depth0.4pt%
  \hskip-8pt
  \vrule width8pt height5pt depth-4.6pt%
  \hskip-8pt
  \vrule width0.4pt height1.5pt depth0.4pt%
  \hskip-0.4pt
  \vrule width0.4pt height5pt depth-2.9pt%
  \hskip7.6pt
  \vrule width0.4pt height1.5pt depth0.4pt%
  \hskip-0.4pt
  \vrule width0.4pt height5pt depth-2.9pt%
}}}

\renewcommand\tt[1]{\leavevmode\text{\textup{\texttt{#1}}}}
\newcommand\itt[1]{\leavevmode\text{\textit{\texttt{#1}}}}
\newcommand\call[2]{\leavevmode\tt{#1 \ensuremath{\rightsquigarrow} #2}}

\newcommand\branches{\mathsf{branches}}
\newcommand\x{{\tt{x}}\xspace}
\newcommand\y{{\tt{y}}\xspace}
\newcommand\hole{\ensuremath{{\scriptstyle\square}}\xspace}
\newcommand\f{{\tt{f}}\xspace}
\newcommand\g{{\tt{g}}\xspace}
\newcommand\C{{\tt{C}}\xspace}
\newcommand\DD[1]{{\tt{.#1}}\xspace}
\newcommand\D{{\DD{D}}\xspace}
\newcommand\m{\ensuremath{^{\tt-}}\xspace}
\newcommand\Cm{{\C\m}\xspace}
\newcommand\p[1]{{\ensuremath{^{#1}}}}

\newcommand\chariot{\texttt{chariot}\xspace}
\newcommand\bstruct{\tt{\symbol{`\{}}}
\newcommand\estruct{\tt{\symbol{`\}}}}
\newcommand\st[1]{%
 \bgroup%
  \renewcommand\DD[1]{{\tt{##1}}\xspace}%
   \renewcommand\D{{\DD{D}}}%
    \bstruct{}#1\estruct%
     \egroup}

\makeatletter
\newlength{\negph@wd}
\DeclareRobustCommand{\negphantom}[1]{%
  \ifmmode
    \mathpalette\negph@math{#1}%
  \else
    \negph@do{#1}%
  \fi
}
\newcommand{\negph@math}[2]{\negph@do{$\m@th#1#2$}}
\newcommand{\negph@do}[1]{%
  \settowidth{\negph@wd}{#1}%
  \hspace*{-\negph@wd}%
}

\DeclareRobustCommand{\dirsup}{%
  \@ifnextchar_{\sub@bigsqcupwithup}{\nosub@bigsqcupwithup}%
}

\newcommand{\nosub@bigsqcupwithup}{%
  \mathop{
    \mathchoice
      {\disp@bigsqcupwithup{}}
      {\nodisp@bigsqcupwithup{\hphantom{\scriptscriptstyle\uparrow}}}
      {\nodisp@bigsqcupwithup{\hphantom{\scriptscriptstyle\uparrow}}}
      {\nodisp@bigsqcupwithup{\hphantom{\scriptscriptstyle\uparrow}}}
  }
}
\def\sub@bigsqcupwithup_#1{%
  \mathop{
    \mathchoice
      {\disp@bigsqcupwithup{#1}}
      {\nodisp@bigsqcupwithup{#1}}
      {\nodisp@bigsqcupwithup{#1}}
      {\nodisp@bigsqcupwithup{#1}}
  }
}

\newcommand{\disp@bigsqcupwithup}[1]{%
  \negphantom{{}^\uparrow}%
  {\mathop{\hphantom{{}^\uparrow}{\bigsqcup}{}^\uparrow}\limits_{#1}}%
}
\newcommand{\nodisp@bigsqcupwithup}[1]{%
  \bigsqcup^{\scriptscriptstyle\uparrow}_{#1}%
}
\makeatother

\newcommand\Rule[2]{\frac{\phantom{\big(}\quad{\displaystyle #1}\quad}{\phantom{\Big(}\quad{\displaystyle #2}\quad}}

\newenvironment{myequation}{%
  \ignorespaces\[\everymath{\displaystyle}\begin{array}{rclr}%
    }{%
  \end{array}\]\ignorespacesafterend%
}

\newcommand\ie{i.e.\ }

\begin{document}

\begin{abstract} %%%<<<1
  This paper shows how to use Lee, Jones and Ben Amram's size-change principle
  to check correctness of arbitrary recursive definitions in an ML / Haskell
  like programming language with inductive and coinductive types.
  Naively using the size-change principle to check productivity and termination
  is straightforward but unsound when inductive and coinductive types are
  nested. We can however adapt the size-change principle to check
  ``totality''~\cite{ph:totality}, which corresponds exactly to correctness with
  respect to the corresponding (co)inductive type.

\end{abstract} %%%>>>1

\maketitle

% \begingroup\footnotesize
% \makeatletter
% \def\l@section{\@tocline{2}{0pt}{1pc}{5pc}{}}
% \def\l@subsection{\@tocline{2}{0pt}{2pc}{6pc}{}}
% \def\l@subsubsection{\@tocline{2}{0pt}{5pc}{6pc}{}}
% \def\l@paragraph{\@tocline{2}{0pt}{6pc}{6pc}{}}
% \makeatother
% \tableofcontents
% \endgroup

\section{Introduction}

One of the goals of strong typing in languages like Caml or Haskell, and the
heart of Hindley-Milner type checking / type inference, is to catch a whole
class of errors before they actually happen: if a piece of code is accepted
(at compile time), evaluation cannot fail (at run time). Of course, the
program can be incorrect but functions can only be applied to arguments they
are ready to accept. The dreaded results \tt{segmentation fault}, \tt{core
dumped} or \tt{NullPointerException} are, in theory, a thing of the past. And
even if, in practice, bindings to libraries written in other languages allow
such errors to creep back into the language, this additional guarantee is a
strong selling point.

In proof assistants based on type theory like Coq~\cite{coq_manual} or
Agda~\cite{agda_manual}, strong typing is even more important as it implies
consistency: no closed element of the empty type can be defined. However,
typing alone cannot prevent such ill-formed definitions as
{\small\begin{alltt}
  val magic = magic
\end{alltt}}\noindent
which is well-typed but belongs to all types. This recursive definition is for
example valid in Haskell and while languages like Caml or SML only allow
recursive definitions for explicit functions they accept the following
variants
{\small\begin{alltt}
  (* Caml syntax *)                        (* SML syntax *)
  # let rec magic x = magic x;;            > val rec magic = fn x => magic x;
  val magic : 'a -> 'b = <fun>             val magic = fn: 'a -> 'b
\end{alltt}}\noindent

Coq and Agda take different approaches to reject such definitions:
\begin{enumerate}

  \item
    Coq restricts a priori the syntax and type system so that only terminating
    functions can be written,

  \item
    Agda forbids some definitions a posteriori by using an external
    termination checker.

\end{enumerate}

Because the halting problem is undecidable~\cite{turing}, the first approach
cannot give a Turing complete language. The language
\tt{charity}~\cite{cockett:charity} takes a similar approach: recursion is
only available in the form of a typed combinators that enforce that some
argument of the recursive function is structurally decreasing.

In Agda however, unrestricted recursion is syntactically possible and as a
programming language, Agda is Turing complete. The termination checker warns
about well-typed definitions that may lead to non termination. The size-change
principle~\cite{lee_jones_benamram:SCT,ph:SCT} is particularly well suited for
this task. Note however that undecidability of the halting problem now implies
that the termination checker will reject some correct functions.
One advantage of this approach is that it is in theory easy to combine several
termination checkers and that this doesn't impact the underlying type theory.
The disadvantage is that the validity of a recursive definition now depends on
some external ``oracle''.

In the presence of coinductive types like streams or infinite trees, the
situation is more complex because we need to prevent infinite computations by
adding some laziness to the evaluation. Both approaches can still be used, but
the second becomes more complex. In simple cases, checking that a definition
is \emph{productive}~\cite{coquand:infinite}\footnote{Productivity of a
recursive definition means that when unfolding the definition, a coinductive
constructor is bound to be output in a finite time. Because of laziness, this
implies that no computation can go on forever.
% The rest of a computation can be done by explicitly inspecting the term
% after the coinductive constructor.
The syntactical notion of \emph{guardedness}, \ie that all recursive call
appear directly below a coinductive constructor is easily checked and implies
productivity but is very restrictive.} is enough to guarantee termination and
validity of the definition.
Unfortunately, as shown by T. Altenkirch and N. A.
Danielsson~\cite[Section~5]{altenkirch_danielsson:nested}, checking
termination and productivity independently is not enough to guarantee that a
recursive definition involving nested (co)inductive types is valid.

This paper presents a provably correct validity checker for recursive
definitions in a first order language. It relies on a characterization of
mixed inductive / coinductive types from a previous paper~\cite{ph:totality}.
I have tried to make this paper as self contained as possible but additional
motivations and references to other works can be found there.
The resulting totality checker is based on the size-change principle and
generalizes both standard termination and productivity. It cannot deal with
the full cornucopia of Agda's dependent types but possible extensions are
mentioned in the conclusion.

\subsection{\chariot}

\chariot is the prototype language developed for experimenting with the
principle described in this paper.\footnote{\chariot is written in Caml and is
freely available: \url{https://github.com/phyver/chariot}} Simply put,
\chariot is a non-strict, first-order, strongly typed, purely functional
language. Like the \texttt{Charity}
language~\cite{cockett:charity,cockett:charitable}, \chariot has both standard
inductive types and coinductive types. Implementation details for the language
\chariot are irrelevant and most ideas can be found in standard
references~\cite{peyton:implementation}.

Unlike \texttt{Charity}, functions are defined by recursion with no
restriction besides standard Hindley-Milner type
checking~\cite{milner:polymorphism}. Writing functions in \chariot is thus
closer to writing functions in ML, Haskell or Agda than it is to writing
functions in \texttt{Charity} or Coq.

\subsubsection*{Examples}

The syntax was formally described in a previous paper~\cite{ph:totality} and
won't be repeated. It should be readable by anyone familiar with Caml or
Haskell.
One thing to remember is that inductive types are given by constructors while
coinductive types are given by destructors. Here are some examples that should
give a taste of what \chariot looks like and what to expect from the totality
checker.
First, a simple example involving only inductive types:
{\small\begin{alltt}
  data nat where  Zero : nat                           --{\it unary natural numbers}
                | Succ : nat  -> nat

  data list('x) where  Nil  : list('x)                 --{\it finite lists}
                     | Cons : 'x -> list('x) -> list('x)

  val length : list(x) -> nat
    | length Nil = Zero
    | length (Cons _ l) = Succ (length l)
\end{alltt}}\noindent
The \tt{stream} datatype is coinductive with destructors giving access to the
head and tail of a stream. We can define a stream using a record notation as
in the \tt{nats} definition below.
{\small\begin{alltt}
  codata stream('x) where  Head : stream('x) -> 'x     --{\it infinite streams}
                         | Tail : stream('x) -> stream('x)

  val nats : nat -> stream(nat)
    | nats x = \st{ Head = x ; Tail = nats (Succ x) }
\end{alltt}}\noindent
The definition of~\tt{length} is structurally decreasing and the definition
of~\tt{nats} is syntactically guarded. It should come as no surprise that they
are accepted by the totality checker.
The next one is more interesting. The function adds the elements of each list
in a stream of lists using an accumulator that is reset for each list. This
definition is a little ad hoc but illustrates some strength\footnote{For
example, Agda doesn't detect that this function terminates.} of the totality
checker.
{\small\begin{alltt}
  val sums : nat -> stream(list(nat)) -> stream(nat)
    | sums acc \st{ Head = Nil ; Tail = s } =
                    \st{ Head = acc ; Tail = sums Zero s }
    | sums acc \st{ Head = Cons(n, l) ; Tail = s } =
                    sums (add acc n) \st{ Head = l ; Tail = s }
\end{alltt}}\noindent
where \tt{add} is the usual addition of natural numbers.
Even though it contains a non guarded recursive call (the second one), the
definition is still productive. Since the head of the stream gets structurally
smaller (it is a list), there can be no infinite sequence of consecutive calls
using only the second clause: the first clause must be used after a finite
time, adding a stream constructor to the result. This makes the definition
productive.

As a last example, here is a non-total recursive definition that is
nevertheless terminating and productive~\cite{altenkirch_danielsson:nested}.
{\small\begin{alltt}
  data stree where Node : stream(stree) -> stree

  val bad_s : stream(stree)
    | bad_s = \st{ Head = Node bad_s ; Tail = bad_s }
\end{alltt}}\noindent
Because the type \tt{stree} is inductive without any base constructor, it is
empty. The definition of~\tt{bad\_s} is however well typed (for
Hindley-Milner) and productive. It would unfold to a stream of non
well-founded infinitary trees. From there, it is easy to construct a ``magic''
value having all types by recursing into~\tt{bad\_s}:
{\small\begin{alltt}
  val lower_left : stree -> 'x
    | lower_left (Node s) = lower_left (s.Head)
  val magic : 'x
    | magic = lower_left bad_s.Head
\end{alltt}}\noindent
Note the important fact that the problem doesn't come from \tt{lower\_left},
which is trivially total (it is structurally decreasing on an inductive type).
It comes from the definition of~\tt{bad\_s}, which will be rejected by the
totality checker described in this paper.

\subsubsection*{Operational semantics}

Totality is a property of the \emph{denotational semantics} of a recursive
definition. Because of that, the operational semantics of \chariot is not very
important. \chariot uses a lazy evaluation of records, which is enough to
guarantee that all computation involving total functions and values are
finite.
% The reason is that totality implies productivity of record constructors, so
% that computing an infinite total value will always produce a record
% constructor, stopping computation until more information is needed. Some
% additional remarks about the denotational semantics can be found in the
% conclusion, on page~\pageref{sub:OS}.

\subsubsection*{Restrictions}

To simplify the presentation, we only consider the first order fragment of the
\chariot language, with the following restrictions:\footnote{Dealing with the
full \chariot language is possible at the cost of additional overhead.}

\begin{itemize}
  \item
    all functions are fully applied,
    % The implementation allows higher order arguments but they are ignored
    % during totality checking. Moreover, the totality checker always gives a
    % negative answer in the presence of a recursive function that is not fully
    % applied, to make it impossible to ``hide'' a recursive call like in
    % {\small\begin{alltt}
    %   val app_zero f = f Zero         --{\it not recursive}
    %   val test n = app_zero test      --{\it non terminating}
    % \end{alltt}}\noindent

  \item
    all functions and constructors take exactly one argument,

  \item
    there are no mutually defined recursive functions,

  \item
    the empty record is forbidden in recursive definitions.

\end{itemize}
The last one is a little ad hoc but makes for a simpler theory. If we insist
that all constructors have a single argument, the empty record becomes the
only atomic term. For example, the constructor~{\Zero} must have type~\tt{unit
-> nat} and is used as~``\tt{Zero\st{}}''. In recursive definitions however,
we can remove it:
\begin{itemize}
  \item
    in the pattern-matching of a clause, it can be replaced by a fresh, dummy
    variable:
    {\footnotesize\begin{alltt}
          | f (Zero \st{}) = Succ ... \end{alltt}}\noindent
    becomes
    {\footnotesize\begin{alltt}
          | f (Zero _x) = Succ ... \end{alltt}}\noindent
    which has the same semantics;

  \item
    in the right hand side of a clause, it can be
    replaced by one of those dummy variables, and
    {\footnotesize\begin{alltt}
          | f (Zero \st{}) = Succ (Zero \st{}) \end{alltt}}\noindent
    simply becomes
    {\footnotesize\begin{alltt}
          | f (Zero _x) = Succ (Zero _x) \end{alltt}}\noindent
    If no such dummy variable is present on the pattern matching side, we can
    rely on an auxiliary ad hoc function
    {\footnotesize\begin{alltt}
          | g (Succ x) = Zero (empty_record x) \end{alltt}}\noindent
    This function \tt{empty\_record}, of type~\tt{nat -> unit} cannot be
    written in our fragment but needs to be considered an outside library
    function that our test cannot inspect. Like all such functions, we'll only
    assume they are total.
\end{itemize}
With that in mind, the definition of \tt{length} we actually deal with in this
paper is
{\footnotesize\begin{alltt}
      val length : list(nat) -> nat
        | length (Nil _x) = Zero _x
        | length (Cons\st{Fst=_; Snd=l}) = Succ (length l)
\end{alltt}}\noindent
Because this transformation can easily be automatized, we will write some
examples using the unrestricted \chariot language and rely on a preprocessor
(the reader) to translate them to the restricted syntax. This will help keep
the example more familiar.

\subsection{Recap of previous work}
\label{sub:recap}

Values are built from structures and constructors but because of coinductive
types, they can be infinite~\cite{ph:totality}.
\begin{defi}\label{def:values}
  The domain~$\V$ (for ``\V{}alues'') is defined \emph{coinductively} by the
  grammar
  \[
    v \qquad::=\qquad
    \bot \quad|\quad
    \C\,v \quad|\quad
    \st{\D_1=v_1; \dots; \D_k=v_k}
  \]
  where
  \begin{itemize}
    \item
      each $\C$ belongs to a finite set of \emph{constructors},

    \item
      each $\tt{D}_i$ belongs to a finite set of \emph{destructors},

    \item
      the order of fields inside records is unimportant,

    \item
      $k$ can be 0. (Empty records are legal values, they are only forbidden
      in recursive definitions.)
  \end{itemize}

  The order on~$\V$ is inductively generated by
  \begin{enumerate}
    \item
      $\bot \leq v$ for all values $v$,

    \item
      ``$\leq$'' is contextual: if $u\leq v$ then $C[\x:=u] \leq C[\x:=v]$ for
      any value~$C$ with variable~$\x$, where substitution is defined in the
      obvious way.

  \end{enumerate}
\end{defi}
Contextuality, together with the fact that we generate an order, implies that
comparing records is done component wise: if~$u_1\leq v_1$ and~$u_2\leq v_2$,
then
\begin{myequation}
  \st{\D_1=u_1; \D_2=u_2}
  &\leq&
  \st{\D_1=v_1; \D_2=u_2}
   & \text{\footnotesize(contextuality, with $C=\st{\D_1=\x; \D_2=u_2}$)}\\
  &\leq&
  \st{\D_1=v_1; \D_2=v_2}
   & \text{\footnotesize(contextuality, with $C=\st{\D_1=v_1; \D_2=\x}$)}\\
\end{myequation}
In many cases, compatible structures are simply not comparable,
like~$\st{\D_1=\bot, \D_2=\st{}}$ and~$\st{\D_1=\st{}, \D_2=\bot}$.

Note that the set of values is defined coinductively but the order is defined
inductively: it is the \emph{least} order subject to some conditions. Because
of that, reasoning about inequalities is usually done with standard inductive
proofs.

Inductive types are interpreted by least fixed points and coinductive types
are interpreted by greatest fixed points~\cite{ph:totality}, but in this case,
they coincide in domains! A value in a given type is \emph{total} when it
belongs to the appropriate fixed point in~$\Set$. For example, the infinite
value~$\tt{Succ}^\infty=\tt{Succ}\,\tt{Succ}\,\tt{Succ}\dots$ is a valid
element of~$\V$
% (it is the limit
% of~$\bot\leq\tt{Succ}\bot\leq\tt{Succ}\tt{Succ}\bot\leq\dots$),
but is not total for the type~\tt{nat}, whose~$\Set$-based interpretation
only contains the finite natural numbers.

The main result of the previous paper~\cite{ph:totality} is an ``untyped''
characterization of total values for a given type as winning strategies for a
parity game constructed from the type. This is done by tagging each type with
a \emph{priority} that is odd for datatypes and even for codatatypes. Those
priorities are taken from a parity game computed from the types involved in
the definition. For~\tt{stream(nat)} and~\tt{stree}, they are

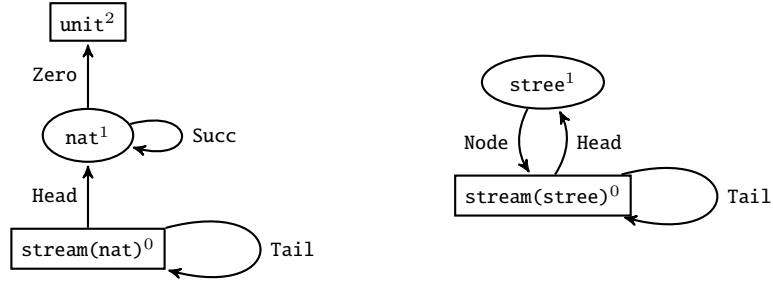
\[
\begin{tikzpicture}[->,>=stealth',shorten >=1pt,auto,node distance=1.5cm,thick,
  mu node/.style={ellipse,draw,aspect=2,font=\nodeSize},
  nu node/.style={rectangle,draw,font=\nodeSize},
  every node/.style={font=\nodeSize},
  % baseline=(stream_nat.base)
  baseline=(current bounding box.center),
  ]
  \node[nu node] (one) {$\tt{unit}^2$};
  \node[mu node] (nat) [below of=one] {$\tt{nat}^1$};
  \node[nu node] (stream_nat) [below of=nat] {$\tt{stream(nat)}^0$};
  \path[every node/.style={font=\transitionSize}]
    (stream_nat) edge node {\tt{Head}} (nat)
    (stream_nat) edge [loop right] node{\tt{Tail}} (stream_nat)
    (nat) edge node{\tt{Zero}} (one)
    (nat) edge [loop right] node{\tt{Succ}} (nat)
    ;
\end{tikzpicture}
\mskip80mu
\begin{tikzpicture}[->,>=stealth',shorten >=1pt,auto,node distance=1.5cm,thick,
  mu node/.style={ellipse,draw,aspect=2,font=\nodeSize}, nu
  node/.style={rectangle,draw,font=\nodeSize}, every
  node/.style={font=\nodeSize},
  % baseline=(stream_stree.base)
  baseline=(current bounding box.center), ]
  \node[mu node] (stree) {$\tt{stree}^1$};
  \node[nu node] (stream_stree) [below of=stree] {$\tt{stream(stree)}^0$};
  \path[every node/.style={font=\transitionSize}]
    (stree) edge [bend right] node [left] {\tt{Node}} (stream_stree)
    (stream_stree) edge [bend right] node [right] {\tt{Head}} (stree)
    (stream_stree) edge [loop right] node{\tt{Tail}} (stream_stree) ;
\end{tikzpicture}
\]
Note that the arcs for inductive constructors are reversed: the \tt{Node}
transition goes from~\tt{stree} to~\tt{stream(stree)} whereas its type
is~$\tt{stream(stree)}\to\tt{stree}$. The reason is that terms are interpreted
as strategies for the corresponding game, which entails making the transitions
deconstruct a term into its subterms.

The constraints are that datatypes [resp. codatatypes] have odd [resp. even]
priorities and that if type~$S$ is a sub-expression of type~$T$, then the
priority of~$S$ is greater than the priority of~$T$. Here, the priority
of~\tt{nat} is indeed greater than the priority of~\tt{stream(nat)}.
Because of that, priorities keep information about the type of fixed point
(inductive or coinductive type) \emph{and} their nesting, which is important
to distinguish between ``inductive of coinductive'' and ``coinductive of
inductive'' definitions.
Values of type~$T$ correspond exactly to strategies starting from~$T$. The
important result is the following.
\begin{propC}[{\cite[Corollary~2.11]{ph:totality}}]\label{prop:totality}
  A value~$v$ of type~$T$ is total iff the corresponding strategy is winning
  for the associated parity game, \ie if no branch of~$v$ contains~$\bot$ and
  along all infinite branches of~$v$, the maximal priority that appears
  infinitely often is even.
\end{propC}

Constructors and destructors appearing in a definition can be tagged with the
corresponding priority during type checking.\footnote{Formally speaking, it is
the record constructor (``curly bracket'') that is tagged with a priority. We
usually put the priority on fields for readability.} For example, the above
definitions become:

{\small\begin{alltt}
  val length : list\p1(x) -> nat\p1
    | length Nil\p1 = Zero\p1\st{}
    | length (Cons\p1\st{Fst\p0=_; Snd\p0=l}) = Succ\p1 (length l)

  val nats : nat\p1 -> stream\p0(nat\p1)
    | nats x = \st{ Head\p0 = n ; Tail\p0 = nats (Succ\p1 x) }

  val sums : nat\p1 -> stream\p0(list\p1(nat\p1)) -> stream\p0(nat\p1)\label{chariot:sums}
    | sums acc \st{ Head\p0 = Nil\p1 ; Tail\p0 = s } =
        \st{ Head\p0 = acc ; Tail\p0 = sums (Zero\p3\st{}) s }
    | sums acc \st{ Head\p0 = Cons\p1 \st{Fst\p0 = n ; Snd\p0 = l} ; Tail\p0 = s } =
        sums (add acc n) \st{ Head\p0 = l ; Tail\p0 = s }

  val bad_s : stream\p0(stree\p1)
    | bad_s = \st{ Head\p0 = Node\p1 bad_s ; Tail\p0 = bad_s }
\end{alltt}}\noindent
For the simple case of \tt{nats}, suppose~$n$ is a total natural number. The
value generated from~\tt{nats $n$} is
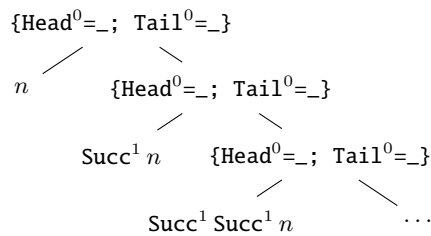
\begin{center}
  \begin{tikzpicture}[baseline=(current bounding box.center), %%%<<<3
       level distance=25pt,
       sibling distance=75pt,
       treenode/.style={inner sep=5pt},
       font=\footnotesize,
       grow=down,
     ]
     \coordinate
     node[treenode] {$\tt{\st{Head\p0=\BLANK ; Tail\p0=\BLANK}}$}
       child {node[treenode] {$n$}}
       child {node[treenode] {$\tt{\st{Head\p0=\BLANK ; Tail\p0=\BLANK}}$}
         child {node[treenode] {$\tt{Succ}\p1\,n$}}
         child {node[treenode] {$\tt{\st{Head\p0=\BLANK ; Tail\p0=\BLANK}}$}
           child {node[treenode] {$\tt{Succ}\p1\,\tt{Succ}\p1\,n$}}
           child {node[treenode] {$\dots$}}
         }
       };
  \end{tikzpicture}     %%%>>>3
\end{center}
For the infinite branch~$\tt{\st{Tail\p0=\st{Tail\p0=\st{Tail\p0=\dots}}}}$,
the maximal priority appearing infinitely often is even. All other infinite
branches would end with an infinite branch in~$n$ which is supposed to be
total. The value~$\tt{nats $n$}$ is thus total in~\tt{stream(nat)}.

\smallbreak
Contrast this with the recursive definition using lists instead of streams:
{\small\begin{alltt}
    val nats_list : nat\p3 -> list\p1(nat\p3)
      | nats_list x = Cons\p1 \st{ Fst\p0 = x ; Snd\p0 = nats_list (Succ\p3 x) }
\end{alltt}}\noindent
Here, \tt{nats\_list $n$} gives
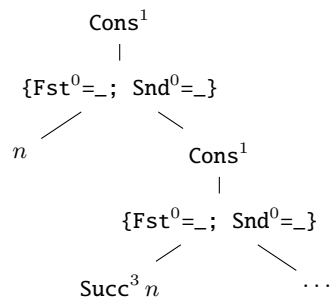
\begin{center}
  \begin{tikzpicture}[baseline=(current bounding box.center), %%%<<<3
       level distance=25pt,
       sibling distance=75pt,
       treenode/.style={inner sep=5pt},
       font=\footnotesize,
       grow=down,
     ]
     \coordinate

     node[treenode] {$\tt{Cons}\p1$} child {node[treenode] {$\tt{\st{Fst\p0=\BLANK ; Snd\p0=\BLANK}}$}
       child {node[treenode] {$n$}}
       child {node[treenode] {$\tt{Cons}\p1$} child {node[treenode] {$\tt{\st{Fst\p0=\BLANK ; Snd\p0=\BLANK}}$}
         child {node[treenode] {$\tt{Succ}\p3\,n$}}
         child {node[treenode] {$\dots$}}
         }
       }};
  \end{tikzpicture}     %%%>>>3
\end{center}
which contains the branch
$\tt{Cons\p1\st{Snd\p0=Cons\p1\st{Snd\p0=Cons\p1\st{Snd\p0=\dots}}}}$
where the maximal priority appearing infinitely often is odd. This value is not
total, as expected.

\smallbreak
The definition of \tt{bad\_s} is non total because it unfolds to
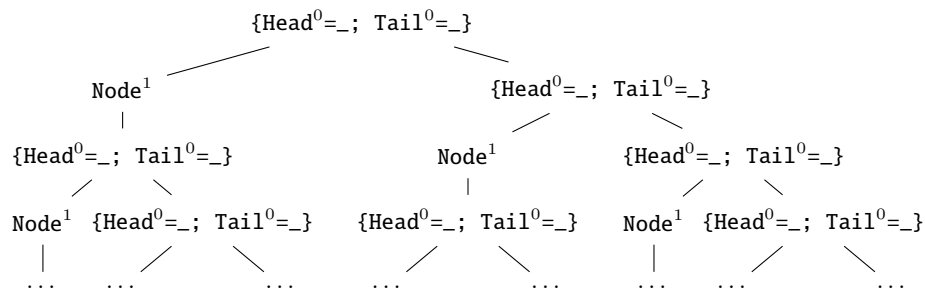
\begin{center}
  \begin{tikzpicture}[baseline=(current bounding box.center), %%%<<<3
       level distance=25pt,
       level 1/.style={sibling distance=180pt},
       level 2/.style={sibling distance=100pt},
       level 3/.style={sibling distance=60pt},
       level 5/.style={sibling distance=30pt},
       emptynode/.style={circle, inner sep=1.5pt, fill=black},
       treenode/.style={inner sep=5pt},
       font=\footnotesize,
       grow=down,
     ]
     \coordinate
     % node[treenode] {$\tt{Node}\p1$}
     %   child
       {node[treenode] {$\tt{\st{Head\p0=\BLANK ; Tail\p0=\BLANK}}$}
         child {node[treenode] {$\tt{Node}\p1$}
           child {node[treenode] {$\tt{\st{Head\p0=\BLANK ; Tail\p0=\BLANK}}$}
             child {node[treenode] {$\tt{Node}\p1$}
               child {node[treenode] {\dots}}
             }
             child {node[treenode] {$\tt{\st{Head\p0=\BLANK ; Tail\p0=\BLANK}}$}
               child {node[treenode] {\dots}}
               child {node[treenode] {\dots}}
             }
           }
         }
         child {node[treenode] {$\tt{\st{Head\p0=\BLANK ; Tail\p0=\BLANK}}$}
           child {node[treenode] {$\tt{Node}\p1$}
             child {node[treenode] {$\tt{\st{Head\p0=\BLANK ; Tail\p0=\BLANK}}$}
               child {node[treenode] {\dots}}
               child {node[treenode] {\dots}}
             }
           }
           child {node[treenode] {$\tt{\st{Head\p0=\BLANK ; Tail\p0=\BLANK}}$}
             child {node[treenode] {$\tt{Node}\p1$}
               child {node[treenode] {\dots}}
             }
             child {node[treenode] {$\tt{\st{Head\p0=\BLANK ; Tail\p0=\BLANK}}$}
               child {node[treenode] {\dots}}
               child {node[treenode] {\dots}}
             }
           }
         }
       }
       ;
  \end{tikzpicture}     %%%>>>3
\end{center}
with leftmost infinite branch
\[
  \tt{\st{Head\p0=Node\p1\st{Head\p0=Node\p1\st{\dots}}}}
\]
which is non-total.

For the rest of this paper, we assume the recursive definitions have been
tagged with priorities, \ie that all constructor [resp. destructor] names come
with an odd [resp. even] priority. All other typing information has been
removed and is irrelevant for totality. Constructors and destructors used in
values (Definition~\ref{def:values}) also carry a priority.
\begin{defi}\label{def:intrinsic_totality}
  An element~$v\in V$ is \emph{total} if it doesn't contain~$\bot$ and for all
  infinite branches of~$v$, the maximal priority appearing infinitely often in
  the branch is even.
\end{defi}
For any given list of recursive definitions, the set of priorities is finite,
so that the definition always makes sense.
We usually call the ``maximal priority appearing infinitely often'' the
\emph{principal priority}.
By Proposition~\ref{prop:totality}, those values correspond precisely to the
total values of the original type.

\bigbreak
The usual semantics of a recursive definition is a function computed using
Kleene's fixed point theorem. We call a function \emph{total} if it sends
total values to total values. This paper describe a computable totality test
taking a recursive definition as argument and answering either ``\tt{YES}, the
semantics of this definition is total'' or ``\tt{I DON'T KNOW} whether the
semantics of this definition is total or not.'' Since checking totality is
undecidable,\footnote{Recall that without coinductive types, totality is just
termination.} nothing more can be expected.

%%% intro <<<1
\subsection{Plan of the Paper} %%%<<<1

We start by giving, in Section~\ref{section:operators}, an interpretation of
recursive definitions that is mathematically simpler than the ordered lists of
clauses used in \chariot. Thanks to the characterization of mixed inductive /
coinductive types~\cite{ph:totality} recalled in Section~\ref{sub:recap}, we
can make this semantics untyped, which means we only have to consider a single
domain of values. The two steps necessary to define this interpretation are
\begin{itemize}
  \item
    instead of only considering the first matching clause, we consider all of
    them and take their non-deterministic sum,

  \item
    since a clause can now be applied to non matching value, we introduce a
    notion of error, which is nothing more than the empty non-deterministic
    sum.
\end{itemize}
Note that errors (and general sums) are just an artifact produced by the
totality checker. They are not part of the \chariot operational semantics,
where Hindley-Milner type checking is precisely meant to prevent their
apparition.

Non-deterministic values are an instance the usual Smyth power domain, but
interpreting definitions and clauses requires more care and technicalities
(Sections~\ref{sub:operators} and~\ref{sub:interpreting_def}).

\smallbreak
The resulting interpretation is still too complex: we thus split each clause
of the recursive definition into the sum of its recursive calls
(Section~\ref{sub:callgraph}). While going from the standard to the
non-deterministic sum left the semantics ``mostly'' unchanged, splitting a
definition into its call-graph results in a very different semantics. That's
not a problem because, as shown in Proposition~\ref{prop:Gt_total}, this
simplification reflects totality.

\smallbreak
The last step before applying the size-change principle, detailed in
Sections~\ref{sub:weight} and~\ref{sub:collapsing}, is to show how we can
collapse the call-graph inside a finitary structure by introducing
approximations that forget about parts of the terms. Doing so in a consistent
way introduces subtle difficulties, like composition of terms becoming
non-associative.

\smallbreak
Everything is then in place to apply the size-change principle from C.~Lee,
N.~Jones and A.~Ben-Amram~\cite{lee_jones_benamram:SCT}. This is done in
Section~\ref{sub:SCP}.

The paper then gives detailed examples to show how the totality checker
reaches its conclusion and some remarks about the actual implementation of the
totality checker.
%%%>>>1

%% vim600:set foldmarker=<<<,>>> foldmethod=marker fileencoding=ascii spelllang=en spell: %%

\section{Non-Deterministic Semantics for Definitions}\label{section:operators}

%%% intro
%%%<<<1
A \chariot recursive definition is a complex object: an ordered set of clauses
accepted by the Hindley-Milner type checking algorithm. We first interpret
them in a standard mathematical structure (a domain) with a simple syntactical
representation. The key ideas are to use a non-deterministic (commutative) sum
of untyped clauses to replace ordered set of clauses and to allow runtime
errors in the model.
%%%>>>1

\subsection{Smyth Power Domain} %%%<<<1

A \emph{domain} will be an \emph{algebraic DCPO}. Basic definitions and
important results about domain theory are recalled in
Appendix~\ref{app:domain}.
An important tool for constructing domains is the \emph{ideal completion}
which transforms any partial order into an algebraic DCPO whose compact
elements are exactly the elements of the original partial order. Refer
to~Appendix~\ref{app:domain} for details.

The Smyth power domain construction \cite{smyth:power_domain} adds a binary
greatest lower bound operation~``$+$'' to any domain. It is similar to adding
arbitrary lower bounds for a partial order by considering upper closed sets
ordered by reverse inclusion: instead of taking all upper closed sets, only
some of them are used. Refer to Appendix~\ref{app:smyth} for some details and
additional references. The important point is that any element of the Smyth
power domain can be seen as a formal sum of elements of the starting domain,
ordered by
\[
  \sum_i u_i \leq \sum_j v_j
  \quad\text{iff}\quad
  \forall j, \exists i, u_i \leq v_j
  \ .
\]
\begin{defi}
  Let~\S (for ``\S{}ums'', or ``\S{}myth'') be the domain obtained from~\V by:
  \begin{enumerate}
    \item taking the Smyth power domain construction over~\V,
    \item adding a greatest element~\Zero, which can be identified with the
      empty sum.
  \end{enumerate}
\end{defi}

\begin{lem}
  The greatest element \Zero is neutral for~``$+$''.
\end{lem}
\begin{proof} % I replaced the \proof ... \qed macros with the proof environment to get a correct spacing. If you do not agree with this, let us know
  Note that adding \Zero as a greatest element to a domain is always possible:
  \begin{itemize}

    \item
      \Zero will be compact because any limit that doesn't contain it exists
      in the original domain, and is thus different from~\Zero;

    \item
      a directed set containing \Zero has \Zero as a limit, and a directed set
      not containing \Zero has a limit in the original domain;

    \item
      any element different from \Zero is the limit of the compact elements
      below it (because that's the case in the original domain) and since
      \Zero is compact, it is the limit of the compact elements below it.

  \end{itemize}
  That \Zero is neutral for~$+$ is a direct consequence of~$+$ being the
  greatest lower bound operation. Refer to Appendix~\ref{app:smyth} for
  details.
\end{proof}

The domain~\S contains all the original values from~\V, which we call
\emph{simple values}. All its elements are ``formal'' sums of elements of~\V.
Those formal sums can be empty (\Zero), unary (simple values), finite
(greatest lower bounds) or infinite. Only infinite sums that can be obtained
as limits of finite sums exist. For example, the sum of all maximal
elements of~\tt{nat}, \ie~$\tt{Zero}+ \tt{Succ Zero} + \tt{Succ Succ Zero} +
\cdots$ is the limit of
\[
  \bot \quad\leq\quad \tt{Zero}+\tt{Succ }\bot \quad\leq\quad \tt{Zero}+\tt{Succ Zero} + \tt{Succ
  Succ }\bot \quad\leq\quad \dots
\]
However, the sum of all \emph{total} elements of~\tt{nat}, \ie the same sum
\emph{without}~$\tt{Succ}^\infty$ doesn't exist in~$\V$.
Totality on~\S is defined in the expected way.
\begin{defi}
  An element~$t\in\S$ is total if all its summands are total in~\V.
\end{defi}
The value~\Zero will be used as the semantics for runtime errors. It may be
surprising that errors are total but because type checking precisely implies
that those error never happen when running actual programs, they can be seen
as be artefacts introduced by the totality checking mechanism.

As previously shown (Lemma~1.8, \cite{ph:totality}), total elements of~$\V$
are maximal. The next lemma is a direct corollary and implies that totality is
compatible with the pre-order on~\S: if $t_1 \approx t_2$, then~$t_1$ is total
iff~$t_2$ is.
\begin{lem}\label{lem:total_closed}
  \leavevmode
  \begin{itemize}
    \item
      If $t_1\leq t_2$ in~$\S$ and if~$t_1$ is total, then so is~$t_2$.
    \item
      if $f\leq g$ in $\S\to\S$ (for the pointwise order) and~$f$ is total, then so is~$g$.
  \end{itemize}
\end{lem}
\begin{proof}
Let $t_1=\sum T_1$ and~$t_2 = \sum T_2$. Writing~$X^\uparrow$ for the upward
closure of~$X$, we have $t_1\leq t_2$ iff~$T_2^\uparrow \subseteq
T_1^\uparrow$. If~$T_1$ is total, it only contains total elements and by
Lemma~1.8 from~\cite{ph:totality}, $T_1^\uparrow = T_1$. As a
result~$T_2^\uparrow$ only contains total elements and~$T_2$ only contains
total elements as well.
The second point follows directly, as a total function is simply a function
sending total elements to total elements. \end{proof}
%%%>>>1

\subsection{Recursion and Fixed Points} %%%<<<1
\label{sub:ndt_semantics}

\subsubsection*{A formula for fixed points} %%%<<<2

Whenever $\varphi: D\to D$ is a continuous function on a domain and~$b\in D$
such that~$b\leq\varphi(b)$, it has a least fixed point greater than~$b$. This
fixed point is equal to (Kleene theorem)
\[
  \fix(\varphi, b) \quad=\quad \dirsup_{n\geq0} \varphi^n(b)
\]
We are interested in fixed points of operators from~$[\S\to\S]$ to itself and
we require that all functions satisfy $f(\Zero) = \Zero$, \ie that errors
propagate.
We write~\Daimon for
\[
  v \quad\mapsto\quad \Daimon(v) = \begin{cases}
    \Zero & \text{if $v=\Zero$}\\
    \bot & \text{otherwise}
  \end{cases}
\]
All the fixed points we are computing are of the form
\[\label{formula:fixed_point}
  \fix(\varphi, \Daimon) \quad=\quad \dirsup_{n\geq0} \varphi^n(\Daimon)
\]
with~$\varphi : [\S\to\S]\to[\S\to\S]$. They will simply be denoted by~$\fix(\varphi)$.
The following is a direct consequence of Kleene's formula.
\begin{lem}\label{lem:kleene_monotonic}\leavevmode
  If $\Daimon\leq\theta(\Daimon)$ and $\theta\leq\phi$ in $[\S\to\S] \to [\S\to\S]$, then
  $\fix(\theta)\leq\fix(\phi)$ in $\S\to\S$.
\end{lem}
%%>>>2

\subsubsection*{Non-deterministic semantics} %%%<<<2

We extend the standard semantics of \chariot functions to accept arbitrary
values in~\V. Overlapping clauses introduce non-determinism and the empty
sum~\Zero naturally arises when no clause matches a value.

Recall that \emph{linear patterns} are inductively generated by the following
grammar
\[
  p \qquad::=\qquad
  \x \quad|\quad%
  \C\,p \quad|\quad%
  \st{\D_1=p_1; \dots; \D_n=p_n}
\]
with the restriction that variables occur at most once.
The standard semantics (\cite[Definition~1.10]{ph:totality}) for a recursive
definition of~\f is the fixed point of the following operator:
\[
    \StdSem_{\rho,\f}(f)\big(v\big) =  \Sem{u[p:=v]}_{\rho,\f:=f}
\]
where ``\tt{\f $p$ = $u$}'' is the first clause from the definition of~\f
where~$p$ matches~$v$. When~\f is of type~$A\to B$, its standard semantics is
a function from the interpretation of~$A$ to the interpretation of~$B$.

The fact that we use the first matching clause means in particular that
clauses are not independent of each other: for example, the pattern
{\small\begin{alltt}
    | f Zero = Zero \end{alltt}}\noindent
doesn't necessarily say anything about the value of~\tt{f} on~\tt{Zero}, as it
could come after
{\small\begin{alltt}
    | f x = Succ Zero
\end{alltt}}\noindent
One way to break this dependency on the order of clauses is to use
non-determinism. Rather than take the first matching clause, we take the sum
of all clauses. Non matching clause evaluate to~\Zero and do not contribute to
the sum and non overlapping clauses do not interfere with each other: at most
one term is non~\Zero. When there are overlapping clauses, some information is
lost. An extreme case would be
{\small\begin{alltt}
  val f : nat -> nat
    | f x =  Zero
    | f Zero = f Zero\end{alltt}}\noindent
The second clause is never used when evaluating~\tt{f Zero} because~\tt{Zero}
matches the first clause. With the non deterministic semantics, \tt{f Zero}
would evaluate to~$\tt{Zero}+\tt{f Zero}$, which would loop: the semantics
of~\tt{f Zero} would thus be~$\bot$. Fortunately such example are very rare in
practice, and the advantages of this simplification more than compensate for
that. The concept of ``operators'' (Section~\ref{sub:operators}) and
call-graph (Section~\ref{sub:callgraph}) are based on this.

Formally, the standard semantics of some~\f of type~$A\to B$ is extended to
the whole of~\S, which contains all terms: those in the interpretations
of~$A$,~$B$ and all possible types.
\begin{defi}\label{def:NdtSem}\leavevmode
  \begin{enumerate}
    \item
      Given a linear pattern~$p$ and a simple value~$v$, the unifier~$[p:=v]$
      is the substitution defined inductively with
      \begin{itemize}
        \item
          $[\y := v] = [\y := v]$ where the RHS is the usual substitution
          of~\y by~$v$,

        \item
          $[\C p:= \C v] = [p:=v]$,

        \item
          $[\st{\D_1=p_1; \dots; \D_n=p_n} := \st{\D_1=v_1; \dots; \D_n=v_n}]
          \ =\ [p_1:=v_1] \cup \cdots \cup [p_n:=v_n]$ (because patterns are
          linear, the unifiers don't overlap),

          \item
          in all other cases, the unifier is the substitution giving \Zero for
          all variables. Those cases are:

          \begin{itemize}
            \item
              $[\C p := \C' v]$ with~$\C\neq\C'$,

            \item
              $[\st{\dots}:=\st{\dots}]$ when the 2 records have different
              sets of fields,

            \item
              $[\C p := \st{\dots}]$ and $[\st{\dots}:= \C v]$.
          \end{itemize}
      \end{itemize}
      If only the first three cases are encountered when computing~$[p:=v]$,
      we say that \emph{the value~$v$ matches the pattern~$p$}.

    \item
      Given a recursive definition for \f and an environment~$\rho$ for all
      other functions, define the \textbf{n}on-\textbf{d}e\textbf{t}erministic
      semantics~$\NdtSem_{\rho,\f}:[\S \to \S] \to [\S \to \S]$ as follows.
      Suppose~$f : \S \to \S$,
      \begin{itemize}
        \item
          $\NdtSem_{\rho,\f}(f)\big(\sum v\big) = \sum\NdtSem_{\rho,\f}(f)(v)$.

        \item
          For $v\in\V$ a simple value, define
          \[
          \NdtSem_{\rho,\f}(f)(v) \quad=\quad \sum_{\tt{f $p$ = $u$}}
          \Sem{u[p:=v]}_{\rho,\f:=f}
          \]
          where the sum ranges over all clauses of the definition.
    \end{itemize}

    \item The non-deterministic semantics of the function~\f is
      then~$\fix\big(\NdtSem_{\rho,\f}\big) : \S \to \S$.
  \end{enumerate}
\end{defi}

Because there is no guarantee that~$v$ matches some clause of the definition,
we can have~$\NdtSem_{\rho,\f}(f)(v) = \Zero$.
Even with matching clauses, this semantics does not necessarily coincide with
the standard one. For example, consider the following two definitions of the
halving function:
{\small\begin{alltt}
  val half1 : nat -> nat
    | half1 (Succ (Succ n)) = Succ (half1 n)
    | half1 (Succ Zero)     = Zero
    | half1 Zero            = Zero
\end{alltt}}\noindent
and
{\small\begin{alltt}
  val half2 : nat -> nat
    | half2 (Succ (Succ n)) = Succ (half2 n)
    | half2 n               = Zero
\end{alltt}}\noindent
Because patterns are disjoint in the first definition, the order of clauses is
not important. On natural numbers, the standard and non-deterministic
semantics coincide. For the second definition however, we get different
semantics:
\[\begin{array}{rcl}
  \StdSem_{\tt{half2}}(f)(\tt{Succ Succ } n)
  &=&
  \tt{Succ} (f(n))\\
  &\neq& \\
  \NdtSem_{\tt{half2}}(f)(\tt{Succ Succ } n)
  &=&
  \tt{Zero} + \tt{Succ} (f(n))
\end{array}\]
The additional ``\tt{Zero}'' comes from the interpretation of clause
``\tt{half2 n = Zero}'' in the definition of~\tt{half2}, which can be applied
to the argument~$\tt{Succ Succ } n$.

To formally compare the two semantics, we extend typed functions to~\S.
\begin{lem}\leavevmode
  If~$f:\Sem{A}\to\Sem{B}$ is a continuous function between the interpretations of 2
  types, define~$\widehat{f}:\S \to \S$ by
  \begin{itemize}
    \item
      $\widehat{f}\big(\sum v) = \sum \widehat{f}(v)$, where $\sum v$ is a sum
      of simple terms;

    \item
      $\widehat{f}(v) = \begin{cases}
        f(v) & \text{if $v\in \Sem{A}$}\\
        \Zero & \text{if $v\notin \Sem{A}$}\\
      \end{cases}$ where~$v$ is any simple term in~\S.

  \end{itemize}
  We have
  \begin{itemize}
    \item
      $\widehat{f}$ is continuous iff $f$ is continuous,

    \item
      $\widehat{f}$ is total iff $f$ is total.
  \end{itemize}
\end{lem}
\noindent
Extending functions in this way doesn't change their standard (typed) fixed
points or totality:
\begin{lem}
  For a definition of \f of type~$A\to B$, an environment~$\rho$ and $f :
  \Sem{A} \to \Sem{B}$, let $\widehat{\StdSem_{\rho,\f}}$ be the lifting of
  the usual semantics of~\f. We have
  \begin{enumerate}
    \item
      $\StdSem_{\rho,\f}(f) = \widehat{\StdSem_{\widehat\rho,\f}}(\widehat
      f)\restrict\Sem{A}$
      (\ie ``the standard semantics is the restriction of its lifting''),

    \item
      $\fix(\StdSem_{\rho,\f}) =
      \fix\big(\widehat{\StdSem_{\widehat\rho,\f}}\big)\restrict\Sem{A}$,
      (``the standard fixed point is the restriction of the
      lifted fixed point''),

    \item
      if $\fix\big(\widehat{\StdSem_{\widehat\rho,\f}}\big)$ is total, then so
      is $\fix(\StdSem_{\rho,\f})$.

  \end{enumerate}
\end{lem}
\begin{proof}
  The first point is straightforward as the lifting of a function gives the
  same (typed) result as the unlifted function on typed values. The second
  point follows from Kleene's formula for computing the fixed point: each
  ${\StdSem_{\rho,\f}}^n(\Daimon)$ is equal to
  $\widehat{\StdSem_{\widehat\rho,\f}}^n(\Daimon)\restrict \Sem{A}$, and their
  limits are thus equal. The third point follows from the fact that outside
  their types, lifting take the value~\Zero, which is total.
\end{proof}

\begin{lem}\label{lem:incr_total}
  Given a recursive definition for \f and environment~$\rho$
  satisfying~$\rho(\g) \geq \Daimon$ for all function names~\tt{g}, we have

  \begin{enumerate}
    \item
      $\Daimon \leq \NdtSem_{\rho,\f}(\Daimon)$,

    \item
      $\NdtSem_{\rho,\f}(f) \leq \widehat{\StdSem_{\rho,\f}}(f)$ for any
      function~$f:\S\to\S$,

    \item
      If $\fix(\NdtSem_{\rho,\f}):\S\to\S$ is total
      then~$\fix\big(\widehat{\StdSem_{\rho,\f}}\big):\S\to\S$ is total as
      well.

  \end{enumerate}
\end{lem}

\begin{proof}

  The first point is straightforward and the third point is a direct
  consequence of Lemma~\ref{lem:kleene_monotonic} and
  Lemma~\ref{lem:total_closed}. For the second point, the only places
  where~$\widehat{\StdSem_{\rho,\f}}$ and~$\NdtSem_{\rho,\f}$ differ are
  \begin{itemize}
    \item
      for values of the appropriate type,~$\widehat{\StdSem_{\rho,\f}}$ only
      uses the first matching clause while~$\NdtSem_{\rho,\f}$ takes the sum
      over all clauses,

    \item
      for values outside the appropriate type,~$\widehat{\StdSem_{\rho,\f}}$
      returns~\Zero.
  \end{itemize}
  In both cases, $\widehat{\StdSem_{\rho,\f}}$ is greater than~$\NdtSem_{\rho,\f}$.
\end{proof}

\smallbreak
As a corollary, we can forget about the standard semantics and
show totality of~$\NdtSem_{\rho,\f}$.
\begin{cor}\label{cor:stdsem_ndtsem}
  For a definition of \f of type~$A \to B$,
  if~$\fix\big(\NdtSem_{\rho,\f}\big) : \S \to \S$ is total, then so
  is~$\fix\big(\StdSem_{\rho,\f}\big) : \Sem{A} \to \Sem{B}$.
\end{cor}
%%%>>>2

%%%>>>1

\subsection{Operators} %%%<<<1
\label{sub:operators}

\subsubsection{Terms} %%%<<<2

The operators~$\NdtSem_{\rho,\f}$ are continuous functions from~$[\S \to\S]$
to itself. This section introduces an inductively generated language
containing them.
\newpage
\begin{defi}\label{def:F0}
  \leavevmode
  \begin{enumerate}
    \item
      $\Op_0$ (for ``\Op{}perators'') is the set of terms inductively
      generated from
      \begin{myequation}
        t &\quad ::= \quad&
          \C\p p t \quad|\quad
          \st{\D_1=t_1; \dots; \D_n=t_n}\p p \quad|\quad \\ &&
            \C\p p\m t \quad|\quad
            \D\p p t \quad|\quad\\&&
            \f\ t \quad|\quad
            \x \quad|\quad\\&&
            \Daimon t\quad|\quad \\
            &&t_1 + \cdots + t_n
      \end{myequation}
      \emph{where~$n>0$, \x is the only possible variable name} and each~\f
      belongs to a finite set of function names.
      Each~\C and~\tt{D} comes from a finite set of constructor and destructor
      names and each~$p$ comes from a finite set of priorities (natural
      numbers). Those priorities are odd for constructors and even for
      destructors.

    \item
      Sums can be empty, in which case they are written~\Zero.

    \item
      Terms are quotiented by associativity, commutativity and idempotence
      of~$+$, together with (multi)linearity of all term constructors (\C,
      \st{\dots; \D=\_; \dots}, \Cm, \D, \f, \Daimon).

    \item
      An element~$t\in\Op_0$ is called \emph{simple} if it contains no
      sum (empty or otherwise).
%
      % We usually write~$\Daimon$ for~$\Daimon\x$.

  \end{enumerate}
\end{defi}
Since all term constructors are linear, any term containing~\Zero and no other
sum is automatically equal to~\Zero, which is not simple.
The full semantics of terms will be given on page~\pageref{def:CSem}, but in
the meantime, it is helpful to keep the following in mind.

\begin{itemize}

  \item
    Constructors $\C t$ and $\st{\D_1=t_1; \dots; \D_k=t_k}$ construct values
    directly, just like in~\V or~\S.

  \item
    Destructors $\Cm t$ and $\D\,t$ deconstruct values (in~\V) respectively
    by:
    \begin{itemize}
      \item
        doing a pattern matching
        %``\tt{match $t$ with C$u$ => $u$}''
        \emph{which fails} if~$t$ doesn't start with constructor~\C,
      \item
        projecting a structure on field~$\tt{D}$ which fails if~$t$ is not a
        structure with field~\tt{D}.
    \end{itemize}

  \item
    ``${}+{}$'' is a non deterministic sum and the empty sum \Zero represents
    an error.

  \item
    \f,~\g, \dots are function names and are either the function being
    recursively defined or a previously defined function meant to be replaced
    by a real function from the environment.

  \item
    Each~$t\in\Op_0$ represents a function depending on~$\x$.

  \item
    If we identify~\f as the recursive function being defined,
    each~$t\in\Op_0$ can be seen as a function on~$[\S\to\S]\to[\S\to\S]$,
    whose fixed point is what interests us.

  \item
    $\Daimon t$ represents the function
  $v \mapsto \begin{cases}
    \Zero & \text{if $t(v)=\Zero$}\\
    \bot & \text{otherwise}
  \end{cases}$
\end{itemize}
Because of the interaction between projections, partial matches, constructors
and records, the order on~$\Op_0$ is more complex than on~\S.
\begin{defi}\label{def:order_F0}
  The order~$\leq$ on~$\Op_0$ is inductively generated from
  \begin{itemize}

    \item
      \Zero is the greatest element: $\forall t\in\Op_0, t \leq \Zero$,
      % and~$\Daimon\x$ is the least element: $\forall t\in\Op_0, \Daimon\x\leq t$,

    \item
      contextuality: if $C\in\Op_0$ is a context, then $t_1\leq t_2 \implies
      C[t_1]\leq C[t_2]$,\footnote{Contexts are terms possibly containing a
      special variable~\hole and~$C[t]$ is the result of
      substituting~$\hole$ by~$t$.}

    \item
      $\Daimon \x \leq \x$,
      and~$\Daimon \x \leq \f\x$ for each function name~\f,

    % \item
    %   $\Daimon$ is the smallest element: $\forall t\in\Op_0, \Daimon\leq t$,

    \item
      $s+t \leq t$,

  \end{itemize}
  together with the following inequalities (``$u\approx v$'' means ``$u\leq v$
  and $v\leq u$''):

  \[ (*)\left\{ \begin{array}{lrcll}
    (1) & \C\m\C t &\approx& t \\
    (1) & \D_{i_0}\st{\dots; \D_i=t_i; \dots} &\geq& t_{i_0}\\
    \noalign{\smallbreak}
    (2) & \C\m\st{\dots} &\approx& \Zero\\
    (2) & \D\C t &\approx& \Zero\\
    (2) & \D\st{\dots} &\approx& \Zero  & \text{if the record has no field~\tt{D}} \\
    (2) & \C\m\C' t &\approx& \Zero & \text{if $\C \neq \C'$} \\
    \noalign{\smallbreak}
    (3) & \Cm\Daimon t &\approx& \Daimon t\\
    (3) & \D\Daimon t &\approx& \Daimon t\\
    (3) & \Daimon\C t &\approx& \Daimon t\\
    (3) & \Daimon\st{\D_1=t_1; \dots; \D_n=t_n} &\geq& \Daimon t_1 + \cdots +
    \Daimon t_n & \text{if $n>0$}\\
    (3) & \Daimon\Daimon t &\approx& \Daimon t\\
  \end{array}\right.\]
\end{defi}
Groups~(1) and~(2) correspond to the intended operational semantics of the
language.
% Even in the case of a strongly typed language like \chariot, errors
% introduced in group~(2) are necessary because we need to deal with applying
% a clause to a non-matching value.
Group~(3) contains (in)equalities that hold semantically and will be justified
a posteriori by Definition~\ref{def:CSem} and Lemma~\ref{lem:CSem}.

Note in particular that projecting a structure on one of its fields (second
inequality) yields a \emph{smaller} term. This accounts for the fact that
projecting may ``hide'' errors that could have occurred in other fields, as in
\begin{myequation}
  \Zero
  &\quad\approx\quad&
  \D_1\st{\D_1=\x; \D_2=\Zero}
  &\text{\footnotesize(linearity)}
  \\
  &\quad\approx\quad&
  \D_1\st{\D_1=\x; \D_2=\C\m\C'...}
  &\text{\footnotesize(fourth rule $(2)$)}
  \\
  &\quad\geq\quad&
  \x
  &\text{\footnotesize(second rule $(1)$)}
\end{myequation}
Lemma~\ref{lem:CSem_reduction} shows this is indeed the only
possibility.

That the pre-order is ``generated'' from the above (in)equalities means that
proving properties of the order can be done by induction. Any~$s\leq t$ either
comes from an (in)equality from the definition, or from reflexivity or
transitivity. It is not obvious at first but
corollary~\ref{cor:leq_non_trivial} will show that~$\leq$ doesn't collapse to
a trivial pre-order.
Here are some simple consequences of the definition.
\begin{lem}\label{lem:simple_consequences}
  We have:
  \begin{enumerate}

    \item
      If for all~$j$, there is an~$i$ s.t. $s_i \leq t_j$, then~$\sum_i s_i
      \leq \sum_j t_j$,

    \item
      $\Daimon s \leq t$ iff $\Daimon s \leq \Daimon t$,

    \item
      $\Daimon t \leq \Daimon\Cm t$,

    \item
      $\Daimon t \leq \Daimon \D t$,

    \item
      for all~$t$, $\Daimon\x \leq t$.
  \end{enumerate}
\end{lem}
\begin{proof}\leavevmode
\begin{enumerate}
  \item
    It is a direct consequence of contextuality and the fact that~$s+t \leq
    t$. The special case where~$\sum_j t_j$ is the empty sum follows from the
    fact that~\Zero is the greatest element.

  \item
    Suppose $\Daimon s \leq t$, we have $\Daimon \Daimon s \leq \Daimon
    t$ by contextuality, and since~$\Daimon\Daimon s \approx \Daimon s$,
    we have~$\Daimon s \leq \Daimon t$. The converse is a consequence of
    transitivity and the fact that~$\Daimon t \leq t$.

  \item
    Because~$t \geq \Daimon t$, we have $\Daimon \Cm t \geq \Daimon \Cm\Daimon
    t$ by contextuality; and since~$\Cm\Daimon t \approx \Daimon t$, we
    have~$\Daimon \Cm t \geq \Daimon \Daimon t\approx\Daimon t$.

  \item
    The third point is proved similarly.

  \item
    The last point is proved by induction on~$t$:
    \begin{itemize}
      \item
        This is obvious if~$t=\x$.

      \item
        If~$t=\Daimon t'$, we have~$\Daimon\x \leq t'$ by induction and the
        result follows from the first point.

      \item
        Similarly, if~$t$ is a sum, the result follows directly from the
        induction hypothesis.

      \item
        If~$t=\C t'$, we have $\Daimon\x  \leq \Daimon t' \leq \Daimon\C t'$
        by induction hypothesis and by definition. Using the first point, this
        implies that~$\Daimon\x \leq \C t'$.

      \item
        The same argument works when~$t=\f t'$.

      \item
        When~$t=\Cm t'$ [resp. $t=\D t'$], we can use the same argument,
        except that~$\Daimon t'\leq \Cm t'$ [resp.~$\Daimon t'\leq \D t'$]
        comes from the second [resp. third] point.

      \item
        If~$t=\st{\dots;\D_i=t_i;\dots}$, we have $\Daimon\x  \leq \Daimon
        t_i$ by induction hypothesis. This shows
        that~$\Daimon\x\leq\sum_i\Daimon t_i \leq
        \Daimon\st{\dots;\D_i=t_i;\dots}$, from which we conclude
        that~$\Daimon\x\leq t$. \qedhere
    \end{itemize}
\end{enumerate}
\end{proof}
Because of inequality~$\sum_i \Daimon t_i \leq \Daimon\st{\dots; \D_i=t_i;
\dots}$, the converse of point~(1) doesn't hold in general and the resulting
domain (Definition~\ref{def:F}) is not a Smyth power domain.
The next lemma, a kind of dual to contextuality, might look obvious but isn't
completely immediate.
\begin{lem}\label{lem:contextuality_left}
  If~$s_1\leq s_2$, then~$s_1[\x:=t] \leq s_2[\x:=t]$.
\end{lem}
\begin{proof}
By induction on the proof of~$s_1\leq s_2$. Most cases are trivial:
\begin{itemize}
  \item
    If~$s_1\leq s_2$ comes from reflexivity [resp. transitivity], we
    have~$s_1[\x:=t]\leq s_2[\x:=t]$ by reflexivity [resp. transitivity] using
    the induction hypothesis.

  \item
    If~$s_2=\Zero$, then~$s_2[\x:=t]=\Zero$ as well so that~$s_1[\x:=t] \leq
    s_2[\x:=t]$ by definition.

  \item
    If~$s_1 = \Daimon\x$ and~$s_2=\x$, we have~$\Daimon(s_1[\x:=t]) \leq
    s_2[\x:=t]$ by contextuality. The case~$s_1=\Daimon\x$ and~$s_2=\f\x$ is
    similar.

  \item
    If~$s_1 = u+v$ and~$s_2=u$, the result holds
    because~$u[\x:=t]+v[\x:=t]\leq u[\x:=t]$.

  \item
    This is trivial for all (in)equalities from~(*).
    For example, if~$s_1= \D_{i_0}\st{\dots; \D_i=u_i; \dots}$
    and~$s_2=u_{i_0}$. By definition, we have~$s_1[\x:=t] = \D_{i_0}\st{\dots;
    \D_i=u_i[\x:=t]; \dots}$ and similarly,~$s_2[\x:=t] = u_{i_0}[\x:=t]$.
    Thus,~$s_1[\x:=t]\leq s_2[\x:=t]$ by definition of~$\leq$.

  \item
    The only interesting case is contextuality: suppose that~$s_1=C[s'_1]$
    and~$s_2=C[s'_2]$ with~$s'_1\leq s'_2$. By induction hypothesis, we know
    that~$s'_1[\x:=t]\leq s'_2[\x:=t]$.
    A straightforward induction on~$C$ shows that~$s'_1[\x:=t]\leq
    s'_2[\x:=t]$ implies~$C[s'_1][\x:=t] \leq C[s'_2][\x:=t]$ for
    all~$C,s'_1,s'_2,t$. \qedhere
  \end{itemize}
\end{proof}

\begin{defi}
  The reduction relation~$\to$ on terms is the contextual closure of the
  left-to-right inequalities~$(*)$ from Definition~\ref{def:order_F0}.
\end{defi}

\begin{lem}\label{lem:SN}\leavevmode
  \begin{enumerate}
    \item
      If~$t\to s$ then~$s\leq t$.

    \item
      The reduction~$\to$ is strongly normalizing.%

    \item
      Simple normal forms are given by the grammar
      \begin{myequation}
        t &\quad ::= \quad&
          \C\p p t \quad|\quad
          \st{\D_1=t_1; \dots; \D_n=t_n}\p p \quad|\quad
          \Daimon \delta \quad|\quad
          \delta
        \\
        \delta &\quad ::= \quad& \C\p p\m \delta \quad|\quad \D\p p \delta
        \quad|\quad \x \quad|\quad \f\,t
      \end{myequation}
      % In particular, two syntactically different normal forms are different in the
      % quotient.
  \end{enumerate}
\end{lem}
\newpage
A typical normal form thus looks like, where $t_1, t_2$, are themselves
in normal form
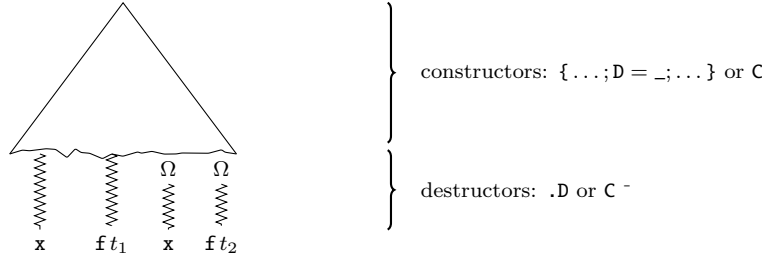
\[
\begin{tikzpicture}[font=\scriptsize]

  \coordinate (a) at (0,0);
  \coordinate (b) at (-1.5,-2);
  \coordinate (c) at (+1.5,-2);
  \coordinate (m) at ($(b)!0.45!(c)$);

  \draw (b) -- (a) -- (c);
  \draw[rounded corners=.5pt]
  (b)
  decorate [decoration={random steps,segment length=4pt,amplitude=2pt}] { -- (m) }
  decorate [decoration={random steps,segment length=4pt,amplitude=2pt}] { --
  (c) };

  \node[text centered] (x2) at ($(b) + (.4,-1.2)$) {$\x$} ;
  \draw ($(b) + (.4,0)$) decorate [decoration={zigzag,segment length=3pt,amplitude=2pt}] {-- (x2)};

  \node[text centered] (x3) at ($(m) + (0,-1.2)$) {$\f\, t_1$} ;
  \draw ($(m) + (0,0)$) decorate [decoration={zigzag,segment length=3pt,amplitude=2pt}] {-- (x3)};

  \node[text centered] (x4) at ($(c) + (-.9,-.2)$) {$\Daimon$} ;
  \node[text centered] (x5) at ($(x4) + (0,-1)$) {$\x$} ;
  \draw ($(x4) + (0,-.2)$) decorate [decoration={zigzag,segment length=3pt,amplitude=2pt}] {-- (x5)};

  \node[text centered] (x6) at ($(c) + (-.2,-.2)$) {$\Daimon$} ;
  \node[text centered] (x7) at ($(x6) + (0,-1)$) {$\f\,t_2$} ;
  \draw ($(x6) + (0,-.2)$) decorate [decoration={zigzag,segment length=3pt,amplitude=2pt}] {-- (x7)};

  \draw[thick, decorate,decoration=brace]
    (3.5,0) --
    node[right] {\quad constructors: $\st{\dots; \D=\BLANK; \dots}$ or $\C$ }
    (3.5,-1.85);

  \draw[thick, decorate,decoration=brace]
    (3.5,-1.95) --
    node[right] {\quad destructors: \D or \Cm}
    (3.5,-3.0);

\end{tikzpicture}
\]

\begin{proof}[Proof of Lemma~\ref{lem:SN}]
  The first point follows from the definition. Reduction is strongly
  normalizing because the depth of the term decreases. For the third point,
  all terms generated by the grammar are obviously in normal form. It is also
  straightforward to check that all simple normal forms are generated by the
  grammar because:
  \begin{itemize}
    \item
      there cannot be a destructor (\Cm or \D) directly above a constructor
      (\C or \st{\dots}),
    \item
      there cannot be a destructor (\Cm or \D) directly above \Daimon, nor a
      constructor (\C or \st{\dots}) directly below \Daimon. \qedhere
  \end{itemize}
\end{proof}
This reduction isn't confluent because terms of the form ``$\D_1\st{\D_1=t_1;
\D_2=t_2}$'' can reduce to~$t_1$ or to~\Zero if~$t_2$ reduces
to~\Zero.\footnote{It is however ``almost'' confluent in that a term can have
at most one non-\Zero normal form. Lemma~\ref{lem:CSem_reduction} is a weaker
version of that fact that is sufficient for our needs.}

\begin{lem}\label{lem:nf}
  Write~$\nf(t)$ for the normal form of a term according to the ``rightmost
  first'' reduction strategy. For any context and term,~$\nf(C[t] =
  \nf(C[\nf(t)])$.
\end{lem}
\begin{proof}
  This is a straightforward induction on the context~$C$.
  \begin{itemize}
    \item
      If $C$ starts with a constructor~\C, \ie $C$ is of the form~$\C\,C'$,
      the result follows from the induction hypothesis: since no reduction
      involves a constructor on the left, we have~$\nf(\C\,C'[t]) =
      \C\,\nf(C'[t]) = \C\,\nf(C'[\nf(t)]) = \nf(\C\,C'[\nf(t)])$.

    \item
      Reasoning is similar if~$C$ starts with a function name or a structure.

    \item
      The result is obvious if~$C$ is the placeholder variable, or~\x.

    \item
      If~$C$ starts with a destructor~\D, \ie $C$ is of the form~$\D\,C'$,
      computing~$\nf(C[u])$ is done by first computing~$v=\nf(C'[u])$
      and then reducing~$\D\,v$. By induction hypothesis, we have~$\nf(C'[t]) =
      \nf(C'[\nf(t)])$, so that reducing~$\nf(C[t])$ and
      $\nf(C[\nf(t)])$ give same results.

    \item
      Reasoning is similar if~$C$ starts with a destructor~$\Cm$ or
      a~$\Daimon$. \qedhere
    \end{itemize}
  \end{proof}
%%%>>>2

\subsubsection{Semantics}   %%%<<<2

Both ``\D'' and ``$\Cm$'' have natural interpretations as continuous
functions:
\[
  v:\S \quad\mapsto\quad \D(v) = \begin{cases}
    \bot & \text{if $v=\bot$}\\
    u & \text{if $v$ is of the form~$\st{\dots; \D=u; \dots}$}\\
    \Zero & \text{otherwise}
  \end{cases}
\]
and
\[
  v:\S \quad\mapsto\quad \Cm(v) = \begin{cases}
    \bot & \text{if $v=\bot$}\\
    u & \text{if $v$ is of the form~$\C u$}\\
    \Zero & \text{otherwise}
  \end{cases}
\]
This allows to define the semantics of any element of~$\Op_0$ as a function
depending on~\x.
\begin{defi}\label{def:CSem}
  Let~$\rho$ be an environment giving, for each functions names, a continuous
  function on~$\S$; let~$t$ be a simple term in~$\Op_0$. We
  define~$\CSem{t}_\rho:\S\to\S$ with
  \begin{enumerate}
    \item[(0)]
      $\CSem{t}_\rho(\sum_i v_i) = \sum_i \CSem{t}_\rho(v_i)$, and in particular
      $\CSem{t}_\rho(\Zero) = \Zero$,

    \item
      $\CSem{\C t}_\rho(v) = \C \big(\CSem{t}_\rho(v)\big)$,

    \item
      $\CSem{\st{\D_1=t_1; \dots}}_\rho(v) = \st{\D_1=\CSem{t_1}_\rho(v);
      \dots}$,

    \item
      $\CSem{\Daimon t}_\rho(v) = \Daimon\big(\CSem{t}_\rho(v) \big) =
      \begin{cases}
        \Zero & \text{if $\CSem{t}_\rho(v) = \Zero$}\\
        \bot & \text{otherwise,}\\
      \end{cases} $
      % $\CSem{\Daimon\{\dots,t_i,\dots\}}_\rho(v) = \Zero$
      % if~$\CSem{t_i}_\rho(v) = \Zero$ for all~$i$, and~$\bot$ otherwise,
      % $\CSem{\Daimon\{\dots,t_i,\dots\}}_\rho = \max_i\big(\Daimon \circ
      % \CSem{t_i}\big)$, \ie \begin{itemize}
      %   \item
      %     $\CSem{\Daimon}_\rho = \Daimon$,

      %   \item
      %     $\CSem{\Daimon\{\dots,t_i,\dots\}}_\rho(v) = \Zero$ if one
      %     of~$\CSem{t_i}_\rho(v) = \Zero$, and~$\bot$ otherwise,
      % \end{itemize}

    \item
      $\CSem{\Cm t}_\rho(v) = \Cm \big(\CSem{t}_\rho(v)\big)$,

    \item
      $\CSem{\D t}_\rho(v) = \D \big(\CSem{t}_\rho(v)\big)$,

    \item
      $\CSem{\x}_\rho(v) = v$,

    % \item
    %   $\CSem{\Zero}_\rho(v) = \Zero$,

    \item
      $\CSem{\g\ t}_\rho(v) = \rho(\g) \big(\CSem{t}_\rho (v)\big)$.
  \end{enumerate}
  $\CSem{\BLANK}$ is extended to all terms in~$\Op_0$ by linearity.

\end{defi}
\noindent
Because~\Daimon is the semantics of~$\Daimon\x$ we sometimes write~$\Daimon$
for~$\Daimon\x$.

\begin{lem}\label{lem:CSem}\hfill
  \begin{enumerate}
    \item
      If $t_1 \leq t_2$, then~$\CSem{t_1}_\rho \leq \CSem{t_2}_\rho$; $\CSem{\_}_\rho$
      is thus compatible with~$\approx$.

    \item
      If $\rho(\g)$ is continuous for any~\g, then~$\CSem{t}_\rho$ is also
      continuous.

    % \item
    %   $T \mapsto \CSem{T}$ is continuous as a function from~$\Op$
    %   to~$[\S\to\S]$.

    % \item\label{point:CSemZero}
    %   If~$t\in\Op_0$ contains~\Zero, then $\CSem{t}_\rho = v\mapsto \Zero$.

    \item\label{point:CSemDaimon}
      $\CSem{T}_\rho \geq \Daimon$.
      provided $\rho(\f) \geq \Daimon$ for all function names,

    \item\label{point:comp1}
      For all terms~$t_1, t_2\in \Op_0$ and environment~$\rho$, we have
      $\CSem{\big.t_1[\x:=t_2]}_\rho = \CSem{t_1}_\rho \circ \CSem{t_2}_\rho$.

  \end{enumerate}
\end{lem}
\begin{proof}
  Checking the first points amounts to checking that all inequations from
  Definition~\ref{def:F0} hold semantically in~$[\S\to\S]$. This is
  straightforward.
  The functions \Cm, \D and~\Daimon are easily shown continuous,
  $\CSem{t}_\rho$ is continuous as a composition of continuous functions.
%
  % Continuity of~$\Sem{\BLANK}$ follows from the definition
  % of~$\Sem{\dirsup\dots}$. The only thing that really needs checking is
  % that~$\Sem{\BLANK}$ is monotonic. This is an easy induction.
%
  Points~(3) follows from linearity of~$\CSem{t}_\rho$, and point~(4) is proved
  by immediate induction.
\end{proof}

\begin{cor}\label{cor:leq_non_trivial}
  The order~$\leq$ on~$\Op_0$ is non trivial.
\end{cor}
\begin{proof}
Any equivalence in~$\Op_0$ gives rise to an equality in~$\S\to\S$, which is non
trivial.
\end{proof}

\begin{lem}\label{lem:CSem_reduction}
  Suppose $t_1 \to t_2$ and let~$v\in\S$, we have~$\CSem{t_1}_\rho(v) =
  \CSem{t_2}_\rho(v)$ or~$\CSem{t_1}_\rho(v)=\Zero$.
\end{lem}
\begin{proof}
This is straightforward:
\begin{itemize}

  \item
    For~$\Cm\C t \to t$: if~$\CSem{t}_\rho(v) = \Zero$, then~$\CSem{\Cm\C
    t}_\rho(v)=\Zero$, and if~$\CSem{t}_\rho(v) \neq\Zero$,
    then by definition,~$\CSem{\Cm\C t}_\rho(v)=\CSem{t}_\rho(v)$.

  \item
    For~$\D_i\st{\dots;\D_i=t_i;\dots}\to t_i$: if
    some~$\CSem{t_j}_\rho(v)=\Zero$
    then~$\CSem{\D_i\st{\dots;\D_i=t_i;\dots}}_\rho(v)=\Zero$ as well and we
    have nothing to prove.
    Otherwise,~$\CSem{\D_i\st{\dots;\D_i=t_i;\dots}}_\rho(v)=\CSem{t_i}_\rho(v)$.

  \item
    For~$\Daimon\st{\dots;\D_i=t_i;\dots}\to\sum_i \Daimon t_i$,
    both~$\CSem{\Daimon\st{\dots;\D_i=t_i;\dots}}_\rho(v)$ and~$\CSem{\sum_i
    \Daimon t_i}_\rho(v)$ can only be equal to~$\bot$ or~\Zero. The only way
    to make the lemma false would be by having~$\CSem{\sum_i \Daimon
    t_i}_\rho(v)=\Zero$
    and~$\CSem{\Daimon\st{\dots;\D_i=t_i;\dots}}_\rho(v)=\bot$. This is
    impossible by point~(1) of the previous lemma.

  \item
    The other reduction rules are all treated similarly. \qedhere
  \end{itemize}
\end{proof}

% We can now define a domain containing all the operators
% coming from recursive definitions. Getting a domain is important because we
% want to compute fixed points using the formula from
% page~\pageref{formula:fixed_point}.
\begin{defi}\label{def:F}\leavevmode
  The domain~$\Op$ is the ideal completion of~$\Op_0$ quotiented
  by~$\approx$.
\end{defi}
This introduces infinite elements like $\C^\infty=\C\C\C\dots = \dirsup
\{\Daimon, \C\Daimon, \C\C\Daimon, \dots\}$. However,
since~$\Daimon\approx\D\Daimon\approx\D\D\Daimon\approx\dots$, we
have~$\D^\infty=\D\D\D\cdots = \dirsup \{\Daimon, \D\Daimon, \D\D\Daimon,
\dots\} \approx \Daimon$. Some infinite sums also appears but~$\Op$ is not a
Smyth power domain because of inequalities of the form~$\st{\dots; \D_i=t_i;
\dots} \geq \Daimon\st{\dots; \D_i=t_i; \dots} \geq \sum_i \Daimon t_i$.
We can extend the semantics of~$\Op_0$ to the whole~$\Op$.
\begin{defi}
  The semantics~$\CSem{\BLANK}$ is extended to~$\Op$ by continuity.
\end{defi}

\bigbreak
From now on, we single out a function name~\f as the recursive function whose
definition we are investigating.
\begin{defi}\label{def:F_Sem}
  Each $T\in\Op$ gives rise to an operator $\Sem{T}$
  from~$[\S\to\S]$ to itself:
  \[
    \Sem{T}_\rho \qquad:\qquad f:\S\to\S  \quad\mapsto\quad  \Sem{T}_\rho(f) = \CSem{T}_{\rho,\f:=f}
  \]
\end{defi}
\noindent
The typical environment~$\rho$ is constructed inductively from previous
recursive definitions and will be omitted in the rest of the paper. In
standard logical terminology, \f is a variable, but all other~\g are
parameters.
Note that dealing with mutually recursive definitions would require the
introduction of several function variables~$\f_1$,~$\f_2$, etc.
%%%>>>2
%%%>>>1

\subsection{Interpreting Recursive Definitions} %%%<<<1
\label{sub:interpreting_def}

\subsubsection{Composition}     %%%<<<2

Elements of~$\Op_0$ will be used to interpret individual clauses from a
recursive definition using Definition~\ref{def:F_Sem}. To use Kleene's
formula, we need a notion of composition. Given~$t_1$ and~$t_2$ in~$\Op_0$, we
want to represent the composition~$\Sem{t_1}\circ\Sem{t_2}$. Using standard
$\lambda$-calculus notation, $\CSem{t}$ is the semantics of~$\lambda\x.t$, of
type~$\S\to\S$ and~$\Sem{t}$ the semantics of~$\lambda\f\lambda\x.t$, of
type~$[\S\to\S]\to[\S\to\S]$. The composition of~$\Sem{t_1}$ and~$\Sem{t_2}$
should thus be
\begin{myequation}
  \Sem{t_1} \comp \Sem{t_2}
  &=&
  \lambda\f. \Sem{t_1}\big(\Sem{t_2}(\f)\big)
  & \text{\footnotesize(definition of composition)}\\
  &=&
  \lambda\f.(\lambda\f\lambda\x.t_1)\big(\Sem{t_2}(\f)\big)
  & \text{\footnotesize(definition of $\Sem{t_1}$)}\\
  &=&
  \lambda\f.\lambda\x.t_1\big[\f:=\Sem{t_2}(\f)\big]
  & \text{\footnotesize($\beta$ reduction)}\\
  &=&
  \lambda\f\lambda\x.t_1[\f:=\lambda\x.t_2]
  & \text{\footnotesize(definition of~$\Sem{t_2}$ and $\eta$ reduction)}\\
  &=&
  \lambda\f\lambda\x.t_1[\f u := (\lambda\x.t_2)u]
  & \text{\footnotesize(because~\f is always fully applied)}\\
  &=&
  \lambda\f\lambda\x.t_1\big[\f u := t_2[\x:=u]\big]\\
  &=&
  \Sem{t_1\big[\f u := t_2[\x:=u]\big]}
  \\
\end{myequation}
The composition of~$t_1$ and~$t_2$, as operators, is thus obtained by
replacing each~$\f u$ in~$t_1$ by~$t_2[\x:=u]$. The next definition
implements that directly.
\begin{defi}\label{def:comp_terms}
  If $t_1,t_2\in\Op_0$, we define $t_1\comp t_2$ by induction
  on~$t_1$:
  \begin{itemize}
    \item
      $\big(\sum t_i\big) \comp t_2 = \sum (t_i \comp t_2)$,

    \item
      $(\C t_1) \comp t_2 = \C (t_1\comp t_2)$,

    \item
      $\st{\D_1=t_1; \dots; \D_k=t_k} \comp t_2 = \st{\D_1=t_1\comp t_2;
      \dots; \D_k=t_k\comp t_2}$,

    \item
      $(\C\m t_1) \comp t_2 = \C\m (t_1\comp t_2)$,

    \item
      $(\D\, t_1) \comp t_2 = \D (t_1\comp t_2)$,

    \item
      $(\Daimon t_1) \comp t_2 = \Daimon(t_1 \comp t_2)$,

    \item
      $\x \comp t_2 = \x$,

    \item
      $(\g\, t_1) \comp t_2 = \g (t_1\comp t_2)$ if $\g\neq\f$,

    \item
      $(\f\, t_1) \comp t_2 = t_2[\x:=t_1\comp t_2]$.

  \end{itemize}
\end{defi}
\noindent
The only interesting case is the last one, where we replace~$\f$ by~$t_2$ and
continue recursively. Because of that, we sometimes abuse the notation and
write~$t_1[\f:=t_2]$.

Recall that to avoid introducing binders, we explicitly assume the argument of
an operator is the function with name~\f. In other words, the only free
symbols in the syntax are~\f and~\x. Any other function name~\g, \tt{h} etc. is
considered a parameter and is bound to some function taken from an implicit
environment.

\begin{lem}\label{lem:comp_associative}
  For any~$t_1, t_2, t_3\in\Op_0$, $t_1\comp (t_2\comp t_3) = (t_1\comp
  t_2)\comp t_3$.
\end{lem}
\begin{proof}
  We first prove that $t[\x:=t_1] \comp t_2 = (t\comp t_2)[\x:=t_1\comp t_2]$
  by induction on~$t$:
  \begin{itemize}
    \item
      if $t=\x$, this is immediate,

    \item
      if $t$ starts with a constructor, record, destructor, non-recursive
      function~\g, or~\Daimon, the result follows by induction,

    \item
      if $t = \f\, t'$, we have
      \begin{myequation}
        (\f\, t')[\x:=t_1] \comp t_2
        &\quad=\quad&
        (\f\, t'[\x:=t_1]) \comp t_2\\
        &\quad=\quad&
        t_2[\x:=t'[\x:=t_1]\comp t_2]
        & \text{definition of $\comp$}\\
        &\quad=\quad&
        t_2[\x:=(t'\comp t_2)[\x:=t_1\comp t_2]] \quad
        & \text{induction}\\
        &\quad=\quad&
        t_2[\x:=t'\comp t_2][\x:=t_1\comp t_2]
        \\
        &\quad=\quad&
        (\f\, t'\comp t_2)[\x:=t_1\comp t_2]
        & \text{definition of $\comp$}\\
      \end{myequation}

  \end{itemize}
  We can now prove that $t_1\comp (t_2\comp t_3) = (t_1\comp t_2)\comp
  t_3$ by induction on any simple~$t_1$:
  \begin{itemize}
    \item
      if $t=\x$, this is immediate,
    \item
      if $t_1$ starts with~$\Daimon$, a constructor, record or destructor, it
      follows by induction,

    \item
      if~$t_1 = \f\, t'_1$, we need to show that
        $t_2[\x:=t'_1\comp t_2] \comp t_3 = (t_2\comp
        t_3)[\x:=t'_1\comp(t_2\comp t_3)]$. By induction, it is enough to
        show that
        $t_2[\x:=t'_1\comp t_2] \comp t_3 = (t_2\comp
        t_3)[\x:=(t'_1\comp t_2)\comp t_3)]$. This
      follows from the previous remark, with $t=t_2$, $t_1=t'_1\comp
      t_2$, and $t_2=t_3$. \qedhere
    \end{itemize}
  \end{proof}

\begin{lem}\label{lem:composition}
  For any~$t_1, t_2 \in \Op_0$,
  $\Sem{t_1\comp t_2} = \Sem{t_1} \comp \Sem{t_2}$.
  If moreover,~$t_2$ doesn't contain~\f, we have~$\CSem{t_1\comp t_2} =
  \Sem{t_1}\big(\CSem{t_2}\big)$.
  In particular,
  \[
    \CSem{\big.\smash{\underbrace{t\comp \cdots \comp t}_{n}} \comp \Daimon\x}
    \quad = \quad
    \Sem{t}^n(\Daimon) \vphantom{\underbrace{t\comp \cdots \comp t}_{n}}
  \]
\end{lem}
\begin{proof}
  This is proved by induction. The
  only non trivial case is~$(\f\, t_1) \comp t_2 = t_2[\x:=t_1\comp t_2]$:
  \begin{myequation}
    \Sem{\big.(\f\, t_1) \comp t_2}_\rho(f)
      &=&
    \CSem{\big.(\f\, t_1)\comp t_2}_{\rho,\f=f}
      & \text{\small definition of $\Sem{\BLANK}$} \\
      &=&
    \CSem{\big.t_2[\x:=t_1\comp t_2]}_{\rho,\f=f}
      & \text{\small definition of $\comp$} \\
      &=&
    \CSem{t_2}_{\rho,\f=f} \circ \CSem{t_1\comp t_2}_{\rho,\f=f}
      & \text{\small point~(\ref{point:comp1}) of Lemma~\ref{lem:CSem}} \\
      &=&
    \CSem{t_2}_{\rho,\f=f} \circ \big(\Sem{t_1\comp t_2}_{\rho}(f)\big)
      & \text{\small definition of $\Sem{\BLANK}$} \\
      &=&
    \CSem{t_2}_{\rho,\f=f} \comp \big(\Sem{t_1}_\rho \circ \Sem{t_2}_\rho(f)\big)
      & \text{\small induction} \\
      &=&
    \CSem{t_2}_{\rho,\f=f} \comp \big(\Sem{t_1}_\rho (\CSem{t_2}_{\rho,\f=f})\big)
      & \text{\small definition of $\Sem{\BLANK}$} \\
      &=&
    \CSem{t_2}_{\rho,\f=f} \comp (\CSem{t_1}_{\rho,\f=\CSem{t_2}_{\rho,\f=f}})
      & \text{\small definition of $\Sem{\BLANK}$} \\
      &=&
    \CSem{\f\, t_1}_{\rho,\f=\CSem{t_2}_{\rho,\f=f}}
      & \text{\small definition of $\CSem{\BLANK}$} \\
      &=&
    \CSem{\f\, t_1}_{\rho,\f=\Sem{t_2}_\rho(f)}
      & \text{\small definition of $\Sem{\BLANK}$} \\
      &=&
    \Sem{\f\, t_1}_\rho \comp \Sem{t_2}_\rho(f)
      & \text{\small definition of $\Sem{\BLANK}$}\\
  \end{myequation}
  The second point is a direct consequence of the first point.
\end{proof}

We also have
\begin{lem}\label{lem:comp_increasing}
  If~$t_1\leq t_2$, then~$s\comp t_1 \leq s\comp t_2$ and~$t_1\comp s \leq
  t_2\comp s$.
\end{lem}
\begin{proof}
The first inequality is proved by induction on~$s$. The only interesting case
is when~$s$ starts with~\f.
\begin{myequation}
  (\f\, s) \comp t_1
  &=&
  t_1[\x:=s\comp t_1]
  &\text{\footnotesize(definition)}
  \\
  &\leq&
  t_2[\x:=s\comp t_1]
  &\text{\footnotesize(because~$t_1 \leq t_2$, by Lemma~\ref{lem:contextuality_left})}
  \\
  &\leq&
  t_2[\x:=s\comp t_2]
  &\text{\footnotesize(by contextuality, because~$s\comp t_1\leq s\comp t_2$ by induction)}
  \\
\end{myequation}
The second inequality is proved by induction on~$s_1\leq s_2$. The proof is
very similar to the proof of Lemma~\ref{lem:contextuality_left} and is omitted.
\end{proof}

% \paragraph{Notation} We omit the semantics brackets~$\Sem{\BLANK}$ in
% expressions of the form~$T^n(\Daimon)$, which always stand
% for~$\Sem{T\comp\dots\comp T\comp\Daimon\x}$.

% The next lemma is important as it makes it possible to use the formula from
% page~\pageref{formula:fixed_point} to define the fixed points of an
% operator~$t$ \emph{inside the domain~\Op}.
% \begin{lem}
%   Suppose~$t$ doesn't contain any function name other than the recursive
%   function~\f itself. Then we have~$\Daimon\x \leq t\circ\Daimon\x$.
% \end{lem}
% \proof
% This is a straightforward induction. We only look at the interesting case,
% when~$t$ is of the form~$\f t'$. In that case, we have~$\f t'\circ\Daimon\x =
% (\Daimon\x)[\x:=t'\circ\Daimon\x] = \Daimon(t'\circ\Daimon\x)$. By induction
% hypothesis,~$t'\circ\Daimon\x \geq \Daimon\x$. By
% Lemma~\ref{lem:simple_consequences}, we have $\Daimon (t'\circ\Daimon\x) \geq
% \Daimon\x$.
% \qed

%%%>>>2

\subsubsection{Interpreting Recursive Definitions} %%%<<<2

We can interpret the operator~$\NdtSem_\f$ (defined on
page~\pageref{def:NdtSem}) by an element of~$\Op_0$. Consider a single clause
``$\f\,p = u$'' of the recursive definition of~$\f$. The pattern~$p$ allows to
``extract'' some parts of the argument of~$\f$ to be used in the right-hand
side~$u$. For example, the clause
{\small\begin{alltt}
  | length (Cons \st{ Fst = e ; Snd = l }) = ...
\end{alltt}}\noindent
introduces 2 variables: \tt{e} and~\tt{l}. If we call the parameter
of~\tt{length} ``\x'', the variable~\tt{e} can be obtained as $\tt{e} =
\tt{.Fst} \, \tt{Cons}\m \, \x$: we remove the leading~\tt{Cons} constructor
and project on field~\tt{Fst}. The variable~\tt{l} is obtained similarly
with~$\tt{l} = \tt{.Snd} \, \tt{Cons}\m \, \, \x$. The following definition
formalizes that by defining, for any pattern~$p$, a substitution~$[p:=\x]$
giving for each variable of~$p$, an element of~$\Op_0$.

\begin{defi}\label{def:sigma_pattern}
  Given a linear pattern~$p$, define the substitution~$[p:=\x]$ as follows:
  \begin{itemize}
    \item
      $[\tt y:=\x] = [\tt y:=\x]$ where the substitution on the right is the
      usual substitution of variable~\tt y by variable~\x,
    \item
      $[\C p:=\x] = [p:=\x] \circ \Cm$,
    \item
      $[\st{\dots; \D_i=p_i; \dots}:=\x] = \bigcup_i ([p_i:=\x] \circ \D_i)$
      (because patterns are linear, the substitutions don't overlap).
  \end{itemize}
  where $\comp$ represents composition. For example,
  $\sigma_p \circ \Cm = [\dots, \tt{y}:=\sigma_p(\Cm\tt{y}), \dots]$.
\end{defi}
\noindent
As another example, consider the last rule from the~\tt{sum} function from
page~\pageref{chariot:sums}
{\small\begin{alltt}
    | sums _ \st{ Head = Cons \st{Fst = n ; Snd = l} ; Tail = s } = ...
\end{alltt}}\noindent
If we call the second argument of~\tt{sums} ``\x'', the corresponding
substitution is
\[
  \left[\ %
  \tt{n} := \tt{.Fst}\ \tt{Cons}\m \; \tt{.Head} \ \x;\quad
  \tt{l} := \tt{.Snd}\ \tt{Cons}\m \; \tt{.Head} \ \x;\quad
  \tt{s} := \tt{.Tail} \  \x;\quad
  \ \right]
\]

\begin{lem}\label{lem:unifier_substitution}
  If $v\in\V$ matches $p$ (Definition~\ref{def:NdtSem}), then
  % $[p:=\x]\circ[\x:=v](\tt y) \neq \Zero$ for all variables~$\tt y$ in~$p$.
  % In that case,
  $[p:=\x]\circ [\x:=v] = [p:=v]$, the unifier of~$p$ and~$v$.
\end{lem}
\begin{proof}
The proof is a simple induction on the pattern.
  \begin{itemize}
    \item
       When~$p=\y$ is a variable, $[p:=\x]$ is the substitution~$[\y:=\x]$,
       and the unifier~$[p:=v]$ is the substitution~$[\y:=v]$. The result is
       obvious.

    \item
      When~$p=\C p'$ starts with a constructor, $[\C p:=\x] = [p:=\x] \circ
      \Cm$. Because~$v$ must match~$p$, it is necessarily of the form~$\C v'$,
      and the unifier~$[p:=v]$ is equal to~$[p':=v']$. We thus have
      \[\begin{array}{rcll}
      [p:=\x] \circ [\x:=v]
        &=&
      [p':=\x] \circ \,\Cm \circ [\x:=\C v']\\
        &=&
      [p':=\x] \circ [\x:=\Cm\C v']\\
        &=&
      [p':=\x] \circ [\x:=v']\\
        &=&
      [p':=v']
        &\text{\footnotesize(induction hypothesis)}\\
        &=&
      [p:=v]\\
      \end{array}\]

    \item Reasoning is similar when~$p$ is a structure. \qedhere
  \end{itemize}
\end{proof}
Any recursive definition can be interpreted by an element of~$\Op_0$ in the
following way:
\begin{defi}\label{def:sem_rec_definition_in_F}
  Given a recursive definition of~\f, define~$T_\f$ with
  \[
    T_\f = \sum_{\tt{\f $p$ = $u$}} u[p:=\x]
  \]
  where the sum ranges over all clauses from the definition of~\f.
\end{defi}
For the \tt{length} function, we get
\[
  T_{\tt{length}} =
    \tt{Succ}\ \tt{length}\ \tt{.Snd}\ \tt{Cons}\m\, \x
    \quad+\quad
    \tt{Zero}\ \tt{Nil}\m\,\x
\]
where
the first summand corresponds to the first clause
{\small\begin{alltt}
    | length (Cons {Fst = _ ; Snd = l}) = Succ (length l)
\end{alltt}}\noindent
and
the second summand corresponds to the second clause
{\small\begin{alltt}
    | length (Nil _x) = Zero _x
\end{alltt}}\noindent

\begin{lem}
  For any environment~$\rho$, we have $\Sem{T_\f}_\rho\leq\NdtSem_{\rho,\f}$.
\end{lem}
\begin{proof}
  Given a value~$v$ and a clause~\tt{\f $p$ = $u$}, we have
  \begin{itemize}
    \item
      if~$p$ matches~$v$, then
      \[\begin{array}{rcll}
        \Sem{u[p:=\x]}(v)
        &=&
        \Sem{u[p:=\x][\x:=v]}
        &\text{\footnotesize(Lemma~\ref{lem:composition})}\\
        &=&
        \Sem{u[p:=v]}\\
      \end{array}\]

    \item
      if~$p$ doesn't matches~$v$, then
      \[\begin{array}{rcll}
        \Sem{u[p:=\x]}(v)
        &\leq&
        \Zero
        &\text{\footnotesize(Definition~\ref{def:NdtSem})}\\
        &=&
        \Sem{u[p:=v]}
        &\text{\footnotesize($p$ doesn't match $v$)}\\
      \end{array}\]
  \end{itemize}
  So that each summand of~$\Sem{T_\f}(v)$ is less than a summand
  of~$\NdtSem_\f(v)$, proving the claim.
\end{proof}

The inequality is strict in general because non-matching clause may introduce
non-\Zero terms: for example, the clause
\[
  \tt{f (} \underbrace{\tt{Cons\st{Fst=Foo ; Snd = l}}}_p \tt{) = Foo}
\]
doesn't match the value~$v=\tt{Cons}\st{\tt{Fst}=\tt{Bar};\ \tt{Snd}=...}$.
The non deterministic~$\NdtSem_{\f}(f)(v)$ is thus~\Zero. But because $[p:=\x]
= [\tt l := \tt{.Snd}\,\tt{Cons}\m\,\x]$, we have~$T_\f(v) = \tt{Foo}$. The
reason is that nothing about the closed pattern ``\tt{Foo}'' is recorded
in~$[p:=\x]$.

By Lemma~\ref{lem:incr_total}, totality of~$\fix(\Sem{T_\f})$ implies totality
of~$\fix(\NdtSem_\f)$.
Because of Lemma~\ref{lem:total_closed}, Lemma~\ref{lem:kleene_monotonic} and
Lemma~\ref{lem:composition}, we have
\begin{cor}\label{lem:operator_semantics}
  To check that~$\fix(\NdtSem_\f)$ is total, is is enough to check that
  \[
    \fix(\Sem{T_\f}) =
    \dirsup_n \Sem{T_\f}^n(\Daimon)
    \quad:\quad \S \to \S
  \]
  is total.
\end{cor}
\noindent
Together with Corollary~\ref{def:sem_rec_definition_in_F}, we finally get
\begin{cor}\label{cor:correctness}
  Given a recursive definition for~\f, we have that $\dirsup_n
  \Sem{T_\f}^n(\Daimon)$ is total implies that the usual semantics of~\f is
  total.
\end{cor}
\noindent
Recall that even though we haven't written them, all constructors~\C /
\st{\dots;\D=\_;\dots} and destructors~\Cm / \D come with a priority and that
totality is defined using those priorities
(Definition~\ref{def:intrinsic_totality}).

%%%>>>2
%%%>>>1

%% vim600:set foldmarker=<<<,>>> foldmethod=marker fileencoding=ascii spelllang=en spell: %%

\section{Call-Graphs and the Size-Change Principle}
\label{section:callgraph}

Except for a few minor differences, $T_\f$ is a faithful representation of the
original recursive definition. We now simplify each~$T_\f$ into a disjoint sum
of independent calls and show that doing so reflects totality.

\subsection{Call-Graphs}\label{sub:callgraph} %%%<<<1

\subsubsection{Call Paths} %%%<<<2

By definition, composition of operators (Definition~\ref{def:comp_terms}) is
linear on the left. When computing~$s\circ(t_1+t_2)$, each occurrence of~\f
inside~$s$ is replaced by~$t_1+t_2$. By linearity, this is a sum of terms
where each occurrence of~\f inside~$s$ is replaced either by~$t_1$ or~$t_2$.
For example, with~$s=\f\f\x$ and~$t_1=\itt{g}\x$, $t_2=\itt{h}\x$, we
get\footnote{recall that \f is the only free function name. Other names like
\itt{g} and \itt{h}, written here in italics, are bound parameters.}
\begin{myequation}
  \f\f\x \comp(\itt{g}\x+\itt{h}\x)
  &=& (\itt{g}\x+\itt{h}\x)[\x:=\f\x \comp(\itt{g}\x+\itt{h}\x)] \\
  &=& (\itt{g}\x+\itt{h}\x)\big[\x:=(\itt{g}\x+\itt{h}\x)[\x:=\x \comp(\itt{g}\x+\itt{h}\x)]\big] \\
  &=& (\itt{g}\x+\itt{h}\x)\big[\x:=(\itt{g}\x+\itt{h}\x)[\x:=\x]\big] \\
  &=& (\itt{g}\x+\itt{h}\x)[\x:=(\itt{g}\x+\itt{h}\x)] \\
  &=& \itt{g}(\itt{g}\x+\itt{h}\x)+\itt{h}(\itt{g}\x+\itt{h}\x) \\
  &=& \itt{g}\itt{g}\x+\itt{g}\itt{h}\x+\itt{h}\itt{g}\x+\itt{h}\itt{h}\x \\
\end{myequation}
To formalize that, we annotate each occurrences of~\f with its index and
write~$\f_1\f_2\x$ for~$\f\f\x$.
\begin{lem}\label{lem:comp_sum_right}
  We have
  \[
  s[\f:=t_1+\cdots+t_n]
  \quad=\quad
  \sum_{ \sigma \,:\, \occ(\f,s) \to \{t_1,\dots,t_n\} } s[\sigma]
  \]
  where $\occ(\f,s)$ represents the set of occurrences of~\f in~$s$, and the
  substitution occurs at the given occurrences. More precisely, $s[\sigma] =
  s\big[\f_i := \sigma(\f_i)]\big]$, which substitutes any~$\f_i\,t$ appearing
  in~$s$ by~$\sigma(\f_i)[\x:=t]$.
\end{lem}
\begin{proof}
This is a straightforward induction.
The most interesting case is when~$s$ is the structure~$\st{\dots;\D_i=s_i;\dots}$.
\begin{myequation}
  &&s[\x:=t_1+t_2]\\
  &=&
  \st{\dots; \D_i=s_i; \dots}[\x:=t_1+t_2]
  \\
  &=& \st{\dots; \D_i=s_i[\x:=t_1+t_2]; \dots}
  \\
  &=& \textstyle\st{\dots; \D_i=\sum_{\sigma_i:\occ(\f,s_i)\to\{1,2\}} s_i[\sigma_i]; \dots}
  &\text{\footnotesize(induction hypothesis)}
  \\
  &=& \textstyle\sum_i \sum_{\sigma_i:\occ(\f,s_i)\to\{1,2\}} \st{\dots; \D_i=s_i[\sigma_i]; \dots}
  &\text{\footnotesize(multilinearity)}
  \\
  &=& \textstyle\sum_{\sigma:\occ(\f,s)\to\{1,2\}} \st{\dots; \D_i=s_i[\sigma]; \dots}
  \\
\end{myequation}
where the last equality comes from the fact that~$\occ(\f,s)$ is the disjoint
sum of all the~$\occ(\f,s_i)$.
\end{proof}

\medbreak
In particular, if~$t = \sum_i t_i$ is a sum of simple terms, then~$t^n =
t\comp\cdots\comp t$ has a very specific shape.
% For~$t = \tt{Zero} + \tt{C}\,\f\,\x + \f\,\D\,\x$, we get
% \begin{itemize}
%   \item $t^1 = \tt{Zero} + \C\,\f\,\x + \f\,\D\,\x$
%   \item $t^2 = \tt{Zero} +
%     \underbrace{\C\,\tt{Zero}+\C\,\tt{C}\f\,\x +
%     \C\,\f\,\D\,\x}_{(\C\,\f\,\x)\comp t} + 
%     \underbrace{\tt{Zero} + \C\,\f\,\D\,\x +
%     \f\,\D\,\D\,\x}_{(\f\,\D\,\x)\comp t}
%     $
% \end{itemize}
Each summand of~$t^n$ is obtained by taking a summand of~$t^{n-1}$ and
replacing each occurrence of \f by some~$t_i$.
More formally:
\begin{defi}\label{def:path}
  Given~$t=t_1+\cdots+t_n$ a sum of simple terms, a \emph{path} for~$t$ is a
  sequence~$(s_k,\sigma_k)_{k\geq0}$ such that:
  \begin{itemize}
    \item
      $s_0 = \f\, \x$,

    \item
      $s_{k+1} = s_k[\sigma_k]$ where $\sigma_k$ replaces each occurrence of~\f
      inside~$s_k$ by some~$t_1$, \dots, $t_n$.
  \end{itemize}
  If some~$s_k$ doesn't contain any occurrence of~\f, then all later~$s_{k+i}$
  are equal to~$s_k$.
  We call such a path \emph{finite}.
\end{defi}
We usually omit the substitution and talk about the path~``$(s_k)$''.
% Note that~$s_1$ is just one of the summands of~$T$.
\begin{lem}
  For any term~$t$ and natural number~$k>0$, we have
  \[
  t^k = t\comp\cdots\comp t\quad=\quad \sum_{\text{$(s)$ path of $t$}} s_k
  \]
\end{lem}
\begin{proof}
  Note that because, for any given~$k$, there are only finitely many
  possible~$s_k$, this sum is actually finite. Suppose
  that~$t=t_1+\cdots+t_n$; the proof is by induction on~$k$:
  \begin{itemize}
    \item
      if $k=1$, $t^1=t_1+\cdots+t_n$, and each~$s_1$ in a path is of the
      form~$s_0=\f\,\x$ where~$\f$ is replaced by one~$t_i$, \ie each~$s_1$ is
      of the form~$\f\,\x\comp t_i = t_i$. Conversely, each~$t_i$ appears as
      some~$s_1$ for some path~$s$.

    \item
      By definition,~$t^{k+1} = t^k\comp t$, and~$t^k$ is the sum of the~$s_k$
      for all path~$(s)$. The term~$t^{k+1}$ is thus equal to the sum over all
      path~$(s)$ of the~$s_k[\f:=t]$. By Lemma~\ref{lem:comp_sum_right}, those
      are precisely the~$s_{k+1}$ of all path~$(s)$ of~$t$. \qedhere
    \end{itemize}
  \end{proof}

\noindent
We can use paths to compute fixed points.
\begin{lem}\label{lem:formula_fix_paths}
  Suppose~$T=t_1 + \cdots + t_n$ is a sum of simple terms, then
  \[
    \dirsup_n T\comp\cdots\comp T\comp\Daimon\x
    \quad=\quad
    \sum_{\text{$(s)$ path of $T$}} \quad \dirsup_{i\geq0} s_i(\Daimon)
  \]
\end{lem}
Note that the infinite sum makes sense because it is a limit of finite sums:
for each~$i$, there are only finitely many possible~$s_i(\Daimon)$.
% Since no ambiguity can arise when doing so, we'll leave the semantics
% brackets~$\Sem{\BLANK}$ inside~$\fix$ implicit and write~$\fix(T)$.
\begin{proof}
  We start by showing that the left-hand side is greater than the right-hand
  side, \ie by showing that any simple term in the LHS is greater than some
  simple term on the RHS.\footnote{That's the order on the Smyth power domain.
  Refer to Appendix~\ref{app:smyth}.}
  Let~$s$ be a simple term in~$\dirsup T^n(\Daimon)$. We want to show
  that~$s$ is greater than some~$\dirsup_{i\geq0} s_i(\Daimon)$.
  For each~$n$,~$T\comp\cdots\comp T\comp \Daimon\x$ is a finite sum of
  elements of~$\Op_0$. Define the following tree:
  \begin{itemize}
    \item
      nodes of depth~$i$ are those summands~$t$ in~$T^i$
      satisfying~$t(\Daimon) \leq s$,

    \item
      a node~$t'$ at depth~$i+1$ is a child of node~$t$ at depth~$i$ if~$t'$
      is the result of substituting all occurrences of~\f in~$t$ by
      one of~$t_1, \dots,t_n$.
  \end{itemize}
  As there are only finitely many possible substitutions from a given node,
  this tree is finitely branching.
  Because~$T^n(\Daimon) \leq \dirsup_i T^i(\Daimon)=\fix(T) \leq s$, each~$T^n$
  contains some simple term~$t$ such that~$t(\Daimon) \leq s$. This tree is
  thus infinite.
  By K\"onig's lemma, it contains an infinite branch~$s_0,s_1, \dots$.
  This sequence is a path of~$T$ and because all~$s_i(\Daimon)$ are less
  than~$s$ by construction, its limit is less than~$s$. We thus have
  \[
    \dirsup T^n(\Daimon)
    \quad\geq\quad
    \sum_{\text{$(s)$ path of $T$}} \quad \dirsup_{i\geq0} s_i(\Daimon)
  \]

  For the converse, it is enough to show that for each path~$(s)$ and natural
  number~$n$, the limit of~$s_k(\Daimon)$ is greater than~$T^n(\Daimon)$. This
  is immediate because each~$s_k(\Daimon)$ is a summand of~$T^k(\Daimon)$.
\end{proof}

\begin{cor}\label{cor:path}
  If $\rho$ is a total environment and~$\fix(\Sem{T})$ is non-total, then
  there is a path~$(s_k)$ for~$T$ such that~$\dirsup_i \Sem{s_i}(\Daimon)$ is
  non total.
\end{cor}
\begin{proof}
  Suppose that~$\fix(\Sem{T})=\dirsup_n \Sem{T}^n(\Daimon)$ is non total, then
  by Lemma~\ref{lem:composition} and the previous lemma, there is a path
  of~$T$ that is non total.
  % Since every finite path is total as a finite
  % composition of total operators, there necessarily exists a non total
  % infinite path.
\end{proof}
%%%>>>2

\subsubsection{Call-Graph} %%%<<<2

Part of the complexity of checking totality of recursive definitions comes
from the fact that clauses can contain nested recursive calls. The Ackerman
function is a well known (but useless) example. The call-graph turns a clause
into the sum of its recursive calls, making recursive calls ``independent''
from each other.
As an illustration, consider the following ad hoc clause
{\small\begin{alltt}\label{_ex:hypothetical_rule}
  | f \st{ D\(\sb1\) = y; D\(\sb2\) = z }  =  C ({f ({f y})}
\end{alltt}}
\noindent
As described in the previous section, it is interpreted by
\[
  \C \big(\underline{\f \big(\underline{\f\, \D_1\, \x)}\big)}\big)
\]
and contains 2 recursive calls (underlined).

It is clear that whenever this clause is used, it adds a~$\C$ constructor just
above the leftmost recursive call, making it guarded. It is also clear that
the rightmost recursive call is structurally decreasing as it uses part of the
initial argument~\x.
% However, this rightmost call isn't guarded and the argument of the leftmost
% call isn't structurally decreasing.
We keep this information and split this clause in two independent calls:
\begin{itemize}
  \item
    the leftmost call gives ``$\C \f (\Daimon\,\D_1\,\x)$'', which we write as
    \call{f x}{C f $(\Daimon\,\D_1\,\x)$}:
    \begin{itemize}
      \item
        this call is guarded by \C,

      \item
        we have no information about the arguments of~\f, except that it
        is~\Zero when~$\D_1\x$ is.

    \end{itemize}

  \item
    the rightmost call gives ``$\C\Daimon \f (\D_1\,\x)$'', which we write
    \call{f x}{\C\Daimon\f ($\D_1\,\x$)}:
    \begin{itemize}

      \item
        besides a topmost~\C, we have no information about constructors
        directly above the call,

      \item
        the argument of~\f is built from part of the initial argument.
    \end{itemize}
\end{itemize}
In general, for each recursive call, we replace all
other function calls (recursive or not) by~\Daimon (this is point (2) in the
definition below) and we split structures into independent fields (this is
point (6) in the definition below).
Visually:
\[
G\left(
\begin{tikzpicture}[font=\scriptsize,baseline=(current bounding box.center)]

  \coordinate (a) at (0,0);
  \coordinate (b) at (-1.5,-2);
  \coordinate (c) at (+1.5,-2);

  \node (f1) at (0.4, -.8) {\f};
  \node (f2) at (0.4, -1.4) {\f};
  \node (f3) at (-0.6, -1.2) {\f};

  \draw[thick] (b) -- (a) -- (c) -- (b);
  \draw (f1) -- ($(f1)+(-0.6,-1.1)$) -- ($(f1)+(+0.6,-1.1)$) -- (f1);
  \draw (f2) -- ($(f2)+(-0.2,-0.4)$) -- ($(f2)+(+0.2,-0.4)$) -- (f2);
  \draw (f3) -- ($(f3)+(-0.2,-0.5)$) -- ($(f3)+(+0.2,-0.5)$) -- (f3);

  \draw[thick,dotted] (a)--(f1);
  \draw[thick,dotted] (f1)--(f2);
  \draw[thick,dotted] (a)--(f3);
\end{tikzpicture}
\right)
=
\begin{tikzpicture}[font=\scriptsize,baseline=(current bounding box.center)]

  \coordinate (a) at (0,0);
  \coordinate (b) at (-1.5,-2);
  \coordinate (c) at (+1.5,-2);

  % \node (f1) at (0.4, -.8) {\f};
  % \node (f2) at (0.4, -1.4) {\f};
  \node (f3) at (-0.6, -1.2) {\f};

  \draw[gray, loosely dashed] (b) -- (a) -- (c) -- (b);
  % \draw (f1) -- ($(f1)+(-0.6,-1.1)$) -- ($(f1)+(+0.6,-1.1)$) -- (f1);
  % \draw (f2) -- ($(f2)+(-0.2,-0.4)$) -- ($(f2)+(+0.2,-0.4)$) -- (f2);
  \draw (f3) -- ($(f3)+(-0.2,-0.5)$) -- ($(f3)+(+0.2,-0.5)$) -- (f3);

  % \draw[thick,dotted] (a)--(f1);
  % \draw[thick,dotted] (f1)--(f2);
  \draw[thick,dotted] (a)--(f3);
\end{tikzpicture}
+
\begin{tikzpicture}[font=\scriptsize,baseline=(current bounding box.center)]

  \coordinate (a) at (0,0);
  \coordinate (b) at (-1.5,-2);
  \coordinate (c) at (+1.5,-2);

  \node (f1) at (0.4, -.8) {\f};
  \node (f2) at (0.4, -1.4) {\Daimon};
  % \node (f3) at (-0.6, -1.2) {\f};

  \draw[gray, loosely dashed] (b) -- (a) -- (c) -- (b);
  \draw (f1) -- ($(f1)+(-0.6,-1.1)$) -- ($(f1)+(+0.6,-1.1)$) -- (f1);
  \draw (f2) -- ($(f2)+(-0.2,-0.4)$) -- ($(f2)+(+0.2,-0.4)$) -- (f2);
  % \draw (f3) -- ($(f3)+(-0.2,-0.5)$) -- ($(f3)+(+0.2,-0.5)$) -- (f3);

  \draw[thick,dotted] (a)--(f1);
  \draw[thick,dotted] (f1)--(f2);
  % \draw[thick,dotted] (a)--(f3);
\end{tikzpicture}
+
\begin{tikzpicture}[font=\scriptsize,baseline=(current bounding box.center)]

  \coordinate (a) at (0,0);
  \coordinate (b) at (-1.5,-2);
  \coordinate (c) at (+1.5,-2);

  \node (f1) at (0.4, -.8) {\Daimon};
  \node (f2) at (0.4, -1.4) {\f};
  % \node (f3) at (-0.6, -1.2) {\f};

  \draw[gray, loosely dashed] (b) -- (a) -- (c) -- (b);
  % \draw (f1) -- ($(f1)+(-0.6,-1.1)$) -- ($(f1)+(+0.6,-1.1)$) -- (f1);
  \draw (f2) -- ($(f2)+(-0.2,-0.4)$) -- ($(f2)+(+0.2,-0.4)$) -- (f2);
  % \draw (f3) -- ($(f3)+(-0.2,-0.5)$) -- ($(f3)+(+0.2,-0.5)$) -- (f3);

  \draw[thick,dotted] (a)--(f1);
  \draw[thick,dotted] (f1)--(f2);
  % \draw[thick,dotted] (a)--(f3);
\end{tikzpicture}
\]
\begin{defi}\label{def:call_graph_4}
  Let $t\in\Op_0$, the \emph{call-graph of~$t$,} $G(t)$, is defined
  inductively as follows:
  \begin{enumerate}
    \item
      $G\big(\sum t\big) = \sum G(t)$,

    % \item
    %   $G(\f\, t) = \f\, t$ if~\f doesn't occur in~$t$,

    \item
      $G(\f\, t) = \f \big(t^\Daimon\big) + \Daimon G(t)$ \emph{where
      $t^\Daimon$ is~$t$ where all function calls
      have been replaced by~$\Daimon$},
      % (note that there is no recursive call to~$G$ in the left summand)

    \item
      $G(\g\, t) = \Daimon G(t)$ if $\g\neq\f$,

    \item
      $G(\x) = \Zero$,

    \item
      $G(\C\p p\, t) = \C\p p\, G(t)$,
      % {\hfill\footnotesize(recall that we count guarding constructor
      % negatively)}

    \item
      $G\big(\st{\dots; \D_i\p p=t_i; \dots}\big) = \sum_i \st{\D_i=G(t_i)}\p
      p$,

    \item
      $G(\Cm t) = \Cm\, G(t)$,

    \item
      $G(\D\, t) = \D\, G(t)$.

  \end{enumerate}
  We write ``\call{f x}{$u$}'' whenever $u$ is a simple term in~$G(T_\f)$.

\end{defi}
\noindent
For example, for~$t = \tt{C\st{Fst=$\f_1$ (C\m x); Snd = $\f_2$ (C ($\f_3$
x))}}$, we obtain
\[
  G(t)
  \quad=\quad
  \underbrace{\C\,\st{\tt{Fst}=\f_1 (\Cm\,\x)}}_{\text{first call}}
  \ +\ %
  \underbrace{\C\,\st{\tt{Snd}=\f_2 (\C\,\Daimon\,\x)}}_{\text{second call}}
  \ +\ %
  \underbrace{\C\,\st{\tt{Snd}=\Daimon\,\C\,\f_3\,\x)}}_{\text{third call}}
\]

The following is a direct consequence of the definition.
\begin{lem}
  For each call~\call{f x}{$u$}, there is exactly one occurrence of~\f
  inside~$u$.
\end{lem}
For mutual recursive definitions (not treated here), this defines an actual
graph:
\begin{itemize}
  \item
    vertices are the function names,
  \item
    arcs from~\f to~\g are the calls~\call{f x}{$u$} where $u$'s only function
    name is~\g.
\end{itemize}

In general, the call-graph of $T$ is not comparable to~$T$. However, the
construction reflects totality.
\begin{prop}\label{prop:Gt_total}
  If $\fix(G(T))$ is total then so is~$\fix(T)$.
\end{prop}
\begin{proof}

  Let~$T=\sum t_i$ is a sum of simple terms, and suppose $\fix(T)$ is
  non-total. By Corollary~\ref{cor:path}, it implies there is a path~$(s_k)$
  of~$T$ and a total element~$u\in\V$ such that~$\dirsup s_i(\Daimon)(u)\in\V$
  is non-total, \ie contains a non total branch~$\beta$. In particular, it
  implies that no~$s_i(\Daimon)(u)$ reduces to~\Zero.
  This branch~$\beta$ is either an infinite branch with odd principal priority
  or a finite branch ending with~$\bot$.

  The path~$(s_i)$ is infinite. Otherwise it becomes constant after a finite
  number of steps and the limit can be obtained by a finite number of
  applications of total operations (including the non recursive~\g, ...) on
  total values (including~$u$).

  \smallbreak
  At each step, we go from~$s_k$ to~$s_{k+1}$ by replacing all occurrences
  of~\f by some~$t_i$.
  Suppose the occurrences of~\f in~$T$ are indexed by natural numbers~$1,
  \dots, n$; we extend that indexing to occurrences of~\f in all the~$(s_k)$
  using lists of natural numbers in~$\{1, \dots, n\}$. For example,
  let~$T=\C\f_1\f_2\x + \D\f_3\f_4\x$ and consider the path that starts with:
  \[
  \f\,\x,\quad
  \C\f_1\f_2\,\x,\quad
  \C\C\f_1\f_2\D\f_3\f_4\,\x,\quad
  \dots
  \]
  % The first substitution is thus~$[\f:=\C\f_1\f_2\x]$ and the second substitution
  % is~$[\f_1:=\C\f_1\f_2\x, \f_2:=\D\f_3\f_4\x]$.
  The new indexing is
  \[
  \f_{[]}\,\x,\quad
  \C\f_{[1]}\f_{[2]}\,\x,\quad
  \C\C\f_{[1,1]}\f_{[1,2]}\D\f_{[2,3]}\f_{[2,4]}\,\x,\quad
  \dots
  \]
  When we replace~$\f_{[2]}$ (in~$ \C\f_{[1]}\f_{[2]}\x$) by~$\D\f_3\f_4\x$, we
  keep the~``$[2]$'' prefix in front of each new occurrence of~\f,
  obtaining~$\D\f_{[2,3]}\f_{[2,4]}\x$.
  Formally,
  \begin{enumerate}
    \item[(0)]
      the only occurrence of~\f in~$s_0=\f\,\x$ is indexed by the empty list

    \item
      given~$k\geq0$, the substitution~$\sigma_k$ replaces each
      occurrence~$\f_L$ in~$s_k$ by some~$t_i$. Each occurrence of~\f
      in~$s_{k+1}$ comes from a single occurrence~$\f_j$ in some summand
      of~$T$. We index such an occurrence by the list~$L,j$.
  \end{enumerate}
  % The lists keep track of the ``genealogy'' of occurrences of~\f appearing
  % in the path.
  %
  An occurrence~$\f_L \in s_k$ is called \emph{infinitary} if the~$s_{k'}$,
  for ~$k'\geq k$, contain infinitely many occurrences of~$\f_{L'}$ with~$L'$
  extending~$L$. Otherwise, an occurrence is called \emph{finitary}.

  By the above remark, the occurrence~$\f_{[]}$ in~$s_0$ is infinitary. We
  construct, by induction, an infinite sequence of infinitary
  occurrences~$\f_{[]}, \f_{[n_1]}, \f_{[n_1, n_2]}, \dots$ in the following
  way: at each step~$k$, choose~$n_k\in\{1,\dots,n\}$ s.t.
  \begin{enumerate}
    \item
      $\f_{[n_1,\dots,n_k]}$ is infinitary (this is always possible
      because~$\f_{[n_1,\dots, n_{k-1}]}$ is infinitary),

    \item
      the branch leading to~$\f_{[n_1, \dots, n_k]}$ in~$\nf(s_k)$ starts with
      a prefix of~$\beta$ of maximal length.
  \end{enumerate}
  At each step~$\f_{[n_1,\dots,n_k]}$ corresponds to the occurrence~$\f_{n_k}$
  in~$T$ and is thus associated to a single call~$\alpha_k$ in the sense of
  Definition~\ref{def:call_graph_4}. The limit~$\dirsup_k \alpha_1\comp \dots
  \comp \alpha_k(\Daimon)(u)$ is non-total:
  \begin{itemize}
    \item
      if for some~$k_0$, the occurrence~$\f_{[n_1,\dots,n_{k_0}]}$
      in~$s_{k_0}$ appears below a function call (recursive or otherwise), all
      the compositions~$\alpha_1\comp\cdots\comp\alpha_{k_0}\comp\cdots$ after
      step~$k$ will be of the form~$\beta_{k_0}\delta\Daimon(\dots)$
      where~$\beta_{k_0}$ is a prefix of~$\beta$ and~$\delta$ a sequence of
      destructors~\Cm/\D, so that their semantics on~$\Daimon$ and~$u$ will be
      equal to~$\beta_{k_0}\bot$. (It cannot be equal to~\Zero as it would
      imply that~$s_k$ is also equal to~\Zero.) The limits is thus equal
      to~$\beta_{k_0}\bot$, which is non total.

    \item
      if for all~$k$, the occurrence~$\f_{[n_1,\dots,n_k]}$ is only below a
      sequence of constructors \C/\st{\D\t=\BLANK} and destructors \Cm/\D,
      this sequence is of the form~$\beta_k\delta_k$ where each~$\beta_k$ is a
      prefix of~$\beta$, and~$\delta_k$ a sequence of destructors.
      \begin{itemize}
        \item
          If the~$\beta_k$s are bounded by some~$\beta_{k_0}$, the limit of
          the compositions~$\alpha_1\comp\cdots\comp\alpha_k$ will be, like
          above, equal to~$\beta_{k_0}\bot$, which is non total.

        \item
          If the~$\beta_k$s are unbounded, the limit of the
          compositions~$\alpha_1\comp\cdots\comp\alpha_k$ will be equal
          to~$\beta$, which is non total by hypothesis.
      \end{itemize}

  \end{itemize}
  None of the compositions~$\nf(\alpha_1\comp\cdots\comp\alpha_k)(\Daimon)(u)$
  can be equal to~\Zero, as it would imply the corresponding~$s_k(\Daimon)(u)$
  is equal to~\Zero as well. By Lemma~\ref{lem:CSem_reduction}, we have
  constructed a non-total branch in~$\dirsup_k \alpha_1\comp\cdots\comp
  \alpha_k(\Daimon)(u)$: this shows that~$G(T)$ is non-total.
\end{proof}

%%%>>>2
%%%>>>1

\subsection{Weights and Approximations} %%%<<<1
\label{sub:weight}

To use the size-change principle, compositions of calls need to be bounded.
In the definition of the~\tt{length} function, the only recursive call is
\[
  \call{length x}{Succ\p1 length (.Snd\p0 Cons\p1\m x)}
\]
Composing it with itself~$n$ times gives
\[\label{ex:unbounded_composition}
  \call{length x}{$\underbrace{\tt{Succ\p1 ... Succ\p1}}_{\text{$n$ repetitions}}$ length ($\underbrace{\tt{.Snd\p0 Cons\p1\m\ \ ...
  .Snd\p0 Cons\p1\m}}_{\text{$n$ repetitions of \tt{.Snd\p0 Cons\p1\m}}}$ x)}
\]
which grows arbitrarily large!
We introduce \emph{approximations} to deal with that. When a term grows too
large, constructors are only counted; and if this counter becomes too big, we
stop counting. Everything is parameterized by two natural numbers defining
what ``too large'' and ``too big'' really mean.

\subsubsection{Weights}

Simply counting constructors isn't enough because we need to keep track
of their priorities.
\begin{defi}[Weights]\label{_def:weight}
  Define the following
  \begin{enumerate}

    \item
      $\Zi = \Z \cup \{\infty\}$ where addition is extended with $w + \infty =
      \infty + w = \infty$ and the order is extended with $w\leq\infty$.

    \item
      \emph{Weights} are tuples of elements of~$\Zi$: $\Coef = \Zi^\P$
      where~$\P$ is the finite non-empty set of priorities. This set is
      ordered pointwise with the \emph{reverse} order of~$\Zi$. Addition
      on~\Coef is defined pointwise.

  \end{enumerate}
  We define the following abbreviations:
  \begin{itemize}
    \item
      $\coef{0} = (0, \dots, 0)$,

    % \item
    %   $\Infty = (\infty, \dots, \infty)$,

    \item
      $\coef{w}\p p$ for the weight~$(w_q)_{q\in\P}$ with $w_p=w$ and $w_q =
      0$ if $q\neq p$,

  \end{itemize}
  We surround weights with the symbols~``$\langle$'' and~``$\rangle$'',
  as in ``$\coef{W}$'' or ``$\coef{W_1+W_2}$''.
\end{defi}
\noindent
Weights can count constructors and destructors (with negative elements
of~$\Zi$). The special value~$\infty$ is a way to stop counting when those
numbers become too big. It does \emph{not} mean that there are infinitely many
constructors.
The next lemma is straightforward.
\begin{lem}\label{lem:coef}
  \leavevmode
  \begin{enumerate}
    \item
      Addition of weights is commutative, associative and monotonic,

    \item
      \coef{0} is neutral for addition,
      % and \Infty is absorbing: $\coef{0} +
      % \coef W = \coef W$ and $\Infty + \coef W = \Infty$,

    \item
      any weight can be written (uniquely) as~$\sum_{p\in
      P} \coef{w_p}\p p$ where~$P\subseteq\P$ and each~$w_p \in \Zi$,

    \item
      whenever~$w_1 \leq w_2$ in~$\Zi$, then~$\coef{w_2}\p p \leq \coef{w_1}\p
      p$ in~\Coef (note the reversal).
  \end{enumerate}
\end{lem}

\subsubsection{Approximations}\label{sub:approx}

An approximation is defined as the sum of all simple terms it is supposed to
approximates. Defining that requires the following notions.
\begin{defi}\label{def:branch_weight}
  A ``shape''~$\Delta\in\Op_0$ is a simple normal form (Lemma~\ref{lem:SN})
  \emph{which contains neither functions names nor~\Daimon}.
  \begin{enumerate}
    \item
      The set of \emph{branches of $\Delta$} is defined inductively
      \begin{myequation}
        \bigg.
        \branches(\x) & = & \{\x\}
        \\
        \bigg.
        \branches\big(\C\p p \Delta\big) & = &
        % \C\p p\cdotp\branches(\Delta)
        \big\{\C\p p \beta\ \mid\ \beta\in\branches(\Delta)\big\}
        \\
        \bigg.
        \branches\big(\D\p p \Delta\big) & = &
        % \D\p p\cdotp\branches(\Delta)
        \big\{\D\p p \beta\ \mid\ \beta\in\branches(\Delta)\big\}
        \\
        \bigg.
        \branches\big(\Cm\p p \Delta\big) & = &
        % \Cm\p p\cdotp\branches(\Delta)
        \big\{\Cm\p p \beta\ \mid\ \beta\in\branches(\Delta)\big\}
        \\
        %\bigg.
        %\branches\big(\f\ t\big) & = & \f.\branches\big(\Delta \big)
        %\\
        %%
        \bigg.
        \branches\big(\st{ \dots; \tt{D}_i\p p=\Delta_i; \dots}\big) & = &
        % \bigcup_i \st{\tt{D}_i=\BLANK}\cdotp\branches(\Delta_i) \\
        \bigcup_i \big\{\st{\D_i\p p=\beta}\ \mid\ \beta\in\branches(\Delta_i)\big\}
        %
        % \bigg.
        % \branches\big(\Daimon\{T\}\big) & = & \bigcup_{\Delta\in T}
        % \Daimon.\branches(\Delta) \\
      \end{myequation}

    \smallbreak
    \item
      If $\beta$ is a branch of~$\Delta$,
      the weight of~$\beta$, written~$|\beta|\in\Coef$ is defined with:
      \begin{itemize}
        \item
          $|\x| = \coef{0}$,

        \item
          $|\C\p p \beta| = {\coef1\p p} + |\beta|$,

        \item
          $|\D\p p \beta| = {\coef{-1}\p p} + |\beta|$,

        \item
          $|\Cm\p p \beta| = {\coef{-1}\p p} + |\beta|$,

        \item
          $|\st{\D=\beta}\p p| = {\coef1\p p} + |\beta|$.

      \end{itemize}

    % \smallbreak
    % \item
    %   If $\beta$ is a branch of~$p$,
    %   the~$p$-weight of~$\beta$, written~$|\beta|_p\in\Zi$ is defined as the~$p$
    %   component of~$|\beta|$.
  \end{enumerate}

\end{defi}
\begin{defi}[Approximations]\label{def:coef}\leavevmode
  \begin{enumerate}
    \item
      Given some~$W\in\Coef$, we put
      \[
      \coef{W}\x
      \quad=\quad
      \sum \left\{\Delta
      \ \middle|\
      \vcenter{\hbox{all branches~$\beta$ of $\Delta$}
               \hbox{satisfy $|\beta| \leq \coef{W}$}}
      \right\}
      \]

    \item
      Given~$t$ in~$\Op$, we write~$\coef{W}t$ for the corresponding sum of
      all~$\Delta[\x:=t]$, for~$\Delta\in \coef{W}\x$.

  \end{enumerate}
\end{defi}
The typical summand of~$\coef{W}t$ looks like:
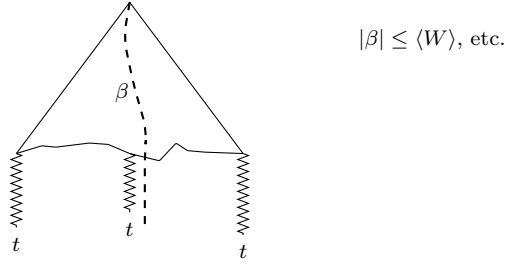
\[
\begin{tikzpicture}[font=\scriptsize]

  \coordinate (a) at (0,0);
  \coordinate (b) at (-1.5,-2);
  \coordinate (c) at (+1.5,-2);
  \coordinate (m) at ($(b)!0.5!(c)$);

  \draw (b) -- (a) -- (c);
  \draw[rounded corners=.5pt]
  (b)
  decorate [decoration={random steps,segment length=8pt,amplitude=4pt}] { -- (m) }
  decorate [decoration={random steps,segment length=8pt,amplitude=4pt}] { --
  (c) };

  % \node[rectangle, rounded corners=1pt, fill=gray!30, text centered] (n1) at (b) {\Daimon} ;
  \node[text centered] (x1) at ($(b) + (0,-1.2)$) {$t$} ;
  \draw (b) decorate [decoration={zigzag,segment length=3pt,amplitude=2pt}] {-- (x1)};

  % \node[rectangle, rounded corners=1pt, fill=gray!30, text centered] (n3) at
  % (m) {$\phantom{\coef{0}}$} ;
  \node[text centered] (x3) at (0,-3) {$t$} ;
  \draw (m) decorate [decoration={zigzag,segment length=3pt,amplitude=2pt}] {-- (x3)};

  \node at ($(a)!.6!(m) + (-.1,0)$) {$\beta$} ;
  \draw[rounded corners=5pt,thick,dashed]
  (a)
  decorate [decoration={random steps,segment length=15pt,amplitude=4pt}] { --
  ($(m)+(.2,0)$) }
  --($(x3)+(.2,0)$);
  \node at (4,-.5) {$|\beta| \leq \coef{W}$, etc.};

  % \node[rectangle, rounded corners=1pt, fill=gray!30, text centered] (n2) at (c) {\Daimon};
  \node[text centered] (x4) at (1.5,-3.3) {$t$} ;
  \draw (c) decorate [decoration={zigzag,segment length=3pt,amplitude=2pt}] {-- (x4)};

\end{tikzpicture}
\]
For example, the approximation~$\coef{\coef{1}\p p}t$ contains, among others,
the following summands: $\st{\tt{Foo}\p p=t}$, $\st{\tt{Foo}\p p=\st{\tt{Foo}\p
p=t}}$ and $\st{\tt{Fst}\p p=t;\tt{Snd}\p p=\C\p q\m t}$.

\begin{lem}\label{lem:approx_finitary}
  Approximations are well defined elements of~$\Op$.
\end{lem}
\begin{proof}
  This relies on the fact that there are only finitely many constructor and
  destructor names.

  Let~$W\in\Coef$, we want to show that~$\coef{W}\x$ can be obtained as the
  limit of a chain of finite sums of simple elements of~$\Op$. Given~$d\in\N$,
  define $\coef{W}\x\restrict d \subset \Op_0$ as the set obtained by
  truncating summands of~$\coef{W}\x$ at depth~$d$. Truncating an
  element~$\Delta$ is done by replacing subterms of~$\Delta$ at depth~$d$
  by~$\Daimon\x$. For example, ``\tt{A B C$\m$ \x}'' truncated at depth~2
  gives~``$\tt{A B \Daimon \x}$''.
  Because there are only finitely many constructors and destructors,
  each~$\coef{W}\x\restrict d $ is finite. Moreover, $\coef{W}\x$ is the limit
  of the chain
  \[
  \coef{W}\x\restrict 1 \quad\leq\quad \coef{W}\x\restrict 2 \quad\leq\quad \cdots
  \]
  Indeed, each element of $\coef{W}\x\restrict d+1$ is either in
  $\coef{W}\x\restrict
  d$ (when its depth is less than~$d$), or greater than an element
  of $\coef{W}\x$ (when its
  depth is strictly greater than~$d$).
  This shows that~$\coef{W}\x$ is a limit of elements of~$\Op_0$.

  This argument works unchanged when~\x is replaced by any term~$t$.
\end{proof}

\begin{lem}\label{lem:order_approx}\leavevmode
  We have
  \begin{enumerate}

    \item
      if $W\leq W'$ in~$\Coef$, then $\coef{W}t \leq
      \coef{W'}t$,

    \item
      $\coef{0} (t) \leq t$,

    \item
      $\coef W \Zero = \Zero$

    \item
      $\coef W\C\p p t = \coef{W+ \coef1\p p}t$,

    % \item
    %   $\coef W\st{\D = t}\p p = \coef{W +  \coef1\p p} t$,

    \item
      $\coef W\st{\D_1=t_1; \dots; \D_k=t_k}\p p \geq \Daimon t_1 + \cdots + \Daimon t_k$,

    \item
      $\C\p p\m \coef{W} t = \coef{W + \coef{-1}\p p}t$,

    \item
      $\D\p p \coef{W} t \geq \coef{W + \coef{-1}\p p}t$,

    \item
      $\Daimon \coef{W} t \geq \Daimon t$,

    \item
      $\coef{W}\Daimon t \geq \Daimon t$,

    \item
      $\coef V\coef{W} t) \geq \coef{V+W}t$.

  \end{enumerate}
\end{lem}
\pagebreak
\begin{proof}\leavevmode
\begin{enumerate}

  \item
    The first point is immediate once you recall the order on~$\Coef$ is the
    \emph{reverse} order on~$\Zi^\P$.

  \item
    The second point follows from the fact that~$t=\x[t]$ is a summand
    of~$\coef{0}t$.

  \item
    Because a summand~$\Delta$ of~$\coef{W}\x$ cannot contain empty structures
    by definition, it must contain the variable~\x. Because of that, all
    summands~$\Delta[\Zero]$ contain~\Zero and are thus equal to~\Zero by
    linearity.

  \item
    Suppose~$\Delta[\C\p p t]$ is a summand in~$\coef{W}\C\p p t$. By
    defining~$\Delta'=\Delta[\C\p p\x]$, we have that~$\Delta'[t]=\Delta[\C\p
    p t]$ is a summand of~$\coef{W + \coef1\p p} t$. This shows that~$\coef
    W\C\p p t \geq \coef{W + \coef1\p p}t$.
    For the converse, suppose~$\Delta[t]$ is a summand in~$\coef{W + \coef1\p
    p}t$. We put~$\Delta' = \Delta[\C\p p\m\x]$ so that~$\Delta'[\C\p p t]
    \geq \Delta[t]$ is a summand in~$\coef{W} \C\p p t$. This shows
    that~$\coef W\C\p p t \leq \coef{W + \coef1\p p}t$.

  \item
    This follows from  points (8) below:
  \[\begin{array}{rcll}
    &&
    \coef W\st{\D_1=t_1; \dots; \D_k=t_k}\p p\\
    &\geq&
    \Daimon\coef W\st{\D_1=t_1; \dots; \D_k=t_k}\p p
    &\text{\footnotesize(by definition of $\leq$)}\\
    &\geq&
    \Daimon\st{\D_1=t_1; \dots; \D_k=t_k}\p p
    &\text{\footnotesize(by points (9) below)}\\
    &\geq&
    \Daimon t_1 + \cdots + \Daimon t_k
    &\text{\footnotesize(by definition of $\leq$)}\\
  \end{array}\]

  \item
    By definition, any summand of $\C\p p\m \coef{W} t$ is also a summand
    of~$\coef{W + \coef{-1}\p p}t$. This implies that $\C\p p\m \coef{W} t
    \geq \coef{W + \coef{-1}\p p}t$.
    For the converse, let~$s=\Delta[t]$ be a summand in~$\coef{W + \coef{-1}\p
    p}t$. Take~$\Cm\p p\C\p p\Delta[t]$: this is a summand of~$\C\p
    p\m\coef{W} t$ which is equal to~$s$. We can conclude that~$s$ is greater
    than a summand of~$t$, which implies that~$s\geq\coef{W + \coef{-1}\p
    p}t$.

  \item
    This is treated similarly, except the second inequality cannot be proved
    because reducing~$\D\st{\D=\delta t}$ yields~$\delta t$, a smaller term
    (second inequality in group~$(1)$ in~$(*)$ in
    Definition~\ref{def:order_F0}).\footnote{We could add this equality
    \emph{for records with a single field}, but it would add yet another case
    to all the proofs involving the order, without any clear gain.}

  \item
    Let~$s$ be a summand in~$\Daimon\coef{W} t$; it is of the
    form~$\Daimon\Delta[t]$. By contextuality, we have that~$\Daimon\Delta[t]
    \geq \Daimon\Delta[\Daimon t]$, and because~\Daimon absorbs constructors
    on its right and destructors on its left, we have
    that~$\Daimon\Delta[\Daimon t] \to^* \Daimon t$. Because terms decrease
    during reduction, we have that~$s \geq \Daimon t$, which implies
    that~$\Daimon\coef{W} t \geq \Daimon t$.

  \item
    We have that $\coef{W}\Daimon t \geq \Daimon \coef{W} \Daimon t$ by
    definition of the order, and $\Daimon \coef{W} \Daimon t\geq \Daimon t$ by
    the previous point (and contextuality). The result follows from
    transitivity.

  \item This is immediate. \qedhere
\end{enumerate}
\end{proof}

\subsubsection{Dual approximations}\label{subsub:dual_weights}

Inductive and coinductive types are dual to each other. Formally,
approximations should come in 2 dual flavors:
\begin{itemize}
  \item
    one that guarantees that at least some number of constructors have been
    \emph{removed}, used to detect that the inductive argument to a recursive
    function gets smaller,

  \item
    one that guarantees that at least some some number of structures have been
    \emph{added}, used to detect that a recursive function is productive.
\end{itemize}
Approximations defined above corresponds to the first kind. Rather than having
two definitions~$\coef{W}^\uparrow$ and~$\coef{W}^\downarrow$, we simply use
negative weights to deal with the second kind. In each call~\call{f x}{$B$ f
$u$}, the branch~$B$ used for checking productivity uses ``negated'' weights,
while the term~$u$ uses the standard weights as defined in this section.
Because calls contain a single occurrence of~\f, there is no ambiguity as to
which weights must be negated.

%%%>>>1

\subsection{Collapsing}\label{sub:collapsing} %%%<<<1

To avoid unbounded compositions like the one shown on
page~\pageref{ex:unbounded_composition} we need to cut off compositions
when they get too big. This is parameterized by 2 natural numbers:
\begin{itemize}
  \item
    $D\geq0$ bounding the depth of terms,

  \item
    $B>0$ bounding the finite weights of approximations.
\end{itemize}
Both bounds can be chosen independently for each recursive definition.

\subsubsection{Calls}

Approximations are already elements of~$\Op_0$, but they correspond to
infinite sums, which cannot be implemented directly. The following extends the
syntax for~$\Op_0$ with ``built-in'' approximations.
\begin{defi}\label{def:gen_pattern}
  The set $\A_0$ (for ``\A{}pproximated terms'') is defined inductively by
  \begin{myequation}
    t &\quad ::= \quad&
      \C\p p\, t \quad|\quad
      \st{\D_1=t_1; \dots; \D_n=t_n}\p p \quad|\quad \\
      &&
      \C\p p\m t \quad|\quad
      \D\p p t \quad|\quad\\
      &&
      \f\, t \quad|\quad
      \x \quad|\quad\\
      &&
      \Daimon t \quad|\quad
      \coef{W} t \quad|\quad
      \\
      &&
      t_1 + \dots + t_n
  \end{myequation}
  where~$n>0$,~\x is a fixed variable name and each~$\coef{W}$ is a weight. As
  previously, \C and~\tt{D} come from a finite set of constructor and
  destructor names, and their priorities come from a finite set of natural
  numbers. They are respectively odd and even.
\end{defi}
% Because the domain~\Op can interpret them, this syntax is only necessary for
% the implementation. It gives a finite syntax for approximations which
% correspond to some infinite sums in~$\Op$. We now show how to deal with the
% order looking only at this finite syntax.
The order is defined as before, adding inequalities from
Lemma~\ref{lem:order_approx}:

\begin{defi}\label{def:order} The order on~$\A_0$ is defined
  inductively using the same rules as the order on~$\Op_0$
  (Definition~\ref{def:order_F0}) together with some additional rules:
  \begin{itemize}
    \item
      if $W\leq W'$ in~$\Coef$, then $\coef{W}t \leq
      \coef{W'}t$,

    \item
      $\coef{0} t \leq t$.

  \end{itemize}
  and
  \[ (*)\left\{ \begin{array}{lrcll}
    (4) & \coef W \Zero &\approx&  \Zero \\
    (4) & \coef W\C\p p t & \approx & \coef{W+ \coef1\p p}t\\
    % (4) & \coef W\st{\D = t}\p p & \approx & \coef{W +  \coef1\p p} t \\
    (4) & \coef W\st{\D_1 = t_1; \dots; \D_k = t_k}\p p & \geq & \Daimon t_1 + \dots + \Daimon t_k\\
    (4) & \C\p p\m \coef{W} t & \approx &
    \coef{W + \coef{-1}\p p}t\\
    (4) & \D\p p \coef{W} t &\geq& \coef{W + \coef{-1}\p p}t\\
    (4) & \Daimon\coef{W} t & \geq & \Daimon t\\
    (4) & \coef{W}\Daimon t & \geq & \Daimon\coef{W}t\\
    (4) & \coef V\coef{W} t & \geq & \coef{V+W}t\\
  \end{array}\right.\]
\end{defi}
By Lemma~\ref{lem:order_approx}, the order on approximations implies the order
on their semantics in~$\Op$. We extend the reduction relation~$\to$ to
approximations.
\begin{defi}
  Reduction on~$\A_0$ extends reduction on~$\Op_0$ by adding rules,
  oriented from left to right, for all inequalities in group~(4).
\end{defi}
%
% The way approximations interact with records is a little ad hoc. We would
% like a reduction of the form
% \[
% \coef{W}\st{\D_1\p p=t_1;\dots;\D_k\p p=t_k}
% \to
% \coef{W+\coef1\p p}t_1 \otimes \cdots \otimes \coef{W+\coef1\p p}t_k
% \]
% where the right side is the sum of all term where any branch leading
% to~$t_i$ has weight~$\coef{W_i}$. Extending approximations in this way is
% possible but makes the theory much more tedious without any clear gain on
% practical examples.

\smallbreak
Just like before (Lemma~\ref{lem:SN}), we have
\begin{lem}\label{lem:order_gen_patterns}\leavevmode
  \begin{enumerate}

    \item
      If $t\to t'$ then~$t\geq t'$.

    \item
      Reduction on~$\A_0$ is strongly normalizing.

    \item
      Normal forms are generated by the following grammar
      \begin{myequation}
        % T &\quad ::= \quad&
        %     t_1 + \cdots + t_n
        %       & \text{$n\geq0$}
        % \\
        t &\quad ::= \quad&
          \C\p p t \quad|\quad
          \st{\D_1=t_1; \dots; \D_n=t_n}\p p \quad|\quad\\
          && \Daimon\delta \quad|\quad
          \coef{W}\delta \quad|\quad
          \delta
        \\
        \delta &\quad ::= \quad& \C\p p\m \delta \quad|\quad \D\p p \delta
        \quad|\quad \x \quad|\quad \f\,t
      \end{myequation}
  \end{enumerate}
\end{lem}
\noindent
The proof is just a slight extension of the proof of Lemma~\ref{lem:SN}.

\medbreak
Recall that~\call{f x}{$u$} is a notation for~$u\in G(T_\f)$. We extend this
to approximations~$u$.
\begin{defi}\label{def:call}
  A \emph{call} from~\f to~\g is a 3-tuple consisting of
  \begin{itemize}
    \item
      a calling function name \f,

    \item
      a called function name \g,

    \item
      an approximation in normal form with the following shape:
      
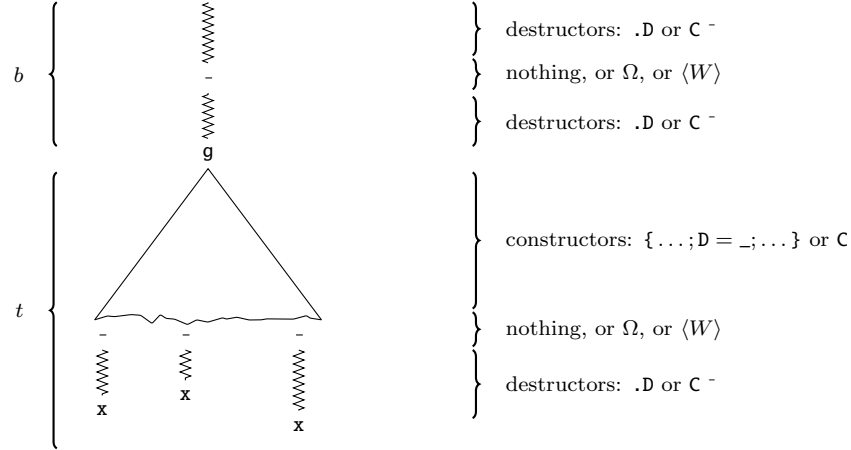
\[
      \begin{tikzpicture}[font=\scriptsize]

        \coordinate (a) at (0,0);
        \coordinate (b) at (-1.5,-2);
        \coordinate (c) at (+1.5,-2);
        \coordinate (m) at ($(b)!0.45!(c)$);

        \draw (b) -- (a) -- (c);
        \draw[rounded corners=.5pt]
        (b)
        decorate [decoration={random steps,segment length=4pt,amplitude=2pt}] { -- (m) }
        decorate [decoration={random steps,segment length=4pt,amplitude=2pt}] { --
        (c) };

        \node[text centered] (y1) at ($(a) + (0,.2)$) {$\g$} ;
        \node[text centered] (y2) at ($(y1) + (0,1)$) {$\_$} ;
        \draw ($(y1) + (0,.2)$) decorate [decoration={zigzag,segment length=3pt,amplitude=2pt}] {-- ($(y2) + (0,-.2)$)};
        \draw ($(y2) + (0,.2)$) decorate [decoration={zigzag,segment length=3pt,amplitude=2pt}] {-- ($(y2) + (0,1)$)};

        \node[text centered] (x2) at ($(b) + (.1,-.2)$) {$\_$} ;
        \node[text centered] (x3) at ($(b) + (.1,-1.2)$) {$\x$} ;
        \draw ($(x2) + (0,-.2)$) decorate [decoration={zigzag,segment length=3pt,amplitude=2pt}] {-- (x3)};

        \node[text centered] (x4) at ($(b) + (1.2,-.2)$) {$\_$} ;
        \node[text centered] (x5) at ($(b) + (1.2,-1)$) {$\x$} ;
        \draw ($(x4) + (0,-.2)$) decorate [decoration={zigzag,segment length=3pt,amplitude=2pt}] {-- (x5)};

        \node[text centered] (x7) at ($(b) + (2.7,-.2)$) {$\_$} ;
        \node[text centered] (x8) at ($(b) + (2.7,-1.4)$) {$\x$} ;
        \draw ($(x7) + (0,-.2)$) decorate [decoration={zigzag,segment length=3pt,amplitude=2pt}] {-- (x8)};

        \draw[thick, decorate,decoration=brace]
            (-2,0.3) --
            node[left] {$b\quad$}
            (-2,2.2);

        \draw[thick, decorate,decoration=brace]
            (-2,-3.7) --
            node[left] {$t\quad$}
            (-2,-0.05);

        \draw[thick, decorate,decoration=brace]
            (3.5,2.2) --
            node[right] {\quad destructors: \D or \Cm}
            (3.5,1.5);

        \draw[thick, decorate,decoration=brace]
            (3.5,1.45) --
            node[right] {\quad nothing, or~\Daimon, or \coef{W}}
            (3.5,1.05);

        \draw[thick, decorate,decoration=brace]
            (3.5,0.95) --
            node[right] {\quad destructors: \D or \Cm}
            (3.5,0.3);

        \draw[thick, decorate,decoration=brace]
          (3.5,-0.05) --
          node[right] {\quad constructors: $\st{\dots; \D=\BLANK; \dots}$ or $\C$ }
          (3.5,-1.85);

        \draw[thick, decorate,decoration=brace]
          (3.5,-1.90) --
          node[right] {\quad nothing, or~\Daimon, or \coef{W}}
          (3.5,-2.35);

        \draw[thick, decorate,decoration=brace]
          (3.5,-2.40) --
          node[right] {\quad destructors: \D or \Cm}
          (3.5,-3.3);

      \end{tikzpicture}
      \]
  \end{itemize}
  In particular, the approximated term contains exactly one occurrence of a
  recursive function name, and no other function name.
  Such a call is written~``\call{f x}{$b\,\g\,t$}''.\footnote{Since we don't
  deal with mutually recursive definitions, \f and \g are equal in our case.}
\end{defi}
\begin{lem}
  For any finite term~$T$,~$G(T)$ is a finite sum of calls.
\end{lem}

\subsubsection{Collapsing}
\begin{defi}
  Given~$B>0\in\N$, the \emph{weight collapsing function}~$\cB B \_$ replaces
  each weight~$\sum_p\coef{w_p}\p p$ (as in point~(3) of Lemma~\ref{lem:coef})
  by $\sum_p\coef{\cB B {w_p}}\p p$ where
  \[
  \cB B w = \begin{cases}
          -B  & \text{if $w < -B$}\\
          w & \text{if $-B \leq w < B$}\\
          \infty  & \text{if $B \leq w$}\\
  \end{cases}
  \]
\end{defi}
To bound the depth, we add~``$\coef{0}$'' below~$D$ constructors and above~$D$
destructors in the calls. When reducing, those weights will absorb the
constructors below~$D$ and the destructors above~$D$. For example, collapsing
$\C_1\C_2\C_3\coef{W}\C_4\m\C_5\m\C_6\m\C_7\m\x$ (where all the constructors
have the same priority) at depth~2 gives
\[
  \underbrace{\C_1\ \C_2}_{D=2}\ \coef{0}\ \C_3\ \coef{W}\ \C_4\m\ \C_5\m\
  \coef{0}\ \underbrace{\C_6\m\ \C_7\m}_{D=2}\ \x
  \quad\to^*\quad
  \underbrace{\C_1\ \C_2}_{D=2}\ \coef{1 + W - 2}\ \underbrace{\C_6\m\ \C_7\m}_{D=2}\ \x
\]
The actual definition is a little tedious.
\begin{defi}
  Let~$t$ be in normal form, given a positive bound~$D\in\N$, the \emph{depth
  collapsing function}~$\cD D \BLANK$ absorbs constructors below~$D$ and
  destructors above~$D$ into weights:
  \begin{myequation}
  \cD{i}{\big(\C\ t\big)} &\eqdef& \C\ \big(\cD{i-1}{t}\big)
     & \text{if $i>0$}\\
    \cD{i}{\st{\dots; \D_k=t_k; \dots}} &\eqdef&
    \st{\dots; \D_k=\cD{i-1}{t_k}; \dots}
     & \text{if $i>0$}\\
    \cD{i}{\delta} &\eqdef& \ccD{D}{\delta}
    % & \text{if $i\geq0$}
    \\
    \cD{i}{\big(\coef W\delta\big)} &\eqdef& \nf\Big(\coef W
    \big(\ccD{D}{\delta}\big)\Big)
     % & \text{if $i>0$}\\
    \\
    \cD{0}{t} &\eqdef& \nf\Big(\ccD{D}{\nf\big(\coef{0} t\big)}\Big) & (*)\\
  \end{myequation}
  and the following are applied to normal forms
  \begin{myequation}
    %
      % \noalign{\medbreak}
      % \ccD{i}{\big(\coef{W_1}t_1 * \dots * \coef{W_n}t_n\big)} &\eqdef&
      % \ccD{i}{(\coef{W_1}t_1)} * \dots * \ccD{i}{(\coef{W_n}t_n)}&
      % (**)\\
      % \ccD{i}{\big(\coef{W}\st{}\big)} &\eqdef& \coef{W}\st{} & (**)\\
      \ccD{i}{\big(\coef{W}\delta\f\,t\big)} &\eqdef&
      \coef{W}\ccD{i}{\delta}\f\,\cD{D}{t} &
      (\dagger)(**)\\
      \ccD{i}{\big(\coef{W}\delta\x\big)} &\eqdef& \coef{W}\ccD{i}{\delta}\x &
      (**)\\
      \noalign{\medbreak}
  \ccD{i}{\big(\delta\C\p p\big)} &\eqdef& \ccD{i-1}{\delta}\C\p p
    \\
  \ccD{i}{\big(\delta\D\p p\big)} &\eqdef& \ccD{i-1}{\delta}\D\p p
    \\
    \ccD{0}{\delta} &\eqdef& \delta\coef{0}
  \end{myequation}
  Note the following.
  \begin{itemize}
    \item
      The clauses are not disjoint and only the first applicable one is used.

    \item
      The innermost normal form in clause~$(*)$ ensures that the
      clauses~$(**)$ cover all cases (since weights absorb constructors on
      their right,~$\coef{0}t$ cannot start with constructors). The outermost
      normal form in clause~$(*)$ ensures the result is in normal form.

    \item
      Clause~$(\dagger)$ allows to collapse both the branch above the call
      to~\f and the argument of~\f. Because calls contain exactly one function
      name, this clause is used exactly once.
  \end{itemize}
\end{defi}
The following is obvious but depends on the fact that there are only finitely
many constructors / destructors.
\begin{lem}
  Given~$B>0$ and~$D\geq0$, the image of~$\cB B {\cD D {(\BLANK)}}$ is finite.
\end{lem}

\subsubsection{Composing calls}

\begin{defi}\label{def:collapse_call}
    \emph{Collapsed composition} is defined by
      \[
      \beta \ccomp_{B,D} \alpha \quad:=\quad
      \cB B {\cD D {\big(\beta \comp \alpha\big)}}
      \]
      Since the bounds are fixed, we usually omit them and write $\beta \ccomp \alpha$.
\end{defi}
% When treating mutually recursive definition, calls of the form~$\call{f x}{$b$
% g $u$}$ can only be composed with calls of the form~$\call{g x}{$c$
% h $v$}$.

\begin{lem}\label{lem:collapse_order}
  For any call~$\alpha, \beta$, we have
  \begin{itemize}
    \item
      $\cB B\alpha \leq \alpha$,

    \item
      $\cD D\alpha \leq \alpha$,

    \item
      $\beta \ccomp \alpha \leq \beta \comp \alpha$.
  \end{itemize}
\end{lem}
\begin{proof}\leavevmode
\begin{itemize}
  \item
    for~$\cB B\alpha$, replacing~$\coef W$ by~$\coef{\cB B W}$ results in a
    smaller term by contextuality and the fact that~$\cB B W\leq W$
    in~$\Coef$.

  \item
    for~$\cD D\alpha$, inserting some~$\coef0$ results in a smaller term by
    contextuality and the fact that~$\coef0t\leq t$. Normalizing can only make
    this smaller. \qedhere
\end{itemize}
\end{proof}

Unfortunately, unless~$B=1$ and~$D=0$, collapsed composition is \emph{not}
associative. For example, using~$B=2$, the calls $\alpha = \call{f x}{\coef{1}
f x}$ and $\beta = \call{f x}{\coef{-1} f x}$ give
\begin{itemize}
  \item
    $\beta \ccomp (\alpha \ccomp \alpha) = \call{f x}{\coef{\infty} f x}$
    because $\cB1{\cB1{1+1}+(-1)} = \cB1{\infty + (-1)} = \infty$,
  \item
    $(\beta \ccomp \alpha) \ccomp \alpha = \call{f x}{\coef{1} f x}$.
    because $\cB1{1 + \cB1{1+(-1)}} = \cB1{1+0} = 1$.
\end{itemize}
Fortunately, just like in previous work on termination~\cite{ph:SCT} the
next property will be sufficient.
\begin{lem}
  If $\sigma_n \comp \cdots \comp \sigma_1 \neq \Zero$, and if $\tau_1$ and
  $\tau_2$ are the results of computing $\sigma_n \ccomp \cdots \ccomp
  \sigma_1$ in two different ways, then $\tau_1$ and $\tau_2$ are
  \emph{compatible}, written $\tau_1 \coh \tau_2$. This means that there is
  some~$\tau\neq\Zero$ such that~$\tau_1 \leq \tau$ and $\tau_2 \leq \tau$.
\end{lem}
\begin{proof}
  Taking~$\tau = \sigma_n \comp \cdots \comp \sigma_1$ works, by
  repeated use of Lemma~\ref{lem:collapse_order}.
\end{proof}
%%%>>>1

%%% vim600:set foldmarker=<<<,>>> foldmethod=marker fileencoding=ascii spelllang=en spell: %%

\section{The Size-Change Principle}
\label{section:SCP}

\subsection{The Size-Change Principle} %%%<<<1
\label{sub:SCP}

Putting Proposition~\ref{prop:Gt_total}, Corollary~\ref{cor:path} and
Lemma~\ref{lem:operator_semantics} together, we get
\begin{cor}
  If all infinite paths in~$G(T_\f)$ have a total semantics,
  then the usual semantics of~\f is total in every total environment.
\end{cor}
Everything is now in place to apply the size-change principle from C. Lee, N.
Jones and A. Ben-Amram~\cite{lee_jones_benamram:SCT}, whose goal is precisely
to deduce some property of all infinite paths of a graph from some property on
its transitive closure. However, because collapsed composition isn't
associative, we need a variant of the combinatorial lemma at the heart of the
size-change principle. The following lemma is a slight generalization of
Lemma~2.1 from previous work on termination~\cite{ph:SCT}.
\begin{lem}\label{lem:ramsey}
  Suppose $(O,\leq)$ is a partial order, and~$F\subseteq O$ is a finite
  subset. Suppose moreover that~$\comp$ is a partial, binary, associative and
  monotonic operation on~$O$ and that~$\ccomp$ is a partial, binary, monotonic
  operation on~$F$ satisfying
  \[
    \forall o_1, o_2 \in F, (o_1 \ccomp o_2) \leq (o_1 \comp o_2)
  \]
  whenever~$o_1\comp o_2$ is defined.\footnote{In particular,~$o_1\ccomp o_2$
  is defined whenever~$o_1\comp o_2$ is.} Then every infinite sequence~$o_1,
  o_2, \dots$ of elements of~$F$ where each finite~$o_1\comp \cdots \comp o_n$
  is defined can be decomposed into
    \[
      \underbrace{o_1,\dots,o_{n_0-1},}_{\hbox{\scriptsize initial prefix}}
      \quad
      \underbrace{o_{n_0}, \dots, o_{n_1-1},}
      \quad
      \underbrace{o_{n_1}, \dots, o_{n_2-1},}
      \quad
      \dots
    \]
    where:
    \begin{itemize}
      \item all the~$(\dots(o_{n_k}\ccomp o_{n_k+1})\ccomp \cdots) \ccomp
        o_{n_{k+1}-1}$ are equal to the same~$r\in F$,
        \item $r$ is \emph{coherent}: there is some~$o\in O$ such that
        $r,(r\ccomp r) \leq o$.
    \end{itemize}
    In particular,
    \[
      \Big(o_{n_0} \comp \cdots \comp o_{n_1-1}
      \comp
      o_{n_1} \comp \cdots \comp o_{n_2-1}
      \comp
      \cdots
      \comp
      o_{n_{k-1}} \comp \cdots \comp o_{n_k-1}\Big)
      \quad\geq\quad
      \underbrace{r \comp r \comp \cdots \comp r}_{\text{$k$ times}}
    \]
\end{lem}
\begin{proof}
  This is a consequence of the infinite Ramsey theorem. Let~$(o_n)_{n\geq0}$
  be an infinite sequence of elements of~$F$. We associate a
  ``color''~$c(m,n)$ to each pair~$(m,n)$ of natural numbers where~$m<n$:
  \[
    c(m,n)
    \quad\eqdef\quad
      (...(o_{m} \ccomp o_{m+1}) \ccomp \cdots) \ccomp o_{n-1}
  \]
  Since~$F$ is finite, the number of possible colors is finite.
  By the infinite Ramsey theorem, there is an infinite set~$I\subseteq
  \N$ such all the~$(i,j)$ for~$i<j\in I$ have the same
  color~$r\in F$.  Write~$I = \{
  n_0<n_1<\cdots<n_k<\cdots\}$. If~$i<j<k \in I$, we have:
  \begin{myequation}
    r
  &=&
      (...(o_{i} \ccomp o_{i+1})\ccomp\cdots)\ccomp o_{j-1}\\
  &=&
      (...(o_{j} \ccomp o_{j+1})\ccomp\cdots)\ccomp o_{k-1}\\
  &=&
      (...((...(o_{i} \ccomp o_{i+1})\ccomp\cdots)\ccomp o_{j})
      \ccomp \cdots)\ccomp o_{k-1}\\
  \end{myequation}
  The first two equalities imply that
  \begin{myequation}
    r\ccomp r
    &=&
      \big((...(o_{i} \ccomp o_{i+1})\ccomp\cdots)\ccomp o_{j-1}\big)
      \ccomp
      \big((...(o_{j} \ccomp o_{j+1})\ccomp\cdots)\ccomp o_{k-1}\big)\\
  \end{myequation}
  If~$\ccomp$ is associative, this implies that~$r\ccomp
  r=r$. If not, we only get
  that both~$r$ and~$r\ccomp r$ are
  smaller than
  $
      o_{i} \comp\cdots\comp o_{j-1} \comp
      o_{j} \comp\cdots\comp o_{k-1}
  $.
\end{proof}

\begin{defi}
  Let~$G$ be a call-graph. Start with~$G^1=G$ and define the edges of
  $G^{n+1}$ to be those of~$G^n$, together with:
  \begin{quote}\it
    if $\sigma$ and $\rho$ are edges from \f to \g and from \g to \tt h
    in~$G^{n}$, then $\rho\ccomp \sigma$ is a new edge from \f to \tt h
    in~$G^{n+1}$.
  \end{quote}
  Finiteness of the set of bounded terms guarantees that this sequence
  stabilizes on some graph, written~$G^*$, called the \emph{transitive closure
  of~$G$}.
\end{defi}
% By construction,~$G^*$ satisfies:
% \begin{lem}
%   For every finite sequence~$s_0, s_1{=}s_0[\f{:=}t_0],\dots,
%   s_k{=}s_{k-1}[\f{:=}t_{k-1}]$ in~$G$, we have~$t\leq t_0\comp \cdots
%   t_{k-1}$ for some simple term~$t$ in~$G^*$.
% \end{lem}
We can now state and prove correctness of the size-change principle. We extend
the notions of branch and weight (Definition~\ref{def:branch_weight}) with a
new clause to deal with approximations:
\begin{itemize}
  \item
    $|\coef{W} \delta| = \coef{W} + |\delta|$.
  % \item
  % $|\coef{W} \varepsilon| =
  % \begin{cases}
  %   \coef{W} + |\varepsilon| & \text{if~$\varepsilon\neq\st{}$}\\
  %   \text{undefined} & \text{otherwise}\\
  % \end{cases}$
\end{itemize}
\begin{thm}[size-change principle]\label{thm:SCP}
  Suppose every loop~$\sigma = \call{f x}{$b$ f $u$}$ in $G^*$ that satisfies $\sigma \coh
  \sigma\ccomp \sigma$ (``$\sigma$ is coherent'') also satisfies one of the
  following two conditions:
  \begin{enumerate}

    \item
    either there is an even priority~$p$ such that:
      \begin{itemize}
        \item[-] the~$p$ component of weight~$|b|$ is strictly negative,
        \item[-] for all~$q>p$, the~$q$ component of~$|b|$ is positive;
      \end{itemize}

    \item
    or there is a branch~$\beta$ of~$u$ and an odd
    priority~$p$ such that:
    \begin{itemize}
      \item[-] the~$p$ component of weight~$|\beta|$ is strictly negative,
      \item[-] for all~$q>p$, the~$q$ component of~$|\beta|$ is positive;
    \end{itemize}
  \end{enumerate}
  then $\fix(G)$ is total.
\end{thm}

\begin{proof}
  By Lemma~\ref{lem:formula_fix_paths}, we only need to check that infinite
  paths are total. Let~$(s_k)$ be an infinite path of~$G$. If any prefix
  composes to~\Zero, the corresponding path is total. If no prefix composes
  to~\Zero, we can use Lemma~\ref{lem:ramsey}: such a path can be decomposed
  into
  \[
    t_0, \ t_1
    \quad
    \dots
    \quad
    t_{n_0} %%% s_0[t_0\comp \cdots \comp t_{n_0-1}]
    \quad
    \dots
    \quad
    t_{n_1} %%% = s_{n_0}[t_{n_0}\comp \cdots \comp t_{n_1-1}]
    \quad
    \dots
    \quad
    t_{n_2}
    \quad
    \dots
  \]
  where:
  \begin{itemize}
    \item
      all the~$t_{n_{k+1}-1}\ccomp\cdots \ccomp t_{n_k}$ are equal to the
      same~$t = \call{\f}{$b \, \f \, u$}$,

    \item
      $t$ is \emph{coherent}: $t\ccomp t\coh t$.
  \end{itemize}
  In particular, we have~$t_{n_{k+1}-1}\comp\cdots \comp t_{n_k}\geq t$.

  \medbreak
  Suppose that~$t$ satisfies the first condition. If we write~$T_0$
  for~$t_0\comp \cdots \comp t_{n_0-1}$, we have
  \begin{myequation}
    \dirsup_k s_k(\Daimon)
    &=&
    \dirsup_k t_0 \comp t_1 \comp \cdots \comp t_k(\Daimon)\\
    &\geq&
    % \dirsup_j t_0 \comp \cdots \comp t_{n_0-1} \comp t^j(\Daimon)\\
    \dirsup_j T_0 \comp \underbrace{t \comp \cdots \comp t}_{\text{$j$ times}} (\Daimon)\\
    &=&
    \dirsup_j T_0 \comp \, (b \f\, u)^j(\Daimon)\\
    &\geq&
    \dirsup_j T_0 \comp \, (b \f\, \Daimon)^j(\Daimon)\\
    &=&
    \dirsup_j T_0 \comp \, b^j \Daimon\\
    &\geq&
    T_0 \comp \dirsup_j b^j \Daimon\\
  \end{myequation}
  where~$b^j$ is simply~$b\,b\,\dots\,b$.

  Now, for any simple value~$v$, $b^k\Daimon(v)$ is either~\Zero or has at
  least~$k$ constructors of priority~$p=2q$ coming from~$b^k$ above any
  constructor coming from~$v$. At the limit, there will be infinitely many
  constructors of priority~$p=2q$, all coming from~$b$. Because~$b$ doesn't
  add constructors of priority greater than~$p=2q$, the limit will be total.

  \medbreak
  Dually, if~$t$ satisfies the second condition. We have
  \begin{myequation}
    \dirsup_k s_k(\Daimon)
    &=&
    \dirsup_k t_0 \comp t_1 \comp \cdots \comp t_k(\Daimon)\\
    &\geq&
    \dirsup_j T_0\comp t^j(\Daimon)\\
    &=&
    \dirsup_j T_0\comp (b\,\f\, u)^j(\Daimon)\\
    &\geq&
    \dirsup_j T_0\comp (\Daimon\,\f\, u)^j(\Daimon)\\
    &=&
    \dirsup_j T_0\comp (\Daimon\,\f\, u^j)(\Daimon)\\
    &\geq&
    T_0 \comp \dirsup_j \Daimon u^j\\
  \end{myequation}
  where~$u^j$ is obtained by replacing all~\x in~$u^{j-1}$ by~$u$: $u^j =
  u[\x:=u^{j-1}]$.
  By hypothesis,~$u$ contains a branch~$\beta$ and there is an odd~$p$ s.t.
  $|\beta|_p < 0$, so that~$u^j$ contains a branch~$\beta\beta\dots \beta$.
  Such a branch globally removes at least~$j$ constructors of
  priority~$p=2q+1$ and doesn't remove constructors of greater priority.
  If~$v$ is a total value, then each~$u^k(v)$ can only be non-\Zero if~$v$
  contains at least~$k$ constructors of priority~$p=2q+1$ and no constructors
  of greater priority. At the limit, the only values such that $\dirsup_k
  u^k(v)$ are non-\Zero are values that contain a branch with an infinite
  number of constructors of priority~$p=2q+1$ and no constructor of priority
  greater than~$p$. This is impossible for total values!

  Note that in both the first and second case, the branch usually contains
  approximations, so that while we may not know exactly which constructors are
  added (in the first case) or removed (in the second case), nor have an exact
  count, we have a bound of how many are added / removed, which is enough for
  the argument.
  \end{proof}

%%%%>>>1

\subsection{Examples} %%%<<<1

Let's apply Theorem~\ref{thm:SCP} to the examples from the introduction. With
explicit priorities, the definition of \tt{nats} is
{\small\begin{alltt}
    val nats : nat -> stream(nat)
      | nats x = \st{ Head\p0 = x ; Tail\p0 = nats (Succ\p1 x) }
\end{alltt}}\noindent
The call-graph contains a single call~$\sigma=\call{nats}{Tail\p0 nats Succ\p1
x}$.
If we use the bound~$D=B=1$, a single step is necessary to build the
transitive closure:
\begin{itemize}

  \item
    the first composition~$\sigma\comp\sigma = \call{nats}{Tail\p0 Tail\p0
    nats (Succ\p1 Succ\p1 x)}$ collapses\footnote{recall that the branch above
    the recursive call uses negated weights
    (Section~\ref{subsub:dual_weights})} to~$\rho=\sigma\ccomp\sigma =
    \call{nats}{Tail\p0 \coef{-1}\p0 nats (Succ\p1 \coef{\infty}\p1 x)}$,

  \item
    after that, all compositions are equal to~$\rho$.

\end{itemize}
The term~$\rho$ satisfies the first property of Theorem~\ref{thm:SCP}. By
Lemma~\ref{lem:ramsey}, all infinite compositions of~$\sigma$ eventually
reduce to an infinite composition of~$\rho$. But the only infinite branch
resulting from an infinite composition of~$\rho$
is~``\tt{Tail}\p0\tt{Tail}\p0\dots'', which has even principal priority. The
recursive definition is total. Note that taking~$D=0, B=1$ would have worked
just as well.

In general, the infinite composition of~$\rho$ could also contain infinite
branches coming from the argument~\x but since totality of a function only
depends on its values on total arguments, we can suppose all infinite branches
coming from~\x have even principal priority.

\medbreak
The \tt{length} function has a call-graph with a single call:
\[
  \sigma = \call{length x}{Succ\p1 length (.Snd\p0 Cons\p1\m x)}
\]
With~$B=1$ and~$D=0$, the transitive closure is reached after one step.
Besides~$\sigma$, it contains
\[
  \rho = \sigma\ccomp\sigma = \call{length x}{\coef{-1}\p1
  length (\coef{\coef{-1}{\p0} + \coef{-1}\p1} x)}
\]
The call~$\rho$ is coherent. It doesn't satisfy the first property of
Theorem~\ref{thm:SCP} but the second. By Lemma~\ref{lem:ramsey}, infinite
compositions of~$\rho$ remove an infinite number of constructors/destructors
of priorities~$0$ and~$1$ from the argument, as seen in the
weight~$\coef{\coef{-1}\p0+\coef{-1}\p1}$. (Those correspond to~\tt{Succ}\p1
and~\tt{.Snd}\p0.) As a result, any argument leading to infinite compositions
cannot be total. The recursive definition is total.

\medbreak
The definition of~\tt{bad\_s} has two recursive calls:
{\small\begin{alltt}
    val bad_s : stream(stree)
      | bad_s = \st{ Head\p0 = Node\p1 bad_s ; Tail\p0 = bad_s }
\end{alltt}}\noindent
Its call-graph contains~$\sigma_1 = \call{bad\_s}{Head\p0 Node\p1 bad\_s}$
and~$\sigma_2 = \call{bad\_s}{Tail\p0 bad\_s}$.
For~$B=D=1$, the transitive closure stabilizes after one step, and it contains
3 calls besides~$\sigma_1$ and~$\sigma_2$:
\begin{itemize}
  \item
    $\rho_{1,1} = \sigma_1\ccomp\sigma_1 = \sigma_1\ccomp\sigma_2=\call{bad\_s}{Head\p0 \coef{\coef{-1}\p0+\coef{-1}\p1} bad\_s}$

  \item
    $\rho_{2,2} = \sigma_2\ccomp\sigma_2 = \call{bad\_s}{Tail\p0 \coef{-1}\p0 bad\_s}$

  \item
    $\rho_{2,1} = \sigma_2\ccomp\sigma_1 = \call{bad\_s}{Tail\p0 \coef{\coef{-1}\p0+\coef{-1}\p1} bad\_s}$

\end{itemize}
Those 3 calls are coherent, but while~$\rho_{2,2}$ satisfies the first
property of Theorem~\ref{thm:SCP} neither~$\rho_{1,1}$ nor~$\rho_{2,1}$ do
because the maximal priority comes from~\coef{-1}\p1. Because there is no
argument to~\tt{bad\_s}, they don't satisfy the second property either. It
means an infinite composition of recursive calls will necessarily generate an
infinite branch with infinitely many constructors of priority~$1$ and~$0$,
which is a non-total branch. The recursive definition is thus rejected by the
totality checker, as it should. Changing~$B$ and~$D$ doesn't make a
difference.

\medbreak
Theorem~\ref{thm:SCP} is strong enough to deal with mixed inductive and
coinductive types. Recall the definition of \tt{sums} from
page~\pageref{chariot:sums}
{\small\begin{alltt}
  val sums : nat\p1 -> stream\p0(list\p1(nat\p1)) -> stream\p0(nat\p1)
    | sums acc \st{ Head\p0 = Nil\p1 ; Tail\p0 = s } =
        \st{ Head\p0 = acc ; Tail\p0 = sums (Zero\p3\st{}\p4) s }
    | sums acc \st{ Head\p0 = Cons\p1 \st{Fst\p0 = n ; Snd\p0 = l} ; Tail\p0 = s } =
        sums (add acc n) \st{ Head\p0 = l ; Tail\p0 = s }
\end{alltt}}\noindent
Because of the second clause, this definition isn't guarded. It is productive
because this second clause cannot occur infinitely many times consecutively.
Agda doesn't detect this definition as total. Provided~$D>0$, this will be
detected by the totality checker and this definition will thus be accepted as
total. With~$B=D=1$, the transitive closure of the call-graph will contains
the following coherent loops:
\begin{itemize}

  \item
    $\rho_1$, coming from compositions of the first call with itself:
    \[
    \call{sums $\x_1$ $\x_2$}%
         {$\underline{\tt{Tail\p0\coef{-1\p0}}}$ %
         sums
         $(\tt{Zero}\p1)$
         $(\coef{-1}\p0\tt{.Tail}\p0\,\x_2)$}
    \]
    where the~\coef{-1}\p0 corresponds to the collapse of~$\tt{Tail}\p0$,

  \item
    $\rho_2$, coming from compositions of the second call with itself:
    \[
    \call{sums $\x_1$ $\x_2$}{sums
    $(\Daimon...)$
    %$\Daimon\{ \x_1, \coef{-1}\p1\tt{.Head}\p0\,\x_2 \}$
    \st{\underline{\tt{ Head\p0=$\coef{-1}\p1\tt{.Head}\p0\,\x_2$}} ;
                       Tail\p0=$\tt{.Tail}\p0\,\x_2$ }}
    \]
    where~\coef{-1}\p1 corresponds to the collapse of
    \tt{Cons\p1\m{}Cons\p1\m},\footnote{The ``$\Daimon...$'' comes from the
    application~\tt{add acc n} and doesn't play any role in this example.}

  \item
    $\rho_3$, coming from compositions of the first and second call:
    \[\begin{array}{rl}
      \tt{sum $\x_1$ $\x_2$} \quad\mapsto\quad
         \underline{\tt{Tail}\p0\coef{-1}\p0} \ \tt{sums} \ (\Daimon...)\ %
         \bstruct & \tt{Head}\p0 = \coef{\coef{-1}\p0{+}\coef{-1}\p1}\tt{.Tail}\p0\,\x_2\ \tt{;}\\
                 & \tt{Tail}\p0 = \coef{-1}\p0\tt{.Tail}\p0\,\x_2 \ \ \estruct
    \end{array}\]
    where \coef{\coef{-1}\p0{+}\coef{-1}\p1} comes from the collapse of
    $\tt{Cons}\p1\m\tt{.Head}\p0$ and~\coef{-1}\p0 from the collapse
    of~\tt{.Tail}\p0.

\end{itemize}
Both~$\rho_1$ and~$\rho_3$ satisfy the first property of
Theorem~\ref{thm:SCP}, while~$\rho_2$ satisfies the second property. This
definition is total.

%%%>>>1

\subsection{Implementing the Totality Checker} %%%<<<1

Implementing the totality checker based on Theorem~\ref{thm:SCP} for a
first-order functional programming language like \chariot is relatively
straightforward.

\begin{enumerate}

  \item
    During type checking / type inference, annotate all constructors appearing
    in the recursive definition with their type.

  \item
    Construct the parity game containing all these types~\cite{ph:totality}.

  \item
    Annotate all constructors and destructors appearing in the recursive
    definition with their priorities.
    The types themselves can be forgotten at this point.

  \item
    The type of calls in normal forms is a simple first-order inductive type.
    Define relevant functions on calls, namely composition, and collapsing.
    Note that composition is implicitly followed by reduction to normal form,
    which must be done during composition.

  \item
    Compute the call-graph of the definition. This is easy for a language like
    \chariot because each clause can be treated independently and each call
    will be in normal form by construction.

  \item
    Compute the transitive closure of the call graph using the previously
    defined functions (composition and collapsing).

  \item
    Loop over all loops of the transitive closure of the call-graph. If a loop
    is coherent, check that it satisfies one of the properties of
    Theorem~\ref{thm:SCP}. If all of them do, the definition is total.

\end{enumerate}
Parallel arcs in the call-graph correspond to non-deterministic sums
and because~$u+v=u$ whenever~$u \leq v$, not all calls need to be added to the
call-graph: if a call is greater than some existing call, it can be ignored.
Doing so requires implementing the order as well. Because calls are 
kept in normal form for collapsing, we can use the syntax directed inductive
relation~$\sleq$ from
Appendix~\ref{app:inductive_order-O} and~\ref{app:inductive_order-A} instead
of~$\leq$.

\medbreak
The last part requires checking coherence for loops in the transitive closure
of the call-graph. In practice, it looks like checking loops for
which~$\sigma=\sigma\ccomp\sigma$ is enough but I haven't tried very hard to
prove this fact.
Devising an inductive characterization of coherence is difficult but we can
simplify things by using a weaker inductive relation that is only implied by
coherence. That means we may end up checking more loop than strictly necessary
but the result is provably correct.\footnote{And I haven't found any
definition where this stricter check changes the result of the totality
checker.} This is described in Appendix~\ref{app:inductive_coherence-A}.

%%%>>>1

%%% vim600:set foldmarker=<<<,>>> foldmethod=marker fileencoding=ascii spelllang=en spell: %%

\section*{Concluding Remarks}

\subsubsection*{Complexity}

Since this totality test extends the termination test described
in~\cite{ph:SCT} and thus the usual size-change termination principle, it is
at least P-space hard. Hardness comes from computing the transitive closure of
the call-graph. It seems to work well in all the examples we tried, but there
are ad hoc examples of very short definitions that lead to exponential
totality checking.
We think (hope) that such examples do not arise naturally. Letting the user
choose the bounds~$B$ and~$D$ (with sane default values) limits the extra
complexity cost to the definitions that really need it. It is nevertheless
difficult to know how this will scale for very big definitions. The situation
is thus not too different from Agda, where the termination checker can become
very slow on big definitions.
This should be contrasted to Coq, where the design choice has always been to
have a simple totality checker with low complexity.

\subsubsection*{Choosing the bounds}\label{sub:bounds}

The totality test is parameterized by the bounds~$B>0$ and~$D\geq0$. In many
simple cases, setting $B=1$ and~$D=0$ is enough but increasing the bounds
locally is interesting in the following cases.

\begin{itemize}

  \item
    Increasing~$D$ helps detecting ``incompatible'' calls. For example, the
    following ad hoc example is accepted with~$D=1$ but rejected with~$D=0$:
    {\small\begin{alltt}
      val f (C1 x) = 0
        | f (C2 x) = f (C1 x) \end{alltt}}\noindent
    The call-graph has a single vertex with a single call~$\alpha=\call{f}{f
    (C1 C2\m x)}$. With~$D=0$, this call is collapsed to \call{f}{f
    (\coef{0}x)} which doesn't pass the totality test because this loop is
    idempotent but doesn't decrease. With~$D=1$, this call is unchanged but is
    not idempotent: $\alpha\ccomp\alpha = \Zero$, and it passes the totality
    test.

    Similarly, if some parts of the argument increase while other parts
    decrease, too small a~$D$ can hide totality:
    {\small\begin{alltt}
      val f \st{Fst\p0=0 ; Snd\p0=x} = x
        | f \st{Fst\p0=Succ\p1 x1 ; Snd\p0=x2} = f \st{Fst\p0=x1 ; Snd\p0=Succ\p1 x2} \end{alltt}}\noindent
    requires~$D>0$ to pass the totality test. With~$B=D=0$, we get the coherent
    loop
    \[
    \call{f}{f$\big(\Daimon($
      \coef{\coef{\infty}\p0 + \coef{-1}\p1} \x
      +
      \coef{\coef{\infty}\p0 + \coef{\infty}\p1} \x
      $)\big)$}
    \]
    which doesn't satisfy the hypothesis of Theorem~\ref{thm:SCP}.
    % With~$B=0$
    % and~$D=1$, we get
    % \[
    % \call{f}{f \st{ Fst = \coef{-1}\p1 x ; Snd = \coef{\infty}\p1 x }}
    % \]
    % which does satisfy the second property of Theorem~\ref{thm:SCP}.

  \item
    Increasing~$B$ isn't as useful. It helps detect totality when some calls
    increase the size of their argument (or dually, remove some output
    constructor). For example, the following ad hoc example of mutually
    inductive definitions is accepted with~$B=2$ and~$D=0$ but rejected
    with~$B=1$ and~$D=0$:
    {\small\begin{alltt}
        val s1 = s2.Tail\p0
        and s2 = \st{ hd\p0 = Zero\p1; Tail\p0 = \st{ hd\p0=1\p1; Tail\p0=s1 }} \end{alltt}}\noindent
    The call-graph has 2 vertices and 2 arcs: \call{s1}{\coef1\p0 s2} and
    \call{s2}{\coef{-2}\p0 s1}. When projecting with~$B=2$, the composition
    gives \call{s1}{\coef{-1}\p0s1} (and similarly for \tt{s2}), which passes
    the totality test. When projecting with~$B=1$, the first arc gives
    \call{s1}{\coef{\infty}\p0s2} which gives compositions
    of~\call{s1}{\coef{\infty}\p0s1} (and similarly for~\tt{s2}), which doesn't
    pass the totality test.

\end{itemize}
In practice, we've found that~$B=2$ and~$D=2$ is enough for most cases. In the
few situations where increasing~$B$ or~$D$ is helpful, the programmer can
change those bounds locally.

Note that none of those examples are detected as correct by the current
termination checker in Agda.

\subsubsection*{Strength of the totality checker}

This paper only proves correctness of the totality checker. It doesn't prove
anything about its strength. Another provably correct totality checker is the
one that always returns ``\tt{I DON'T KNOW}''.
The only argument in favor of the totality checker is of a practical nature:
experimenting with \chariot shows that it is enough for many recursive
functions. General results like ``all structurally recursive definitions are
total'', ``all syntactically guarded definitions are total'', etc. are
certainly true but not investigated here.

\subsubsection*{Higher order types}
\label{sub:higher_order}

The implementation of \chariot deals with some higher order
datatypes. With $\tt b$-branching trees (coinductive) defined as
{\small\begin{alltt}
  codata tree('b, 'n) where
      child : tree('b, 'n) -> ('b -> tree('b, 'n))
\end{alltt}}\noindent
or (inductive)
{\small\begin{alltt}
  data tree('b, 'n) where
      root : unit -> tree('b, 'n)
    | fork : ('b -> tree('b, 'n)) -> tree('b, 'n)
\end{alltt}}\noindent
the corresponding \tt{map} functions passes the totality test. The theory should
extend straightforwardly to account for this kind of datatypes.

\subsubsection*{Dependent Types}
\label{sub:dependent_types}

This totality checker could deal with dependent types by simply tagging any
definition involving dependent types with ``\tt{I DON'T KNOW}''. Of course,
extending it to do something interesting on dependent types would be
preferable.
Many useful dependent types like ``lists of size $n$'' can be embedded in
bigger non dependent datatypes like (``lists'' in this case). Because the
totality checker is essentially untyped, those types, together with dependent
sums could be dealt with by using the totality checker unchanged. That, and
the extension to some higher order as described above would go a long way to
provide a theoretically sound totality checker for a dependent languages like
Agda.\footnote{Types that cannot be analyzed with this totality checker
probably include things like functions types with non constant arity.}

\bibliographystyle{alphaurl}
\bibliography{chariot}

%%%>>>1

\appendix

\section{Inductive order}   %%%<<<1
\label{app:inductive_order}

This section describes some simple inductive relations that are much easier to
implement than the order~$\leq$ on~$\Op_0$ and~$\A_0$, and the
coherence~$\coh$. This is possible because the implementation only needs to
deal with terms in normal form.

\subsection{Inductive order on normal forms in~\Op}     %%%<<<2
\label{app:inductive_order-O}

\begin{defi}\label{def:inductive_order_NF}
  The relation~$\sleq$ on normal forms of~$\Op _0$ is inductively generated by the
  following rules:
  \[\begin{array}{cc}
    \multicolumn{2}{c}{
    \Bigg.\Rule{\forall j, \exists i, s_i \sleq t_j}{\textstyle\sum_i s_i \sleq \sum_j t_j}{\scriptscriptstyle\sleq_+}
    }\\
    \Bigg.
    \Rule{}{\x\sleq\x}{\scriptscriptstyle\sleq_\x}
    &
    \Rule{s\sleq t}{\f\,s\sleq\f\,t}{\scriptscriptstyle\sleq_\f}
    \\
    \Bigg.
    \Rule{s\sleq t}{\C\,s\sleq\C\,t}{\scriptscriptstyle\sleq_\C}
    &
    \Rule{s_1\sleq t_1\qquad\dots\qquad s_n\sleq t_n}
         {\st{\D_1=s_1; \dots; \D_n=s_n}\sleq\st{\D_1=t_1; \dots; \D_n=t_n}}{\scriptscriptstyle\sleq_{\st{\D_1; \dots; \D_n}}}
    \\
    \Bigg.
    \Rule{s\sleq t}{\Cm s\sleq\Cm t}{\scriptscriptstyle\sleq_\Cm}
    &
    \Rule{s\sleq t}{\D\,s\sleq\D\,t}{\scriptscriptstyle\sleq_\D}
    \\
    \Bigg.
    \Rule{s\sleq t}{\Daimon s\sleq \Daimon\delta t}{\scriptscriptstyle\sleq_{\Daimon1}}
    &
    \Rule{\Daimon s \sleq \nf(\Daimon t)}{\Daimon s\sleq t}{\scriptscriptstyle\sleq_{\Daimon2}}
    \text{\small\emph{(provided~$t$ doesn't start with \Daimon)}}
    \\
  \end{array}\] where in rule~$\sleq_{\Daimon1}$,~$\delta$ is any sequence of
  destructors \Cm/\D and function names \f, \g, etc.
\end{defi}
The last two clauses are the most interesting. The rule~$\sleq_{\Daimon2}$
requires normalizing~$\Daimon t$ before the inductive step,
and~$\sleq_{\Daimon1}$ requires finding the appropriate prefix~$\delta$. This
is decidable as there are only finitely many prefixes to check. The first rule
is computationally the most complex one, as it potentially requires
checking~$s_i \sleq t_j$ for all~$i,j$. Here again, there are only finitely
many possibilities.
Note for any given~$s$ and~$t$, there is at most one rule that can be used to
derive~$s\sleq t$, which makes the~$\sleq$ relation straightforward to
implement.

\begin{lem}\label{lem:order_NF}
  $s\sleq t$ implies~$s\leq t$.
\end{lem}
\begin{proof}
This is a straightforward induction on~$s\sleq t$. For example, suppose~ the
last rule was~$\sleq_\C$, with conclusion~$\C s\sleq\C t$ and premise~$s\sleq
t$. By induction, we have~$s\leq t$, which implies that~$\C s\leq \C t$ by
contextuality of~$\leq$.
All other cases are treated similarly, with an extra bit of reasoning for
rules~~$\sleq_+$, $\sleq_{\Daimon1}$ and~$\sleq_{\Daimon2}$:
\begin{itemize}
  \item
    For $\sleq_+$, we use point~(1) of Lemma~\ref{lem:simple_consequences}.

  \item
    for $\sleq_{\Daimon1}$, we have $s\leq t$ by induction. From there, we have
    \[\begin{array}{rcll}
      \Daimon s & \leq & \Daimon t
      &\text{\footnotesize(definition of~$\leq$: contextuality)}\\
      &\approx&
      \Daimon \Daimon t
      &\text{\footnotesize(definition of~$\leq$)}\\
      &\leq&
      \Daimon \delta \Daimon t
      &\text{\footnotesize(Lemma~\ref{lem:simple_consequences})}\\
      &\leq&
      \Daimon \delta t
      &\text{\footnotesize(definition of~$\leq$: contextuality)}\\
    \end{array}\]

  \item
    For $\sleq_{\Daimon2}$, we have that $\Daimon s\leq \nf(\Daimon t)$ by
    induction. Since~$\nf(\Daimon t) \leq \Daimon t \leq t$, we get
    that~$\Daimon s \leq t$ by transitivity. \qedhere
\end{itemize}
\end{proof}  % I inserted this \end{proof}, because there was no qed. Please tell us, if this is wrong
The rest of this section will show that~$\sleq$ is in fact equivalent
to~$\leq$ restricted to normal forms

\begin{prop}\label{prop:order_NF}
  For all terms~$s$ and~$t$,~$s\leq t$ implies~$\nf(s)\sleq \nf(t)$.
\end{prop}
The proof is a little tedious. We decompose it into several auxiliary lemmas
with straightforward proofs.

\begin{lem}\label{lem:inductive_order_aux}\leavevmode
  \begin{enumerate}

    \item
      If~$s\sleq t$ are simple normal forms, then~$\nf(\D s) \sleq \nf(\D t)$.

    \item
      The same holds when~$\D$ is replaced by~$\Cm$ or~\f.

    \item
      The same holds when~$\D$ is replaced by~$\Daimon$.
  \end{enumerate}
\end{lem}
\begin{proof}\leavevmode
\begin{enumerate}
  \item
    We prove the first point by induction on~$s\sleq t$, looking at the last
    rule.

    \begin{itemize}
      \item
        Rule~$\sleq_+$: by induction hypothesis, we have that~$\forall j,
        \exists i, \nf(\D s_i)\leq \nf(\D t_j)$. This implies that~$\nf(\D s)
        = \sum_i\nf(\D s_i) \sleq \sum_j \nf(\D t_j)=\nf(\D t)$.

      \item
        Rule $\sleq_\x$: the result holds by definition.

      \item
        Rule~$\sleq_\f$: because~$\f s$ and~$\f t$ are in normal form,
        $\nf(\D\f s) = \D\f s$ and~$\nf(\D\f t)=\D\f t$. The result holds by
        definition of~$\sleq$. (No induction necessary.)

      \item
        Rule~$\sleq_\C$, or~$\sleq_\st{\dots}$ when the record doesn't contain
        the~\tt{D} field: $\nf(\D s)=\nf(\D t)=\Zero$. The result follows from
        rule~$\sleq_+$.

      \item
        Rule~$\sleq_\st{\dots; \D; \dots}$: because~$\nf(\D\st{\dots; \D=u;
        \dots} = \nf(u)$, the result holds by definition.

      \item
        Rule~$\sleq_{\D'}$: because~$\D' s$ is in normal form by hypothesis,
        we have~$\nf(\D\D' s) = \D\D' s$. Similarly,~$\nf(\D\D't) = \D\D't$,
        and the result holds by definition.

      \item
        The same reasoning works for rule~$\sleq_\Cm$.

      \item
        Rule~$\sleq_{\Daimon1}$: we have~$\nf(\D\Daimon s) = \Daimon s$
        and~$\nf(\D\Daimon\delta t) = \Daimon\delta t$. The result hold
        trivially.

      \item
        Rule~$\sleq_{\Daimon2}$ is the most complex case. Suppose we
        have~$\Daimon s \leq t$ because~$\Daimon s \sleq \nf(\Daimon t)$.

        Because~$\Daimon s$ is already in normal form by hypothesis, we
        have~$\nf(\D \Daimon s) = \Daimon s$. We thus need to check
        that~$\Daimon s\sleq \nf(\D t)$.
        Using~$\sleq_{\Daimon2}$, we need to show that~$\Daimon s \sleq
        \nf(\Daimon\nf(\D t))$. By Lemma~\ref{lem:nf}, $\nf(\Daimon\nf(\D
        t))=\nf(\Daimon\D t)$. We do a case analysis on~$t$:

        \begin{itemize}
          \item
            if $t=\x$, $t=\f t'$, $t=\Cm t'$ or~$t=\D' t'$: because~$t$ in
            already in normal form, we have~$\nf(\Daimon \D t) = \Daimon\D t$.
            By hypothesis,~$\Daimon s \sleq \nf(\Daimon t)$, which an only be
            proved using rules~$\sleq_+$ (if~$\nf(\Daimon t)$ involves sums)
            followed by~$\sleq_{\Daimon1}$. So all summands of~$\nf(\Daimon
            t)$ are of the form~$\Daimon\delta t'$ with~$s\sleq t'$.

            But when~$t$ has one of the above shapes,~$\nf(\Daimon t) =
            \Daimon t$, so that~$t$ itself is of the form~$\delta t'$
            with~$s\sleq t'$. This implies that~$\D t$ is also of the
            form~$\delta t'$ with~$s'\sleq t'$. From that, we get
            that~$\Daimon s \sleq \Daimon\D t$ using rule~$\sleq_{\Daimon1}$
            and thus that~$\Daimon s \sleq \D t$ using
            rule~$\sleq_{\Daimon2}$.

          \item
            If $t=\C t'$, or if~$t$ is a record without the~\tt{D} field,
            then~$\nf(\D t)=\Zero$ and the result holds trivially.

          \item
            If~$t=\st{\dots; \D=t'; \dots}$, and because~$t$ is in normal
            form, we have that~$\nf(\D t)=t'$. By hypothesis, we now
            that~$\Daimon s\sleq \nf(\Daimon t)$, which can only be proved
            using rules~$\sleq_+$ and~$\sleq_{\Daimon1}$. This implies that
            all summands of~$\nf(\Daimon t)=\sum_j \nf(\Daimon t_j)$ are of
            the form ~$\delta t''$ with~$s\sleq t''$. Since all summands
            of~$\nf(\Daimon t')$ are summands of~$\nf(\Daimon t)$, we can
            conclude that~$\Daimon s \sleq \nf(\D t)$.

          \item
            If~$t=\Daimon t'$, then~$\nf(\Daimon \D t) = t$. The result holds
            trivially.
        \end{itemize}

    \end{itemize}

  \item
    The same reasoning applies to the case where we replace~$\D$ by~$\Cm$
    or~\f.

  \item
    The third point is proved similarly:
    \begin{itemize}
      \item
        rule~$\sleq_+$: follows directly from the induction hypothesis and
        linearity of~\Daimon.

      \item
        rule~$\sleq_\x$: the result holds by definition.

      \item
        rule~$\sleq_\f$: because~$\f s$ and~$\f t$ are in normal form, we
        have~$\nf(\Daimon\f s) = \Daimon\f s$ and~$\nf(\Daimon\f t) =
        \Daimon\f t$. The result holds by definition of~$\sleq$
        (rule~$\sleq_{\Daimon1}$).

      \item
        rule~$\sleq_\C$: because~$\nf(\Daimon\C s) = \nf(\Daimon s)$
        and~$\nf(\Daimon\C t) = \nf(\Daimon t)$, the result holds by
        induction.

      \item
        rule~$\sleq_\st{\dots;\D;\dots}$: because~$\nf(\Daimon\st{\dots;
        \D_i=u_i; \dots} = \sum_i \nf(\Daimon u_i)$, the result follows from
        the induction hypotheses and rule~$\sleq_+$.

      \item
        rule~$\sleq_\Cm$: because both~$\Cm s$ and~$\Cm t$ are in normal form,
        we have~$\nf(\Daimon\Cm s) = \Daimon\Cm s$ and similarly for~$t$. The
        result hold by definition (rule~$\sleq_{\Daimon1}$).

      \item
        The same reasoning works for rule~$\sleq_\D$.

      \item
        rule~$\sleq_{\Daimon1}$: both~$\Daimon s$ and~$\Daimon t$ are in
        normal form, so that we have~$\nf(\Daimon\Daimon s) = \Daimon s$ and
        similarly for~$t$. The result hold by hypothesis.

      \item
        rule~$\sleq_{\Daimon2}$: $\Daimon s$ is in normal, so
        that~$\nf(\Daimon\Daimon s) = \Daimon s$. We need to prove
        that~$\Daimon s \sleq \nf(\Daimon t)$, which is precisely the premise
        of rule~$\sleq_{\Daimon2}$. \qedhere
    \end{itemize}
  \end{enumerate}
\end{proof}
Note that the fact that the terms are in normal form is crucial in the proof.

\begin{lem}\label{lem:inductive_order_contextual}
  The relation~$\sleq$ is contextual: if~$s\sleq t$ for some~$s$, $t$ simple
  normal forms, and if~$C$ is a context, then~$\nf(C[\y:=s]) \sleq
  \nf(C[\y:=t])$.
\end{lem}
\begin{proof}
This is done by induction of the context~$C$, where the difficult inductive
steps are taken care of by the preceding lemma.
\begin{itemize}
  \item
    If~$C=\x$ or~$\C=\y$, the result holds trivially.

  \item
    If~$C=\st{\dots; \D_i=C_i; \dots}$, then we know that
    each~$\nf(C_i[\y:=s]) \sleq \nf(C_i[\y:=t])$ by induction.
    Since~$\nf(C[\y:=s]) = \st{\dots; \D_i=\nf(C_i[\y:=s]); \dots}$ and
    similarly for~$C[\y:=t]$, the results holds by definition of~$\sleq$.

  \item
    The same reasoning works for the cases~$C=\f C'$ or~$C=\C C'$.

  \item
    If~$C=\D C'$, then we know that~$\nf(C'[\y:=s]) \sleq \nf(C'[\y:=t])$ by
    induction. Lemma~\ref{lem:inductive_order_aux} implies
    that~$\nf(C[\y:=s])\sleq\nf(C[\y:=t])$.

  \item
    The cases~$C=\Cm C'$ is treated similarly, with the help of
    Lemma~\ref{lem:inductive_order_aux}.

  \item
    If~$C=\Daimon C'$ then we know that~$\nf(C'[\y:=s]) \sleq \nf(C'[\y:=t])$
    by induction. The previous lemma implies that~$\nf(\Daimon
    C'[\y:=s])\sleq\nf(\Daimon C'[\y:=t])$. \qedhere
\end{itemize}
\end{proof}

\begin{lem}\label{lem:inductive_order_transitive}
  The relation~$\sleq$ is reflexive and transitive.
\end{lem}
\begin{proof}
Reflexivity is an obvious induction.
For transitivity, suppose~$s\sleq t$ and~$t\sleq u$. We proceed by induction
on~$t\sleq u$ and case inspection on~$s\sleq t$.

\begin{itemize}
  \item
    if both~$t\sleq u$ and~$s\sleq t$ come from~$\sleq_+$, each~$u_j$ is
    greater than some~$t_i$ which is greater than some~$s_k$. By induction,
    each~$u_j$ is thus greater than some~$s_k$, implying than~$\sum_k s_k\sleq
    \sum_i u_i$ by rule~$\sleq_+$.

  \item
    if only~$t\sleq u$ comes from~$\sleq_+$, then~$t$ is not a sum. Each~$u_j$
    is thus greater than~$t$, and induction implies that each~$u_j$ is greater
    than~$s$. Rule~$\sleq_+$ implies that~$s \sleq \sum_i u_i$.

  \item
    rule~$\sleq_\f$: we have~$t=\f t'$ and~$u=\f u'$, together with~$t'\sleq
    u'$. Because~$s\sleq t$, we also have~$s=\f s'$ with~$s'\sleq t'$. By
    induction hypothesis, we get that~$s'\sleq u'$, and thus that~$s\sleq u$ by
    rule~$\sleq_\f$.

  \item
    The rules~$\sleq_\x$, $\sleq_\C$, $\sleq_\st{\dots; \D; \dots}$,
    $\sleq_\Cm$ and~$\sleq_\D$ are all treated similarly.

  \item
    If~$t\sleq u$ comes from rule~$\sleq_{\Daimon1}$, reasoning is similar,
    as~$s\sleq t$ also comes from~$\sleq_{\Daimon1}$.

  \item
    The last case is when~$t\sleq u$ comes from~$\sleq_{\Daimon2}$, \ie
    when~$t=\Daimon t'$ and~$u$ doesn't start with~\Daimon. The premise of
    this rule is~$\Daimon t' \sleq \nf(\Daimon u')$. Because~$s\sleq \Daimon
    t'$, $s$ is necessarily of the form~$\Daimon s'$, and we get that~$\Daimon
    s'\sleq \nf(\Daimon u)$ by induction. We conclude that~$\Daimon s' \sleq u$
    by rule~$\sleq_{\Daimon2}$. \qedhere
\end{itemize}
\end{proof}

We can now put everything together to prove the main proposition of this
section. \begin{proof}[Proof of Proposition~\ref{prop:order_NF}] The order~$\leq$ is
generated inductively by reflexivity, transitivity, commutativity,
associativity, idempotence of~$+$, (multi) linearity,
contextuality,~$t\leq\Daimon t$, $t\leq\Zero$, $s+t\leq t$ and
the(in)equalities from Definition~\ref{def:order_F0}. The proof of
Proposition~\ref{prop:order_NF} is an induction on~$s\leq t$.

\begin{itemize}
  \item
    If~$s\leq t$ by reflexivity, \ie~$t$ syntactically equal to~$s$, then
    we have~$s\sleq t$.

  \item
    If~$s\leq t$ holds by transitivity~$s\leq u$ and~$u\leq t$. By induction
    hypothesis, we get that~$\nf(s)\sleq\nf(u)$
    and~$\nf(u)\sleq\nf(t)$. We thus get~$\nf(s)\sleq\nf(t)$ by
    transitivity of~$\sleq$ (Lemma~\ref{lem:inductive_order_transitive}).

  \item
    If~$s\leq t$ holds by commutativity, associativity, idempotence of~$+$ or
    (multi)-linearity, the result follows from rule~$\sleq_+$.

  \item
    Similarly, if~$s\leq t$ holds by contextuality, we get that~$s\sleq
    t$ by induction and contextuality of~$\sleq$
    (Lemma~\ref{lem:inductive_order_contextual}).

  \item
    If~$s$ is equal to~$\Daimon t$, we need to check that~$\nf(\Daimon t)
    \sleq \nf(t)$. This is obvious if~$t$ starts with a~$\Daimon$. Otherwise,
    we need to use rule~$\sleq_{\Daimon2}$: it is enough to show
    that~$\nf(\Daimon t) \sleq \nf(\Daimon\nf(t))$.
    Since~$\nf(\Daimon\nf(t))=\nf(\Daimon t)$, the result holds by reflexivity
    of~$\sleq$.

  \item
    If~$t=\Zero$, we have~$\nf(s)\sleq \Zero$ using rule~$\sleq_+$.

  \item
    If~$s=s'+t$, we have~$\nf(s)= \nf(s') + \nf(t) \sleq \nf(t)$ using
    rule~$\sleq_+$.

  \item
    If~$s \leq t$ using one (in)equality from Definition~\ref{def:order_F0}, we
    have that~$s$ reduces to~$t$ or that~$t$ reduces to~$s$. In either
    case,~$\nf(s)=\nf(t)$ so that~$\nf(s) \sleq \nf(t)$. \qedhere
\end{itemize}
\end{proof}

%%%>>>2

\subsection{Inductive order on normal forms in~\A}     %%%<<<2
\label{app:inductive_order-A}

We can extend~$\sleq$ to approximations:
\begin{defi}
  We extend Definition~\ref{def:inductive_order_NF} to normal forms of~\A by
  adding the following rules:
  \[
    \Rule{s\sleq t \qquad \nf\big(\coef{W}\delta\coef{0}t\big) = \coef{W'}t \qquad
    \coef{V} \leq \coef{W'}}{\coef{V} s\sleq \coef{W}\delta
    t}{\scriptscriptstyle\sleq_{\coef{}1}}
    \qquad
    \Rule{\coef{W} s \sleq \nf\big(\coef{0} t\big)}{\coef{W} s\sleq
    t}{\scriptscriptstyle\sleq_{\coef{}2}} \text{*}
  \]
  where, in rule~$\sleq_{\coef{}2}$, $t$ doesn't start with an approximation;
  and in rule~$\sleq_{\coef{}1}$,~$\delta$ is any sequence of destructors
  $\Cm$/\D.

\end{defi}
Note that approximations behave very similarly to~\Daimon.
Checking~$\sleq$ is still decidable, as the rules are syntax directed, and we
still have
\begin{lem}
  For all approximated terms~$s$ and~$t$ in normal form, if~$s\sleq t$,
  then~$s\leq t$.
\end{lem}
\begin{proof}
All the rules remain valid, and we only have to check that the two new rules
are correct.
\begin{itemize}

  \item For $\sleq_{\coef{}1}$, suppose that~$s\leq t$,
    $\nf(\coef{W}\delta\coef{0}t) = \coef{W'}t$ and~$\coef{V}\leq\coef{W'}$
    in~$\Coef$. We need to show that~$\coef{V} s\leq \coef{W}\delta t$.
    We have
    \[\begin{array}{rcll}
      \coef{W}\delta t
      & \geq &
      \coef{W}\delta\coef{0} t
      &\text{\footnotesize(contextuality, because $\coef{0}t\leq t$)}\\
      & \geq &
      \nf(\coef{W}\delta\coef{0} t)
      &\text{\footnotesize(terms decrease along reduction)}\\
      & = &
      \coef{W'} t
      &\text{\footnotesize(hypothesis)}\\
      & \geq &
      \coef{W'} s
      &\text{\footnotesize(contextuality, because $s\leq t$)}\\
      & \geq &
      \coef{V} s
      &\text{\footnotesize(because~$\coef{V} \leq \coef{W}$)}\\
    \end{array}\]

  \item For $\sleq_{\coef{}2}$, we have that~$\coef{W} s \leq \nf(\coef{0} t)$
    by hypothesis. We also know that~$\nf(\coef{0} t) \leq \coef{0}t \leq t$,
    so that we have~$\coef{W}\leq t$ by transitivity. \qedhere
\end{itemize}
\end{proof}
%%%>>>2

\subsection{Weak coherence on normal forms in~\A} %%%<<<2
\label{app:inductive_coherence-A}

\begin{lem}
  \leavevmode
  \begin{enumerate}
    \item 
      Define an inductive binary relation~$\sqcoh$ with:
      \begin{enumerate}

        \item
          $\x \sqcoh \x$,

        \item
          $\C u \sqcoh \C v$ iff $\Cm u \sqcoh \Cm v$
          iff
          $\D u \sqcoh \D v$ iff $\f u \sqcoh \f v$ iff $u \sqcoh v$,

        \item
          $\st{\D_1=u_1 ; \dots ; \D_k=u_k} \sqcoh \st{\D_1=v_1 ; \dots ;
          \D_k=v_k}$ iff $\forall i, u_i \sqcoh v_i$,

        \item
          $\Daimon u \sqcoh \Daimon v$ iff
          \begin{itemize}
            \item there is a sequence of destructors~$\delta$ s.t. $u = \delta u'$
              with~$u' \sqcoh v$,

            \item or there is a sequence of destructors~$\delta$ s.t. $v = \delta v'$
              with~$u \sqcoh v'$,
          \end{itemize}

        \item
          $u \sqcoh \coef{W}v$
          iff~$\coef{W} u \sqcoh v$
          iff~$u \sqcoh \Daimon v$
          iff~$\Daimon u \sqcoh v$
          iff~$\nf(\Daimon u)\sqcoh\nf(\Daimon v)$.

        \item
          In all other cases,~$u \not\sqcoh v$.

      \end{enumerate}

    \item For all terms in normal form~$u$ and~$v$, if~$u\coh v$,
      then~$u\sqcoh v$.

  \end{enumerate}
\end{lem}
\begin{proof}
We prove that~$u\sleq t$ and~$v\sleq t$ implies $u\sqcoh v$ by induction
on~$u\sleq t$ and~$v\sleq t$. Using Proposition~\ref{prop:order_NF}, this
implies point~(2) above.

\begin{enumerate}

  \item[(a)]
    If $u=v=\x$, then we obviously have~$t=\x$ and thus~$u\sqcoh v$.

  \item[(b)]
    If $u=\C u'$ and~$v=\C v'$, then we necessarily have~$t=\C t'$,
    with~$u'\sleq t'$ and~$v'\sleq t'$, which implies by
    Lemma~\ref{lem:order_NF} that~$u'\coh v'$. By induction, we have~$u'\sqcoh
    v'$, and thus~$\C u' \sqcoh \C v'$.

    The other cases with~$\Cm$, $\D$ and~$\f$ are treated similarly.

  \item[(c)]
    The case~$u=\st{\dots; \D_i=u_i; \dots}$ and~$v=\st{\dots; \D_i=v_i;
    \dots}$ is treated similarly.

  \item[(d)]
    If~$u=\Daimon u'$ and~$v=\Daimon v'$, we have~$\Daimon u'\sleq \nf(\Daimon
    t)$ and~$\Daimon v'\sleq \nf(\Daimon t)$. It implies that~$\nf(\Daimon t)$
    is of the form~$\delta_1\delta_2 t'$ with~$u'\sleq \delta_2 t'$
    and~$v'\sleq t'$ (or vice versa).

    This implies that~$\Daimon u' \sleq
    \Daimon\delta_2 t'$ and~$\Daimon v' \sleq \Daimon\delta_2 t'$, and thus
    that~$\Daimon u' \sqcoh \Daimon v'$ by induction.

  \item[(e)]
    If~$v=\Daimon v'$, and~$u$ not of the form~$\Daimon$, we have~$\Daimon
    v'\leq t$ and~$u\leq t$. This implies (with no need of the induction
    hypothesis) that

  \item[(f)]
    No other cases are possible. \qedhere

\end{enumerate}
\end{proof}
%%%>>>2

%%%>>>1

\section{Basic domain theory}   %%%<<<1
\label{app:domain}

Here are the definitions and basic results on domain theory that are used in
this paper. Individual references are given for readers who want additional
details / proofs.

\begin{defi}
\leavevmode
\begin{itemize}
  \item
    If~$(O, \leq)$ is a partial order, a subset~$X\subset O$ is
    \emph{directed} if it is non-empty and if every pair of elements of~$X$
    has an upper bound in~$X$. Important examples of directed sets are
    increasing chains~$x_0\leq x_1 \leq \dots$.

  \item
    A \emph{directed-complete partial order} (\emph{DCPO}) is a partial
    order~$(D,\leq)$ for which every directed set~$X$ has a least
    upper-bound~$\dirsup X$ in~$D$.

  \item
    An element~$k$ of a DCPO is \emph{compact} if whenever~$k\leq\dirsup X$,
    then~$k\leq x$ for some~$x\in X$. Compact elements are usually associated
    with a notion of ``finite approximation''.

  \item
    A DCPO~$(D, \leq, \dirsup)$ is \emph{algebraic} if for all~$x\in D$, we have
    \[
      x = \dirsup \{ k \ |\ \text{$k$ is compact and $k\leq x$}\}
    \]
    \ie if every element is the limit of its finite approximations.

  \item A function between DCPOs is continuous if it is monotonic and if it
    commutes with directed least upper bounds.

\end{itemize}
\end{defi}

An important tool in domain theory is the notion of ideal completion.

\begin{defi}
  An \emph{ideal} for the partial order~$(O, \leq)$ is a non empty directed
  set~$I\subset O$ that is downward closed: if~$y\in I$ and~$x\leq y$
  then~$x\in I$.

  The ideal completion of~$(O,\leq)$ is the set of ideals of~$(O,\leq)$
  ordered by inclusion.
\end{defi}

\begin{prop}\leavevmode
  \begin{enumerate}
    \item
      The ideal completion of a partial order is an algebraic DCPO.
      \cite[Proposition~2.2.22]{abramsky_jung:domain}

    \item
      The ideal completion of the compact elements of an algebraic DCPO~$(D,
      \leq, \dirsup)$ is isomorphic to~$(D,\leq,\dirsup)$.
      \cite[Proposition~2.2.25]{abramsky_jung:domain}

    \item
      The ideal completion has the following universal property: if~$D_1$
      and~$D_2$ are two DCPOs, then any monotonic function from the compact
      elements of~$D_1$ to~$D_2$ can be uniquely extended to a continuous
      function from~$D_1$ to~$D_2$. \cite[Corollary~2.2.26]{abramsky_jung:domain}
  \end{enumerate}
\end{prop}

The next lemma is straightforward.
\begin{lem}
  If~$D_1$ and~$D_2$ are DCPOs, then the set of continuous functions
  from~$D_1$ to~$D_2$ ordered pointwise is also a DCPO. We write~$[D_1\to
  D_2]$.
  % \footnote{Sadly, continuous functions between algebraic DCPOs needs not
  % be algebraic.}
\end{lem}

%%%>>>1

\section{Smyth power domain}   %%%<<<1
\label{app:smyth}

Here are the important facts about the Smyth power domain. More details and
proofs of the results given below can be found in \cite[Section~6.2.2
and~6.2.3]{abramsky_jung:domain}.
If~$D$ is an algebraic DCPO, the Smyth power domain can be described in
several equivalent ways:
\begin{enumerate}
  \item
    the free algebraic DCPO for the binary operation~``$+$'' with
    \[
      x+y = y+x \qquad
      (x+y)+z = x+(y+z) \qquad
      x+x = x \qquad
      x+y \leq x
    \]
    \cite[Definition~6.2.7]{abramsky_jung:domain}

  \item
    the ideal completion of the following order on finite sets of compact
    elements of~$D$:
    \[
      X \leq Y\qquad\text{iff}\qquad
      \forall y\in Y,\exists x\in X, x \leq_D y
    \]
    \cite[Proposition~6.2.12]{abramsky_jung:domain}

  \item
    the algebraic DCPO of all compact saturated for the Scott topology on~$D$,
    ordered by reverse inclusion. \cite[Theorem~6.2.14]{abramsky_jung:domain}

\end{enumerate}
The fact that~$x+y$ turns out to be the greatest lower bound of~$x$ and~$y$
(\cite[Proposition~6.2.8]{abramsky_jung:domain}) is interesting to note but
not important in this paper.

From that, we get the following.
\begin{itemize}

  \item
    From point~(3): the Smyth power domain of~$(D, \leq, \dirsup)$ is a set of
    formal sums of elements of~$D$.

  \item
    From point~(1): it contains all the finite sums of elements of~$D$.

  \item
    From point~(2): it is generated from finite sums of compact elements
    of~$D$.
\end{itemize}
In order to show that some infinite sum belongs the Smyth power domain, rather
than unfolding the definition of ``compact-saturated set for the Scott
topology'', we can simply show how this infinite sum is a limit of finite sums
of compact elements of~$D$.

\bigbreak
In practice, most sets from point~(3) are infinite, which is difficult to work
with. In the Scott topology, a set is saturated precisely when it is upward
closed for the order. We thus allow using non saturated sets and instead of
reverse inclusion, use the order
\[
  X \leq_{\text{Smyth}} Y
  \quad\text{iff}\quad
  X^\uparrow \supseteq Y^\uparrow
\]
where~$X^\uparrow$ is the upward closure, \ie~$\{z\ |\ \exists x\in X x\leq
z\}$. Unfolding the definition, we get
\[
  X \leq_{\text{Smyth}} Y
  \quad\text{iff}\quad
  \forall y\in Y, \exists x\in X, x \leq y
\]
which will serve as our definition of the order on the Smyth power domain.

To summarize, we have
\begin{cor}
  Given a domain~$D$, the corresponding Smyth power domain is obtained with:
  \begin{itemize}
    \item
      elements are sets of elements of~$D$, seen as formal sums,

    \item
      two such sets are ordered by~$S\leq T$ iff $\forall t\in T, \exists s\in S, s\leq_D t$
      (note that is only defines a pre-order),

    \item
      compact elements are precisely the finite sums of compact elements
      of~$D$,

    \item
      all finite sums are in the Smyth power domain,

    \item
      infinite sums are in the Smyth power domain if they can be obtained as
      directed limits of finites sums.

  \end{itemize}

\end{cor}

%%%>>>1

%% vim600:set foldmarker=<<<,>>> foldmethod=marker fileencoding=ascii spelllang=en spell: %%

\end{document}